# MODELING AND SIMULATION OF ADAPTIVE CRUISE CONTROL SYSTEM

**MESF 6950(C)**

**YUNCHENG JIANG**

**ID:20566438**

**SUPERVISOR: LILONG CAI**

**5/4/2019**

## Contents

**Abstract**

In this report, linear quadratic regulator is used to design adaptive cruise control system. In the regulator, Q and R parameters vary with time according to current traffic situations. Phase-plant method is used to give constraints on Q and R parameters, and coefficient descent method is applied to solve for the constrained optimization problem. Meanwhile, data based controller design method is also introduced in this paper, where the time vary vehicle dynamic parameters are no longer considered. Q-function, which consists of Markovian state and action penalty, is introduced to indicate the cost function. According to current traffic states, Q-function is generalized and minimized by directly using least error method, whose stability is ensured by nonlinear regression theory. Simulation is conducted and results show the advantages of using time varying parameter linear quadratic regulator over other controller discussed in this paper. Vehicle tests are also conducted to ensure the feasibility and efficiency of the controller.

## Introduction

Advanced Driver Assistance System (ADAS) has become a hot topic since 1990s. ADAS provides the drivers not only warning signals i.e. lane departure warning (LDW) signals when the vehicle departs from the current lane and forward collision warning signals (FCW) when there is danger of car crash, but also subjective vehicle control functions like Lane Keeping Assistant (LKA), and Adaptive Cruise Control (ACC) by controlling throttle valve, brake and steering wheel. ACC system is one of the most important active control functions in ADAS. Equipped with ACC, a vehicle can not only cruise at a preset speed, but also maintain a desired distance with a longitudinal preceding vehicle. ACC system, therefore, relax drivers in both of long-time highway driving and urban traffic congestion.

ACC system is a update of cruise control system(CCS). Unlike CCS which can only cruise at a preset speed, ACC system can not only cruise at a constant speed like CCS, it can also maintain a safety distance with the preceding vehicle by using sensors to detect preceding vehicles. ACC system consists of measurement sensors, controller, and vehicle actuators. Measurement sensors are used to detect vehicle surroundings. Typically, smart camera (which can measure distance and velocity) and millimeter-wave radar, by data fusion method, are used as measurement sensors, providing longitudinal vehicle information like relative distance and relative velocity to controller. Desired acceleration is calculated by controller to maintain a desired distance. Finally, vehicle actuators i.e. throttle valve and brake realize the desired acceleration.

In this report, the first method introduced in the design of ACC controller is 2 DOF robust controller. While due to the inertial characteristic of vehicle dynamics, robust control does not work well with ACC system. Then main focus is devoted to the design and optimization on Linear Quadratic Regulator (LQR). LQR has been applied to ACC upper controller design long ago[1]. In terms of real situation application, the advantage of using LQR controller is (1) the removal of what we consider in current MPC approaches to be tedious tuning process, that is, the control horizon N (the number of future controls moves in the current optimal control step[2]. (2) It is also shown that LQR is less computationally expensive than MPC. However, when applying LQR method in ACC controller design, engineers still need to face the tuning problem, but few well tackles with the LQR parameters tuning issue, deals with vehicle nonlinearity, parametric uncertainty, and measurement noise in real situations. Up till now, few researchers have put forward effective and feasible tuning methods in ACC controller design that uses LQR method. [3] used LQR method in ACC controller design, but the author directly chose constant Q and R without explaining how the constant parameters are determined. [4] also used LQR method. The author, however, goes only through several different Q parameters, by simulation comparison, to determine the most appropriate parameters. [5] noticed that the importance of driver comfortability increases with the increase of vehicle speed. The authors, therefore, fixed all parameters except the one corresponds to vehicle acceleration which indicates driver comfortability. By tuning this parameter in a constrained range, they tried to find the optimal choice under different vehicle speed. But this tuning method neither considered the effects of the other parameters, nor explained how parameter initialization is determined. In this paper, the author tried to optimize the design of LQR so as to improve system performance of ACC.

In the design and optimization of model based LQR controller, the main contribution of this paper is: (1) improvement of dynamics state space which considers time delay and gain $K$ effect. (2) Due to the real situations where there are measurement noise and process noise, the state can be inaccurately calculated or measured, the "optimal" result, accordingly, may be suboptimal. The authors, therefore, introduced LQG, where noise is simply considered as Gaussian white noise. By using LQG, real situation noise is greatly compensated by Kalman filter. (3) In order to better improve the parameter combination of Q and R parameters, the authors introduced time varying LQG where Q and R parameters vary with time according to current traffic situations. First, phase-plant method is used to give constraints on Q and R parameters based on current traffic situations, then coefficient descent method is applied to solve for the constrained optimization problem. In simulation analysis, automatically-tuned Linear Quadratic Regular(ALQG) has better performance over LQR and MPC, and it is also less computationally expensive than MPC. Real vehicle test results are also released to verify the feasibility of ALQG controller in real time application.

In the design and optimization of data based controller, the main contribution of this paper is: (1) introduction of Q function to illustrate cost for all actions with respect to each state. (2) nonlinear regression is applied to achieve optimal vehicle action based on current vehicle states. (3) regularization and feature scaling are used to ensure numerical stability of using linear regression.

This report is organized as follows. The vehicle dynamic model is described in section ***Vehicle Dynamics***. The weakness of some current vehicle dynamic models in use are also discussed. In section ***Model based upper controller design and optimization***, the design of ACC controller based on LQR and LQG methods are explored, ALQG method is given at the end of this section. Data based model is designed and optimized in section ***Data based upper controller design and optimization***. In section ***Lower controller design and optimization***, brake controller and throttle controller design are explored separately. In section xx, In section ***Switching logic***, switching logics between CCS and ACC, and switching logics between brake and throttle are discussed. In section ***Simulation and vehicle test***, the simulation results which compares between LQR, MPC, and ALQG are verified. The simulation results of data based model is also shown in this section. Real time vehicle test result is given at the end of this section.

## Related Work

The history of ACC dates back to 1970s. In 1971, an American company EATON has been devoted in ACC research. Later on, MITSUBI in Japan developed preview distance control(PDC) which is the first ACC system. Over the past 50 years. The development of ACC is three-folded. The first part is high-speed ACC, whose function mainly focus on highway driving. It can provide cruise control and distance maintenance functions. The second part is stop&go ACC system, which not only has the function of high speed ACC system, but also can be used in crowded urban areas where the host vehicle can "stop and go" so that it can follow the preceding vehicle even when there are traffic light ahead. The third part is what we call intelligent ACC, which takes into consideration of fuel economy, driver comfortability, and function robustness, etc. what's more, ACC system is still improving. For example, ACC system can combine with lane keeping assistance system(LKA) so that vehicle can automatically follow the preceding vehicle at a safety distance, it can also stay at the middle line of the current lane[6]. This function combination realizes both lateral and longitudinal vehicle control. ACC also combines with automatic emergency braking system, which means vehicle can automatically brake in emergency[7]. Some other example is that ACC is equipped with lane changing assistant system(LCA) which can automatically change line with the command of the driver [8].

## Environment Detection

As mentioned in previous section, ACC system consists of measurement sensors, controller, and vehicle actuators. Measurement sensors are used to detect vehicle surroundings. Typically, smart camera (which can measure distance and velocity) and millimeter-wave radar, by data fusion method, are used as measurement sensors, providing longitudinal vehicle information like relative distance and relative velocity to controller.

Recently, researches on environment detection mainly focus on data fusion of both millimeter-wave radar and smart camera, through which detection accuracy can increase. Some other hot topics include adaptive detection algorithm to adapt to curve driving scenarios, and multi-vehicle environment detection.

[9] uses millimeter-wave radar to detect preceding vehicle, and applied Kalman filter to denoise. In order to deal with the disadvantage of millimeter-wave radar that it cannot detect static objects, [10] tried to use data fusion method to fuse both millimeter-wave radar and smart camera so as to detect preceding static objects. [11] and [12] focus on tracking algorithm of curve driving scenarios. [13] put forward a multi-vehicle environment detection method, which models the surroundings based on the information detected by millimeter-wave radar, smart camera, and lidar. [14] also tried to select from multi-vehicle situations the primary target. The scheme of primary target selection algorithm is as follows:

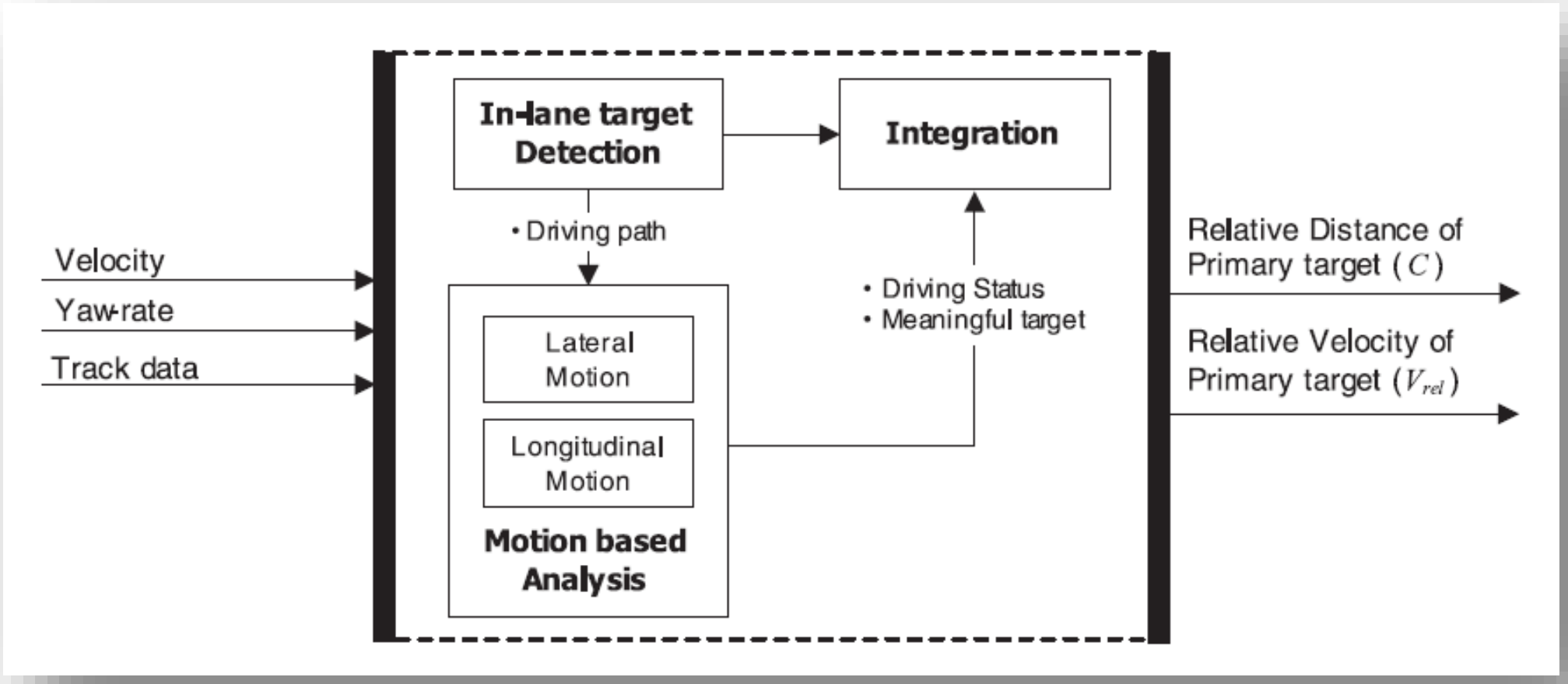

Fig.1 The scheme of primary target selection algorithm

Moreover, due to the complexity of vehicle surroundings: bad weather, vehicle and pedestrian cutting in, curve driving, channels, etc. sensors have to deal with multiple unpredictable situations, and even with data fusion method, there are still many works to be done to ensure the robustness and accuracy of sensor detection.

## Headway Time

As mentioned in previous section, the host vehicle should maintain a safety distance with the preceding vehicle. The safety distance should consider both traffic efficiency and safety, which means it cannot be too long or too short.

Many models have been put forward in determining safety distance. [15] made a summary of three types of headway time models, which are as follows:

$$D_{S,brake} = v_f t_d + \frac{v_i^2}{a_{fmax}} + d_0 \tag{1}$$

$$D_{S,h} = v_f t_h + d_0 \tag{2}$$

$$D_{s,pre} = v_{rel} t_g - \frac{a_p t_g^2}{2} + X_{lim} \tag{3}$$

Where $D_{S,brake}$ is desired distance based on the braking process; $v_f$ is host vehicle velocity; $t_d$ is time delay of drivers and vehicle braking system; $a_{fmax}$ is the maximum acceleration that host vehicle can achieve; $d_0$ is the distance between host vehicle and preceding vehicle when the host vehicle stops; $D_{S,h}$ is desired distance based on headway time; $t_h$ is headway time; $d_0$ is a constant distance to ensure two vehicles have a minimum distance; $D_{s,pre}$ is desired distance based on driver prediction; $v_{rel}$ is relative velocity; $t_g$ is prediction time; $a_p$ is preceding vehicle acceleration; $X_{lim}$ is the limit of the distance that drivers can accept.

The above three different safety distance have different characteristics:

1. The first safety distance model give safety to first priority. By using braking process as dynamic model, the safety distance is calculated in a inverse way so that it is guaranteed that when the vehicle stops, there will be a constant distance remained.
2. The second safety distance model is relatively much easier, but still very effective and reasonable. Actually, in this paper, we, for the convenience of calculation and simulation, we simply choose this safety distance model, which takes into consideration of only host vehicle velocity.
3. The third safety distance model is based on driver prediction; therefore, it can be more efficient, but due to the variation of different drivers, some of the parameters can be hard to determine.

We should notice an important parameter: headway time in the second safety distance model. It can be further classified as changeable headway time[16] and unchangeable headway time[17]. Unchangeable headway time is simple, yet it cannot adapt to different traffic situations. Changeable headway time, on the other hand, is more adaptive to different traffic situations while some more parameters are pending for tuning. [18] and [19] considers host vehicle velocity and relative velocity, giving two different changeable headway time which are as follows:

$$t_h = h_2 + h_3 v_f \tag{4}$$

$$t_h = t_0 - c_v v_{rel} \tag{5}$$

Where $h_2, h_3, t_0$, and $c_v$ are parameters to be tuned, and $v_{rel}$ is relative velocity. As can be seen in the equation: headway time is linear to relative velocity.

Furthermore, safety distance can be both linear[20] and nonlinear[21]. For example, [22]used highway driving data to fit the safety distance in exponential form. [23] also made safety model in a quadratic form as follows:

$$d_{des} = a v_f^2 + b v_f + c \tag{6}$$

Where $\mathrm{d_{des}}$ is desire distance; a, b, and c are parameters to be tuned.

Since a, b, and c do not have physical meaning in the equation, [3] suggested that we do a Taylor expansion at average velocity, and the equation is rearranged as:

$$d_{des} = r v_f \left(v_f - v_{fmean}\right) + \tau_h v_f + d_0 \tag{7}$$

Where r is a coefficient for the quadratic term; $\tau_h$ is headway time; $v_{fmean}$ is the average velocity of the host vehicle.

To some degree, a quadratic form safety distance can more explain the traffic situations indicating the driver following expectations. While the problem is, again, that the parameter tuning process can be time consuming.

## Controller Design

The objective of ACC controller is to realize cruise control and adaptive control(maintaining a desired distance away from the preceding vehicle). Based on the variation of different traffic situations, we can classify the traffic scenarios as: steady following, preceding vehicle emergency braking, preceding vehicle emergency accelerating, vehicle cutting in, vehicle cutting out, and host vehicle emergency braking. Apparently, it is required that vehicle equipped with ACC should be able to deal with the mentioned scenarios. And it is also required that the switching process should be smooth, making the driver as comfortable as possible, and driver safety should meanwhile be guaranteed.

Usually, ACC system is designed in a structure containing multilayer controllers: upper controller and bottom controller which is shown in Fig.2. Upper controller calculates desired acceleration based on current road situations, and lower controller translates the desired acceleration into brake and throttle so that vehicle actuator can realize.

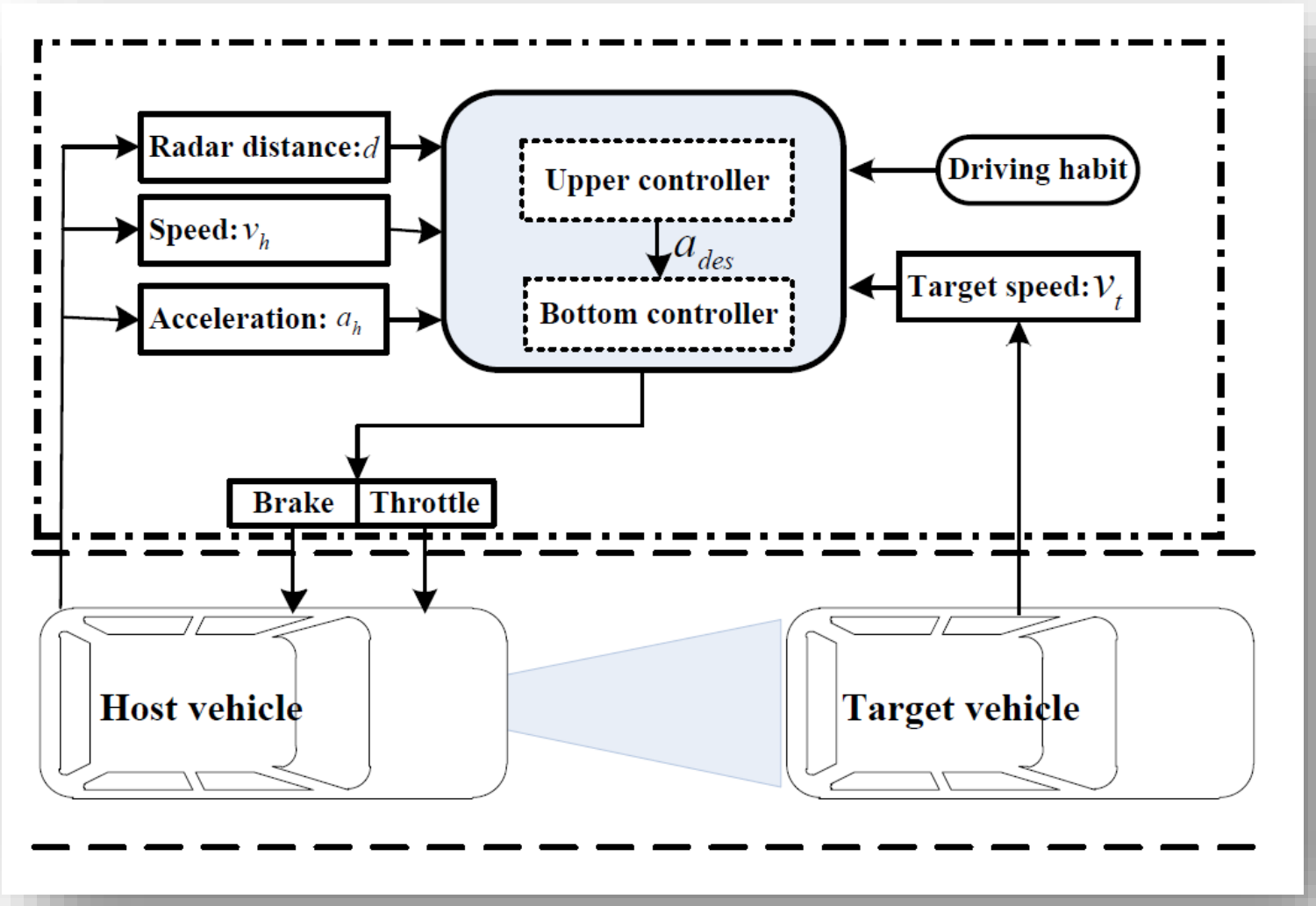

Fig. 2 Structure of ACC system

### PID Controller

Many optimization methodologies have been applied to different controllers to ensure vehicle safety, increase traffic efficiency, and improve driver comfortability. One of the most popular controllers is Proportional-Integrate-Derivative(PID) controller [24], which use relative distance, relative velocity and relative acceleration as inputs, and desired acceleration as output. As the most popular controller, PID is the first method applied in cruise control system design. [25] used PID to design ACC upper controller, through which distance error and relative velocity can be adjusted, and by using pole placement method, appropriate parameters are designed. [26] combined PI controller with feedforward controller in ACC upper controller design. Response time was reduced while the robustness of the system was affected. [27][28] used fuzzy-PID method to optimize the PID controller, through which error between desired distance and real distance is better minimized. But the weakness of using PID controller is the large amount of time needed for parameter tuning, and PID cannot predict the future motion of vehicle. Therefore, due to the vehicle inertial, the control always has a phase lag between the motion of the vehicle. And according to vehicle test, PID controller is not very robust to external perturbation.

### Optimal Control

Optimal control theory deals with the problem of finding a control law for a given system such that a certain optimality condition is achieved. In ACC system, the goal of the controller is to achieve the best vehicle following performance, and meanwhile guarantee driver comfortability, vehicle safety, traffic efficiency and fuel economy. [11] converted ACC control problem into an finite time optimal control problem, and by using dynamic programming, the author tried to find the optimal feedback control rule. [29] picked distance error and relative velocity as states, and host vehicle acceleration as input to design a LQR controller, so that the cost function is minimized.

### Fuzzy Logic Control

Some researchers also used fuzzy logic as a optimization method to design ACC controller[30][31]. Based on the drivers driving habits, a bunch of rules are generated to describe the relation between inputs and outputs as shown in Fig. 3. Some other more advanced model include neuro-fuzzy control[32] which combined single neural network with fuzzy logic to improve fuel economy. One of the advantages of applying fuzzy logic method is that it is suitable for multi-parameters and nonlinear control problems. Another advantage is that the transfer function of the system is not necessary, because it makes use of human empirical control reaction to the surrounding environment. Many control methods strongly depend on the accuracy of transfer function, once the transfer function is inaccurate, the system performance of that controller may decrease. Meanwhile, it is not guaranteed that the vehicle dynamic parameters do not change over time. Therefore, data based method like fuzzy logic control can be somehow promising. On the other hand however, a large amount of data is needed to determine the fuzzy logic rules, which again costs much time. Also, since driver driving habits are needed for generating control rule, the variation of driving habits of different drivers becomes a big issue to deal with in rule design.

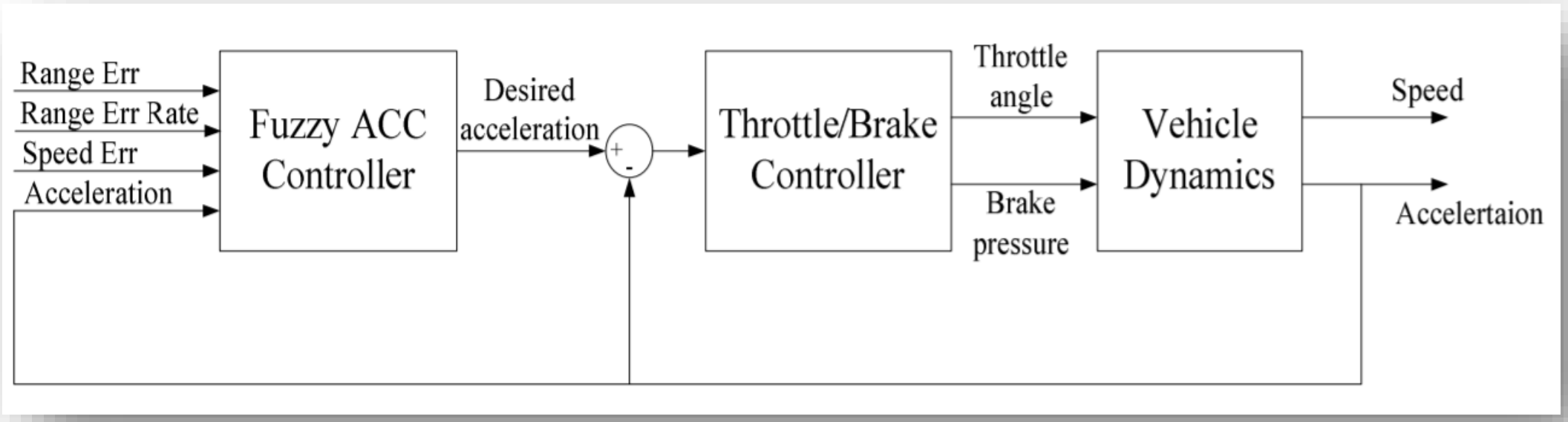

Fig. 3 Fuzzy logic based ACC system

### Model Predictive Control

Model predictive control (MPC) is an advanced method of process control that is used to control a process while satisfying a set of constraints. The main advantage of MPC is the fact that it allows the current timeslot to be optimized, while keeping future timeslots in account. This is achieved by optimizing a finite time-horizon, but only implementing the current timeslot and then optimizing again, repeatedly, thus differing from Linear-Quadratic Regulator (LQR). Also, MPC has the ability to anticipate future events and can take control actions accordingly. PID controllers do not have this predictive ability. [33] introduced adaptive Neuro-Fuzzy Predictive Control(ANFPC) where predictive control law is derived by Fuzzy Neural Networks (FNNs) to optimize predictive control. [34] used combined- MPC in controller design, and [35] applied nonlinear MPC in vehicle following controller design. But MPC is computationally expensive, and the tuning process is still time consuming compared with other controllers.

### Data based control

Except for model based controller design and optimization, researchers are putting more efforts into data based controller design and optimization. As driving habits change among drivers and over time in ACC system, an intelligent ACC system should adapt to different driving habits. Otherwise a driver would intervene even in situations that ACC is able to manage, typically when it could not meet the driver’s expectations. Therefore, an ACC system which takes the individual driving habit into consideration would, to a great extent, contribute to its attraction and put into commercial production. Most of the traditional controller design is based on fixed driving habit model, and do not consider the change of driving habits, while data based controller model is designed to adapt to varying driving habits. Therefore, much work has also been done in this field.

## Vehicle Dynamics

ACC system design is illustrated in Fig. 4. ACC controller can be divided into upper controller and lower controller. Upper controller calculates desired acceleration based on current road situations, and lower controller translates the desired acceleration into brake and throttle so that vehicle actuator can realize. In Fig.1, the preceding vehicle is denoted by lowercase *p*, and the following vehicle is denoted by lowercase *f*. The inputs of upper controller are following vehicle velocity $v_f$, following vehicle acceleration $a_f$, real distance between two vehicles *d*, desired distance $d_{desire}$, preceding vehicle velocity $v_p$, and preceding vehicle acceleration $a_p$. The output of upper controller is desired acceleration $a_{f,des}$. The input of lower controller is desired acceleration $a_{f,des}$, and it is translated to brake and throttle as outputs. The objective of ACC system is to minimize the error between desired distance and real distance, relative velocity, and acceleration of the following vehicle. For the sake of simplicity, desired distance with fixed headway time $\tau_h$ and constant safe distance $d_0$ are used:

$$d_{desire} = \tau_h v_f + d_0 \quad (8)$$

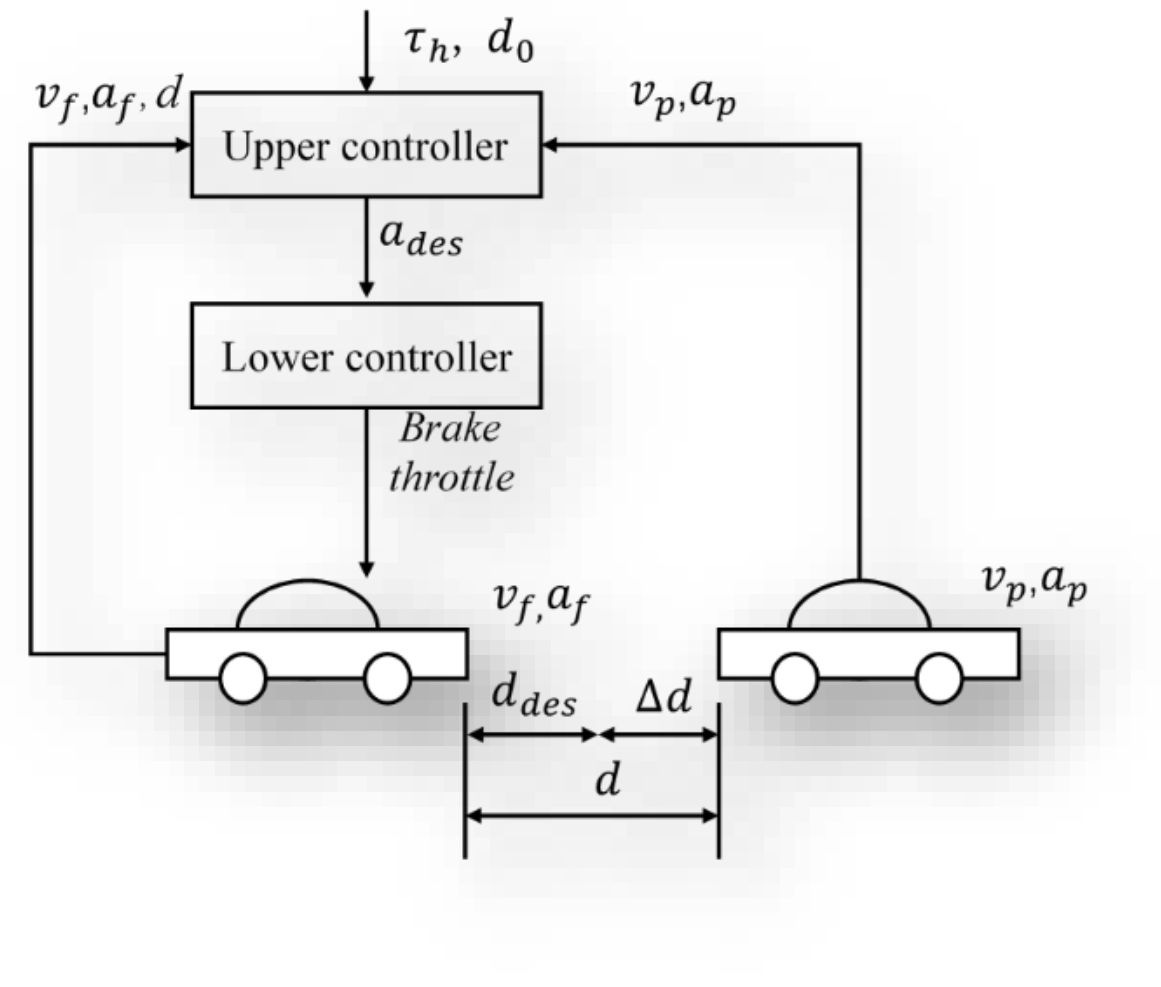

Fig. 4: structure of ACC system

This paper deals mainly with the ACC upper controller optimization, the lower controller and vehicle, therefore, is combined together and simplified as a first order system with time constant $T_L$ and gain $K_L$. Time constant $T_L$ and gain $K_L$ well simulate real vehicle dynamics which always has a time delay and different between real acceleration and desired acceleration. The relationship between desire acceleration $a_{f,des}$ calculated by upper controller, and real acceleration $a_f$ actuated by lower controller and vehicle is as follows:

$$a_f = \frac{K_L}{T_L S+1} a_{f,des} \tag{9}$$

To rewrite the vehicle dynamic system in a state space form, distance error $d_{error}$, relative velocity $v_{rel}$ and acceleration of following vehicle $a_f$ are chosen as the three states, Distance error is a parameter indicating safety. Relative velocity is directly related to traffic efficiency, and acceleration of following vehicle denotes driver comfortability. They are as follows:

$$d_{error} = d_{desire} - d \tag{10}$$

$$v_{rel} = v_p - v_f \tag{11}$$

$$a_f = \frac{d}{dt} v_f \tag{12}$$

then the state space is written as follows:

$$\dot{x} = Ax + Bu + \Gamma w \tag{13}$$

where

$$\mathrm{x} = \begin{bmatrix} d_{error} \\ v_{rel} \\ a_f \end{bmatrix}, \mathrm{A} = \begin{bmatrix} 0 & -1 & \tau_h \\ 0 & 0 & -1 \\ 0 & 0 & -\frac{1}{T_L} \end{bmatrix}, \mathrm{B} = \begin{bmatrix} 0 \\ 0 \\ \frac{K_L}{T_L} \end{bmatrix}, \Gamma = \begin{bmatrix} 0 \\ 1 \\ 0 \end{bmatrix}, \; w = a_p$$

Preceding vehicle acceleration $a_p$ is seen as disturbance to the system, and the output $y$ is the three states defined above:

$$y = Cx \tag{14}$$

where

$$C = \begin{bmatrix} 1 & 0 & 0 \\ 0 & 1 & 0 \\ 0 & 0 & 1 \end{bmatrix}$$

It is noted that, in previous paper, researchers also used similar state space form before applying other controller design methods, but some of them made obvious mistakes in developing vehicle dynamics, thus resulting unreasonable and unreliable results. [36] chose the same states as this paper, i.e. distance error $d_{error}$, relative velocity $v_{rel}$ and acceleration of following vehicle $a_f$. But in developing control matrix *A*, the author made obvious mathematic mistakes. [37] also used LQR methods in upper controller design. But the author chose only two states, distance error $d_{error}$, and relative velocity $v_{rel}$. This vehicle dynamic model considers neither the time delay and gain between desired acceleration and actual acceleration, nor driver comfortability which can be represented by acceleration change rate of following vehicle $\dot{a}_f$.

As in real cases, the controller design is usually based on discrete state space, the continuous form, therefore, by using zero-order-hold method, is converted into discrete form with sample time T:

$$x_{k+1} = Gx_k + Hu_k + Lw_k \tag{15}$$

## Model based upper controller design and optimization

### Robust control

In controller design, we may first choose controller which are easy to design, and its robustness is ensured. Therefore, robust control by pole placement sounds very reasonable right now. it is not only easy to design by placing appropriate poles, but also its robustness is ensured so that the controller is stable. Therefore, a 2 DOF feedback controller is design with structure as follows:

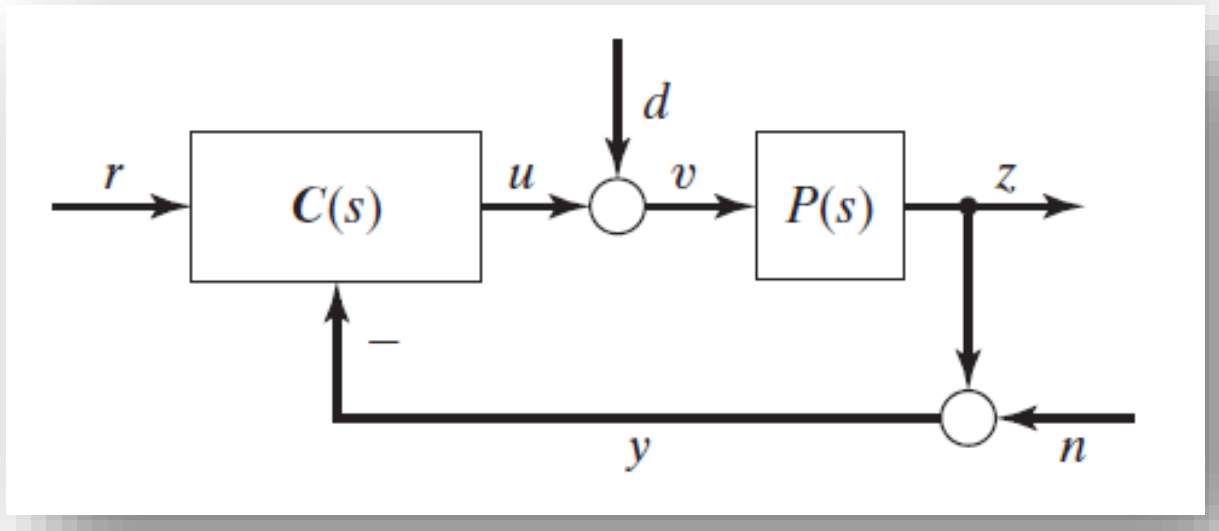

Fig.5 2 DOF feedback control system

Before we apply pole placement method, the prerequisite is that the system is controllable, which mean that the controllability matrix made by A and B are nonsingular:

$$rank[B \quad BA \quad BA^2]\text{=3, where } A = \begin{bmatrix} 0 & -1 & \tau_h \\ 0 & 0 & -1 \\ 0 & 0 & -\frac{1}{T_L} \end{bmatrix} \text{ and } B = \begin{bmatrix} 0 \\ 0 \\ \frac{K_L}{T_L} \end{bmatrix}.$$

Fortunately, the controllability matrix has full rank, and therefore, the pole placement method can be applied. Meanwhile, the choice of poles is arbitrary as long as we place poles on left hand side of y axis.

Here, the 2DOF controller shown in Fig. 5 is given as ***C****(s)* = $[C_1(s)\ C_2(s)]$, which consist of two parts: the first part $C_1(s)$ is feedforward part, and the second part $C_2(s)$ is feedback part. When considering the 2DOF control system above, we first assume that input *r(t)* is step signal, which denotes the desired distance, and *d(t)* and *n(t)* are impulse signals which denote noise. The form of the noise n(t) is assumed to be a unit impulse, and the form of the noise d(t) is assumed to be either impulse signal or unit step signal. Also, we assume that the vehicle dynamic plant P should be at least type 1, and therefore it is strictly proper:

$$p(s) = \frac{b(s)}{a(s)} = \frac{b_1 s^{n-1} + \ldots + b_n}{a_0 s^n + a_1 s^{n-1} \ldots + a_{n-1} s} \tag{16}$$

Where a(s) and b(s) are coprime, $a_0 > 0$, and $b_n \neq 0$.

Since the purpose of using 2DOF control is to minimize the distance error between the desire distance and real distance, the energy tracking error is given by:

$$\|e(t)\|_2^2 = \|r(t) - z(t)\|_2^2 \tag{17}$$

We also add control input u(t) into the system performance index.

$$\left(\|e(t)\|_2^2 + \|u(t)\|_2^2\right)\Big|_{\substack{r(t)=\delta(t)\\ d(t)=0\\ n(t)=0}} + \left(\|e(t)\|_2^2 + \|u(t)\|_2^2\right)\Big|_{\substack{r(t)=0\\ d(t)=\delta(t)\\ n(t)=0}} + \left(\|e(t)\|_2^2 + \|u(t)\|_2^2\right)\Big|_{\substack{r(t)=0\\ d(t)=0\\ n(t)=\delta(t)}} \tag{18}$$

Then Gang of six is given by:

$$\begin{bmatrix} 1 - P(s)C_1(s)S(s) & -P(s)S(s) & T(s) \\ C_1(s)S(s) & -T(s) & C_2(s)S(s) \end{bmatrix}$$

Where the sensitivity matrix and complementary sensitivity functions are given by

$$S(s) = \frac{1}{1+P(s)C_2(s)},\ T(s) = \frac{P(s)C_2(s)}{1+P(s)C_2(s)}$$

By using Parseval theorem, the cost function is given by

$$J = \left\|(1 - P(s)C_1(s)S(s))\frac{1}{s}\right\|_2^2 + \left\|C_1(s)S(s)\frac{1}{s}\right\|_2^2 + \left\|-P(s)S(s)\frac{1}{s}\right\|_2^2 + \|-T(s)\|_2^2 + \|T(s)\|_2^2 + \|C_2(s)S(s)\|_2^2 \qquad (19)$$

The optimal problem is to find the 2DOF controller ***C****(s)* = $[C_1(s)\ C_2(s)]$ that minimizes cost function *J*. And the algorithm is well developed [38].

***Algorithm*** **(design of the optimal 2DOF controller)**

***step 1***: find a stable polynomial d(s) such that:

$$a(-s)a(s)+b(-s)b(s) = d(-s)d(s);$$

***step 2***: the feedback part of the optimal controller $C_2(s) = \frac{q_{2(s)}}{p(s)}$ has the same order as the plant and is the unique strictly proper pole placement controller with

$$a(-s)a(s) + b(-s)b(s) = d(s)^2;$$

And the feedforward part of the optimal controller is $C_1(s) = \frac{d(s)}{p(s)}$;

***step 3*** *the optimal performance index is given by*

$$J^* = \left\|(1 - P(s)C_1(s)S(s))\frac{1}{s}\right\|_2^2 + \left\|C_1(s)S(s)\frac{1}{s}\right\|_2^2 + \left\|-P(s)S(s)\frac{1}{s}\right\|_2^2 + \|-T(s)\|_2^2 + \|T(s)\|_2^2 + \|C_2(s)S(s)\|_2^2$$

According to the algorithm, the robust controller is developed as follows:

$$\boldsymbol{C}(s) = [C_1(s)\ C_2(s)] = \left[\frac{s^3+0.6582s^2+0.2064s+0.0214}{s^3+1.1731s^2+0.6785s+1.8137} \quad \frac{-10s^2+0.2928s+0.0214}{s^3+1.1731s^2+0.6785s+1.81137}\right],$$

and $\boldsymbol{P}(s) = \frac{0.12s+0.02143}{s^3+0.1429s^2}$

The simulation model is designed as follows:

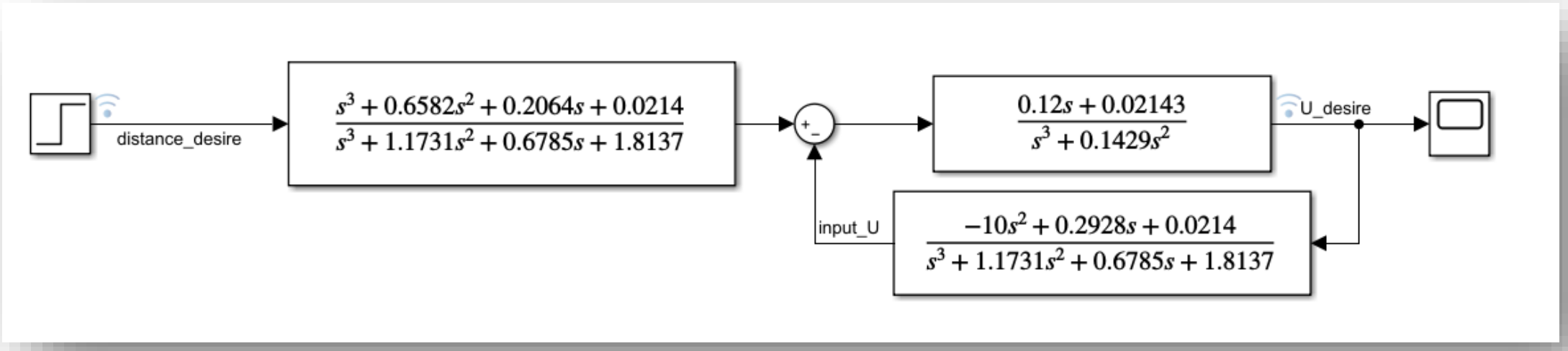

Fig. 6 Robust control system simulation structure

As can be seen in the simulation result Fig.7 that even though the robustness of robust controller is ensured theoretically, the system performance is not satisfactory. When the input is a desired distance at 20m, the system finally can asymptotically track the desired distance, but it takes too much time to converge. In real cases, the desired distance can vary with time frequently, if the response time is too big, the system may not perform well in real cases. One of the reason why the system does not perform well is that the same weights are put to distance error and input u, yet we can give different weight to the two variables, which will be shown in the next part, by introducing weight parameters, more degree of freedom is allowed in controller design, hence the design becomes more flexible. But again, similar to the tuning problem when using LQR, the problem is that it is time consuming to decide how the weights are chosen. Therefore, we just given up using robust control theory to design the ACC upper controller, and instead, we introduce a similar optimal design method LQR which will be explored in detail in the next section.

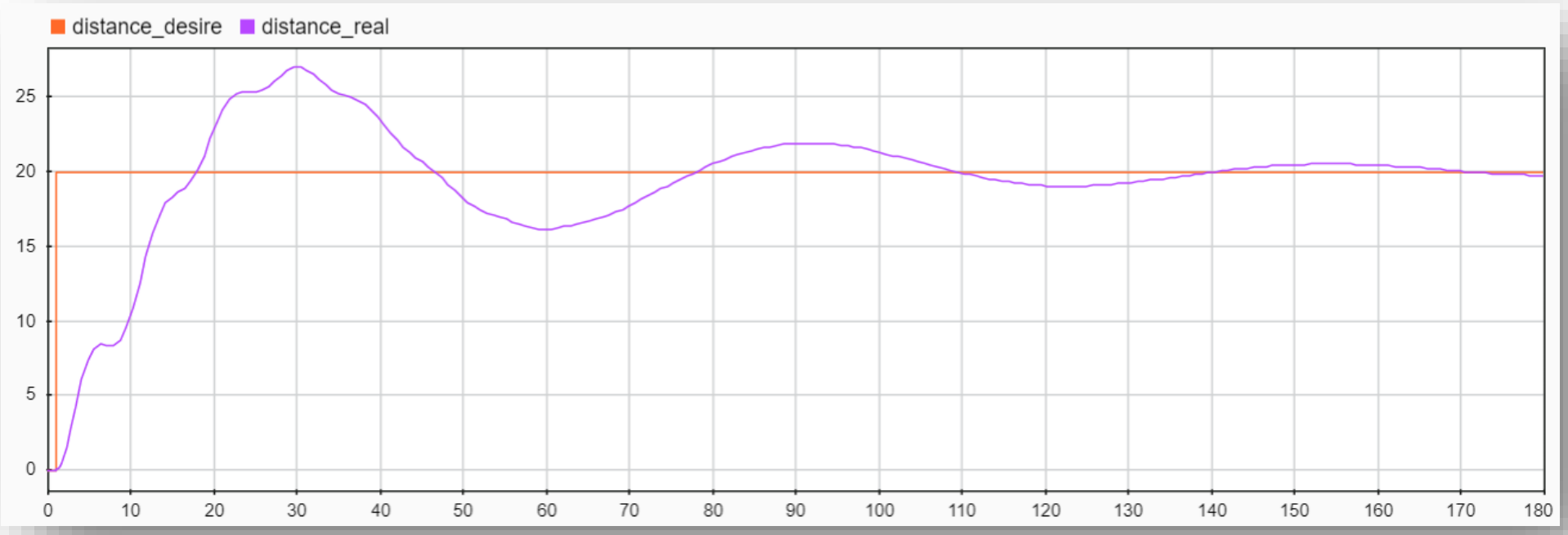

Fig. 7 Simulation result of robust controller

## Linear Quadratic Regulator

In real cases, the controller is designed in discrete form. The continuous state space is therefore discretized, by zero order hold method, into discrete form with sampling time *T*:

$$x_{k+1} = f(x_k, u_k, w_k) \tag{20}$$

which is supposed to have equilibrium state at $x = 0$, $u = 0$, and $w = 0$. In order to simplify this nonlinear problem, the system is linearized at equilibrium state, and rewritten as:

$$\tilde{x}_{k+1} = A\tilde{x}_k + Bu_k + g_k \tag{21}$$

In (9), by checking the observability matrix of $A$ and $C$, system states $\tilde{x}$ are ensured observable and measurable (full rank). System inputs are constrained in a safe range. According to national regulation on ACC system, the input $u$ is constrained in a range of [-0.25$g$,0.25$g$]. Process noise $g$ is considered as Gaussian white noise with zero mean.

After linearization, the problem of search for an optimal feedback controller can be seen as a linear quadratic optimization problem. The control objective, as mentioned in previous section, is to minimize the error between desired distance and real distance, relative velocity, and acceleration of the following vehicle. Even though the cost function, theoretically, is infinite horizon, a finite yet long enough time period is chosen for feasibility of cost function calculation in real cases. The cost function, therefore, in discrete form, is given:

$$J = \frac{1}{N}\sum_{k=0}^{N-1}[x^T(k)Qx(k) + u^T(k)Ru(k)] \tag{22}$$

where Q matrix is positive semi-definite, and R matrix is positive definite. The two weighting matrices are written as:

$$\mathrm{Q} = \begin{bmatrix} \rho_1 & 0 & 0 \\ 0 & \rho_2 & 0 \\ 0 & 0 & \rho_3 \end{bmatrix}, \ \mathrm{R} = [r]$$

The optimal output is then written in the form of $u = -Kx$, where $K$ can be obtained by:

$$K = R^{-1} B^T P \tag{23}$$

$P$ can be obtained by solving the famous Algebraic Riccati Equation (*ARE*), which is written as:

$$A^T P + PA + PBR^{-1}B^T P + Q = 0 \tag{24}$$

Once the Q and R matrices are determined, optimal $u$ can be achieved. Only when the states are perfectly measured, the optimal $u$ is globally optimal. In real scenarios, however, due to the process gaussian noise $g$, the optimal $u$ is only locally optimal. In order to compensate for the process noise effect, the input states of upper controller should be well estimated and filtered by Kalman Filter(KF).

As shown in Fig. 8, state $y$ is first filtered by Kalman filter and then be future used. LQR with KF is what we call LQG controller. How Kalman Filter works is shown in the next subsection.

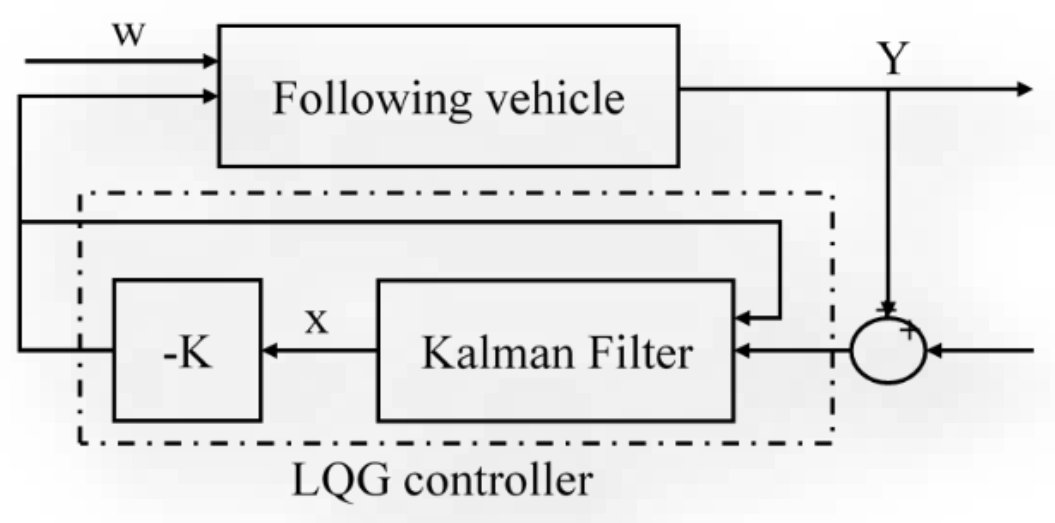

Fig.8: Structure of LQG controller

## Kalman Filter

Based on the state space of dynamic model, initial states and covariance matrices at previous step *k-1*, a new state and covariance matrix can be updated at current step $k$, which is written as:

$$x_{kp} = Ax_{k-1} + Bu_k + w_k \tag{25}$$

Then Kalman gain $K$ can be calculated with process covariance, and measurement covariance matrices, which is written as:

$$P_{kp} = AP_{k-1}A^T + Q_K \tag{26}$$

$$K = \frac{P_{kp}H}{H \cdot P_{kp}H^T + R} \tag{27}$$

where $P$ is process covariance, Q is process noise covariance, R is sensor noise covariance matrix (measurement error).

Then new states $X_k$ can be updated by using Kalman gain $K$, previous step states $X_{kp}$, and measurement input $Y$, which is written as:

$$Y_k = CX_{km} + Z_k \tag{28}$$

$$X_k = X_{kp} + K(Y - HX_{kp}) \tag{29}$$

Finally, process covariance matrix $P_k$ is updated by Kalman gain $K$ and previous step process covariance matrix $P_{kp}$, which is written as:

$$P_k = (I - KH)P_{kp} \tag{30}$$

With updated states $X$ and process covariance matrix $P_k$, the same filtering process can be repeated continuously.

### Real scenario risks

Till now, it seems that by applying LQG method is reasonable, and it is supposed to work well with real scenarios. But one of the most important things we have neglected in previous analysis is the real cases constraints on acceleration and change rate of acceleration, which may have great effect on the system performance on LQR controller. A simple analytical simulation is conducted so that the effect of constraints is explored. Assuming that there is no noise and the vehicle dynamics is developed accurately, LQR and MPC controllers are designed based on the mentioned rules above. Fig. 8.1 shows system performance of unconstrained LQR.

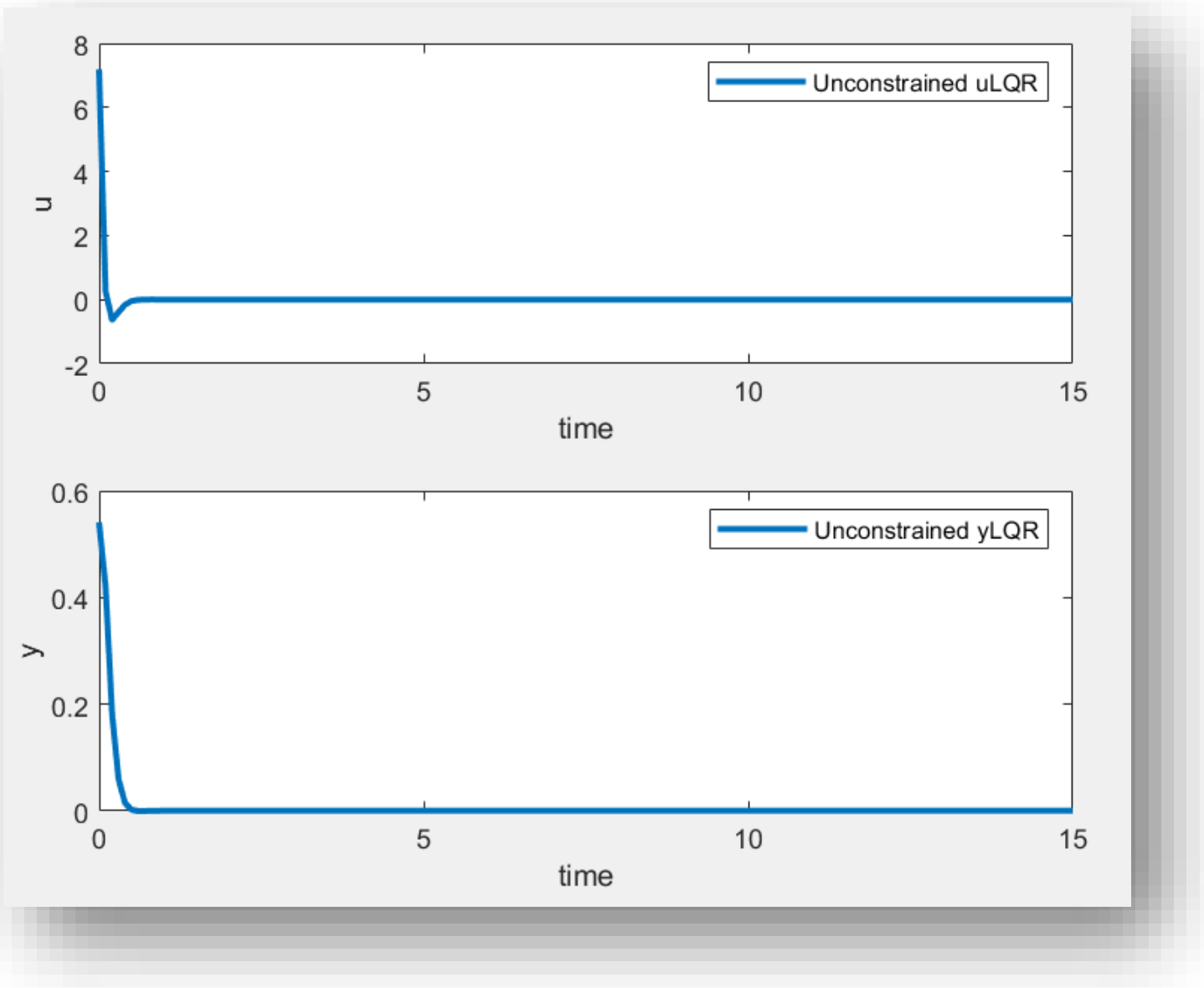

Fig. 8.1 Unconstrained LQR system performance

It is shown in the figure that when there is no constraints on vehicle acceleration, the maximum number can reach upper to nearly 0.8 g, which in real cases is impossible. Even though the distance error figure turns out to be very satisfactory in which it takes no more than 1 sec to re-stabilize the system, the result, however, makes no sense.

Then it is reasonable to add a boundary limit to the vehicle accelerate. Let us assume that a boundary limit [-0.1g, 0.1g] is added. And MPC controller is also designed so that we can make a comparison between constrained LQR and constrained MPC controllers. Fig. 8.2 and Fig. 8.3 show the comparison between the two controllers.

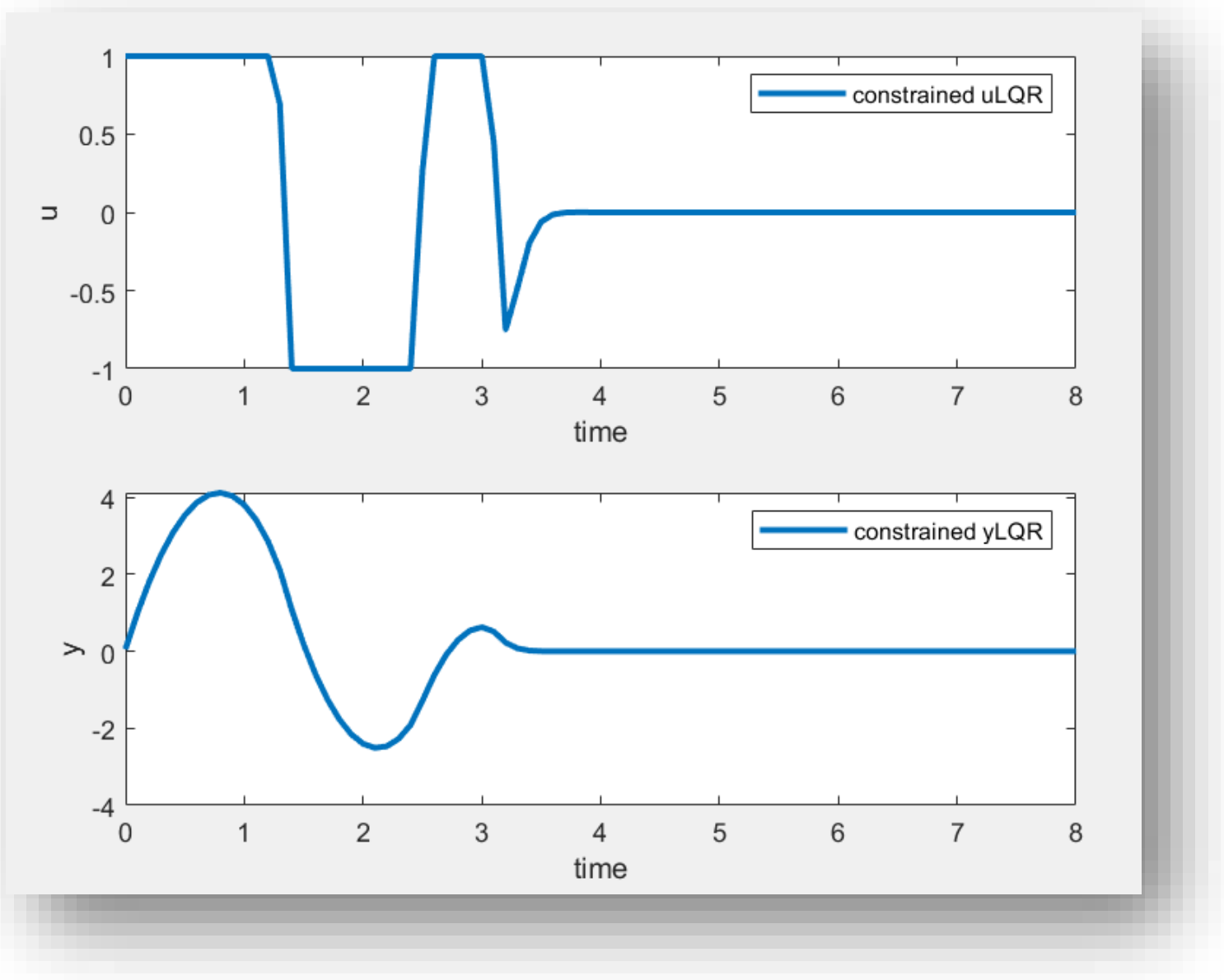

Fig. 8.2 Constrained LQR system performance

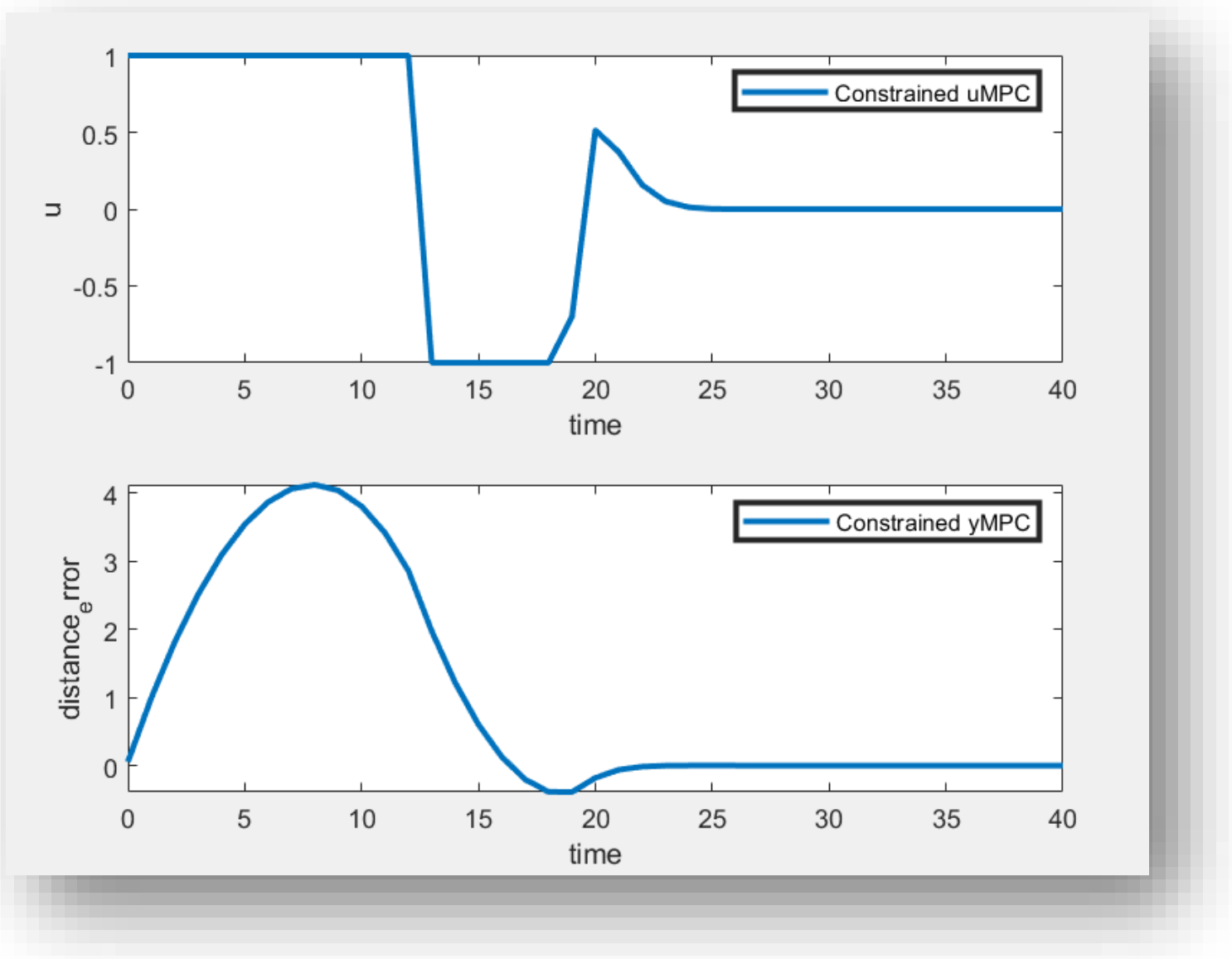

Fig. 8.3 Constrained MPC system performance

From Fig. 8.2 and Fig. 8.3, the effect of constraints is obvious, both LQR and MPC controllers have constrained input, and therefore, it takes more time for the system to re-stabilize again. There is not much difference in input figures, since both LQR and MPC have constrained inputs periods. While when looking through the distance error figures, MPC may have a little bit advantage over LQR. As can be seen from the distance error figures, LQR varies in a much wide range that MPC. Even though the upper limit of both controllers reaches to 4, the lower limit is -2 and -0.5 for LQR and MPC respectively. Apparently, MPC controller is more robust to disturbance, while the weakness of MPC is that it is updated online, which means it is more computationally expensive.

We are not in a dilemma: constrained LQR does not have satisfactory system performance, and the feasibility of constrained MPC is not guaranteed in real scenario application. But remember that we are now using fixed parameter LQR controller. What if we use time varying parameter LQR? It is not guaranteed that we can ensure the stability of time varying parameter LQR, since there is still not complete theory and strict proof time such method, but as long as we successively apply it into real cases, and the results turn out to be satisfactory, we can, at least, claim that the time varying parameter LQR works well with our ACC system on this vehicle.

### Phase-plant method

In most cases, we will choose constant Q and R parameters in LQR design. However, in ACC controller design, as mentioned above, in order to achieve satisfactory following performance without feeling uncomfortable, the parameters should be appropriately designed and tuned. And we are trying to use time varying parameter LQR method. Before we design the LQR controller, it is emphasized that the weights $\rho_1, \rho_2, \rho_3$ have the following effects on system behavior:

1. If $\rho_1$ is larger, ACC controller focuses more on minimization of distance error
2. If $\rho_3$ is larger, large host vehicle acceleration is constrained, and the system response will be slower.
3. If $\rho_1$ and $\rho_1$ are large, similarly, the system response will be further constrained due to additional constrains on relative velocity.

A phase plant is designed to illustrate how to set parameters under different occasions. In the phase plant, the x-axis is relative velocity, and the y-axis is distance error. Under different traffic situation, the vehicles may be in any of the states of the phase plant, and the purpose of the controller is to drive current state to origin, which is #8 state. The phase plant is shown in Fig. 9.

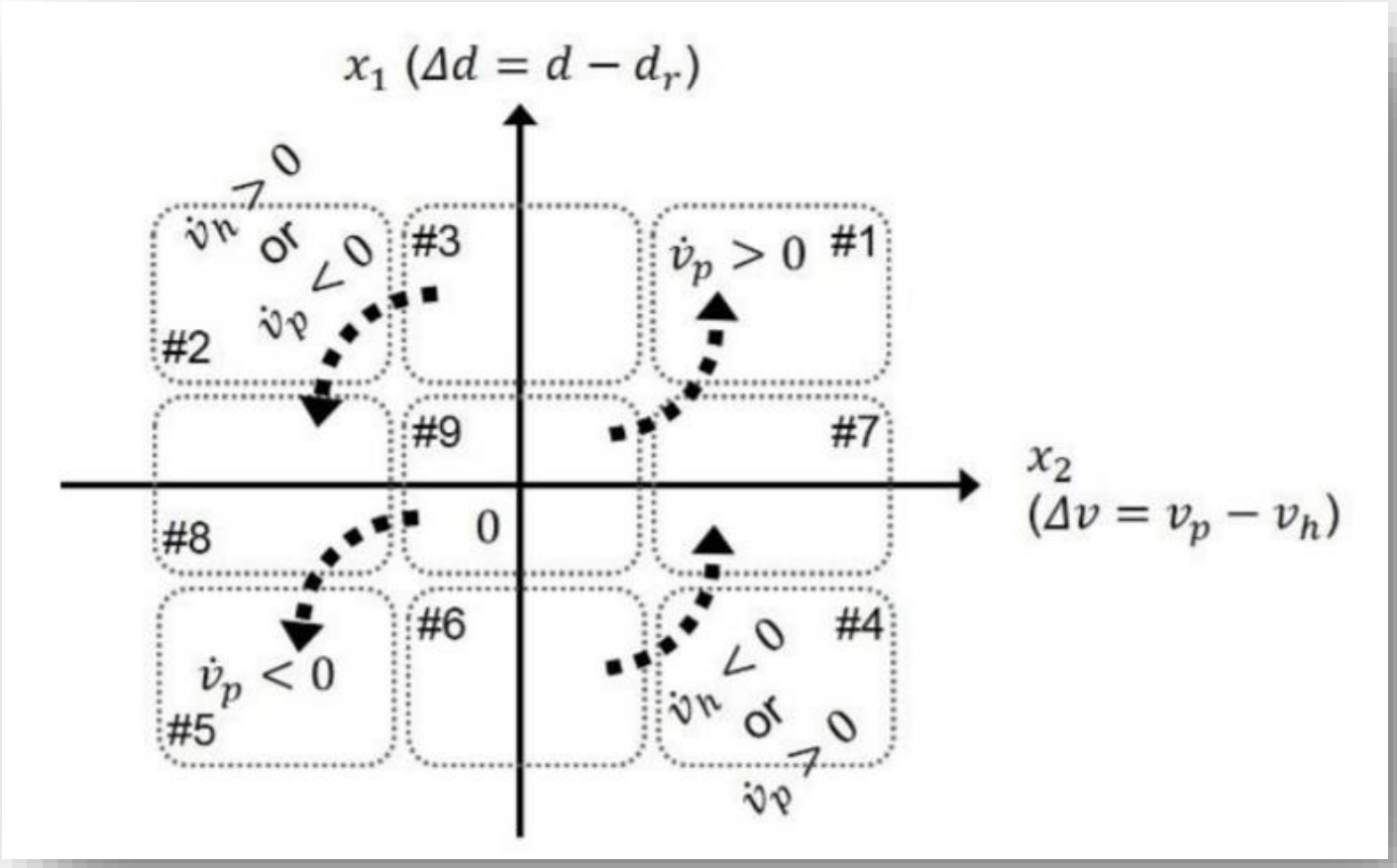

Fig. 9 the operating region division in phase plane for the weight design

According to the phase plant, we give the following parameter setting rules:

1. In #9 state, set $\rho_3$ as a comparatively larger value to moderate accelerating or braking manipulation and avoid uncomfortable feeling in the steady state. In this state, it is obvious that both distance error and relative distance are relatively small, and this is the reason why we called it steady state. In this state, even though, due to external noise, the relative velocity or distance error may shift a little bit away from the origin, we still do not expect too much intervention to system so as to ensure driver comfortability.
2. Similarly, we define the area between $y = \delta$ and $y = -\delta$ as dead zone ($\delta$ is a small positive real number), and in this deal zone, we assume that the state is still steady. Therefore, we constrain all controller output when the vehicles are in this dead zone. Another very important function of deal zone is to avoid frequent shifting between brake and throttle. In some cases, when host vehicle is in this state. In order to perfectly track the preceding vehicle. The host vehicle will keep adjusting its power by shifting between brake and throttle, which greatly decrease driver comfortability.
3. In #1-3 states, set $\rho_1$ and $\rho_2$ as comparatively smaller values to prevent the host vehicle from accelerating with full throttle when a preceding vehicle is detected. In these three states, distance error is positive, which ensures the safety of vehicle states. Therefore, in these three states, we can care more about driver comfortability instead of velocity response.

4. In #9 state, set $\rho_3$ as comparatively small values to quickly respond to sudden deceleration of preceding vehicle. And in #6 state, reduce $\rho_3$ value further to reduce the response time of host vehicle.
5. In #4, #5 and #7 states, set $\rho_1$ and $\rho_2$ as comparatively smaller values to avoid host vehicle accelerating or decelerating too fast.

Now, let us discuss some typical scenarios so that we have a better understand on the parameter setting in this phase plant. Suppose we are, at the very beginning, in a steady state. Then, we will talk about the following two situations:

The preceding vehicle suddenly accelerates, then the vehicles states shift from #8 state to #7 state. In this state, we apply comparatively small $\rho_1$ and $\rho_2$ values, as this is a relatively safe state where preceding vehicle speed is bigger then host vehicle speed, which means there is no possibility for car crash. Further, if we do not have a good control on tracking the preceding vehicle, the only result is increase of distance error, and we shift from #7 state to #1 state. Again, this is a safe region.

The preceding vehicle suddenly decelerates, then the vehicles state shifts from #8 state to #9 state. In this state, we apply comparatively small $\rho_3$ value. This is a danger region, as the relative velocity is negative even though distance error is not that big. But if we do not have a good control in this state, we will shift from #9 state to #6 state, which is the most dangerous state, where both distance error and relative velocity are negative. Therefore, in # 9 state, we give safety the first priority so that host vehicle can quickly responds to the speed change of the preceding vehicle.

A vehicle from the adjacent lane cut in and becomes the new preceding vehicle. In this situation we will be in #4 state when relative velocity is positive or #6 state when relative velocity is negative. We have discussed the situation in #6 state. And in #4 state, since the relative velocity is still positive, it is a safe state, therefore, we apply more flexible policy that we do not allow too big host vehicle deceleration in this state.

### ALQG based on coefficient descent

According to (18), cost function is denoted as $J = \frac{1}{N}\sum_{k=0}^{N-1}[x^T(k)Qx(k) + u^T(k)Ru(k)]$. More generally, we can rewrite the cost function as a function of Q and R, which is given as:

$$J = J(Q_1, Q_2, Q_3, R) \tag{31}$$

And the object is to minimize the cost function. Meanwhile, constraints are also applied to cost functions in each phase, which can be generally written as:

$$s_1 \leq Q_1 \leq s_2;\ p_1 \leq Q_2 \leq p_2;\ q_1 \leq Q_3 \leq q_2;\ t_1 \leq R \leq t_2\ \ and$$

$$Q_1 + Q_1 + Q_1 + R \geq C$$

Here each parameter is constrained in a range based on the phase-plant method positioning. You may notice that, as long as the minimum parameters are chosen, we can get the minimum cost function, therefore, another constraint is given so that lower boundary cannot be chosen simultaneously. The constrained optimization problem is therefore given as:

$$u_k = K' x_k = arg \min_{u_k} (J(Q_1, Q_2, Q_3, R)) \tag{32}$$

$$subject\ to\ \ s_1 \leq Q_1 \leq s_2;\ p_1 \leq Q_2 \leq p_2;\ q_1 \leq Q_3 \leq q_2;\ t_1 \leq R \leq t_2\ \ and$$

$$Q_1 + Q_1 + Q_1 + R \geq C$$

To solve the constrained optimization problem online, we should not only consider the feasibility of optimization method, but also efficiency of that method. Therefore, coefficient descent method(COD) is chosen. The principle of COD is that the function is optimized with respect to one parameter at one time, and by each parameter is optimized iteratively. Computationally, this method is more efficient than genetic based algorithms. Assume that we have the following parameters to be optimized, and they are initialized as:

$$x^0 = (x_1^0, \ldots, x_n^0) \tag{33}$$

In iteration k+1, $x^{k+1}$ is defined from $x^k$ by iteratively solving the single variable optimization problem:

$$x_i^{k+1} = arg \min_{u_k} (f(x_1^{k+1}, \ldots, x_{i-1}^{k+1}, \ldots, x_{i+1}^{k}, \ldots, x_n^{k})) \tag{34}$$

After one iteration, all parameters are updated one by one, and update will stop until stopping criteria is satisfied.

### Data based upper controller design and optimization

Except for model based controller design and optimization, researchers are putting more efforts into data based controller design and optimization. As driving habits change among drivers and over time in ACC system, an intelligent ACC system should adapt to different driving habits. Otherwise a driver would intervene even in situations that ACC is able to manage, typically when it could not meet the driver's expectations. Therefore, an ACC system which takes the individual driving habit into consideration would, to a great extent, contribute to its attraction and put into commercial production. Most of the traditional controller design is based on fixed driving habit model, and do not consider the change of driving habits, while data based controller model is designed to adapt to varying driving habits. The weakness of model based models, if any, is that it heavily depends on the accuracy of physical model, and its parameters, in real cases, may vary with time. Data based models do not reply on accurate physical model, and it can also adapt to the changing of parameters. But we still did not develop theories to ensure the stability of the controllers designed by this method.

### Q-Function

The system dynamics converts the ACC design problem into a regulator problem. As mentioned in previous section, the design objective is to keep track of preceding vehicle at a desired distance, and at the same time minimize relative velocity and following vehicle acceleration. Mathematically, the design purpose is to regulate the states of the system to the equilibrium state $x = [0,0,0]^T$. Like the linear quadratic problem discussed above, we again define similar cost function in quadratic form:

$$J(x_k) = \frac{1}{N}\sum_{i=0}^{N-1}[x^T\,(k+i)Qx(k+i) + u^T(k+i)Ru(k+i)] \qquad (35)$$

Where Q is positive semi-definite matrix, and R is positive definite matrix. $x_k$ is the state vector at time step $k$. according to Bellman's principle of optimality, the optimal cost function $J^*(x_k)$ is time invariant and satisfies the discrete time Bellman's function shown below. Bellman's function consists of two part: the first term is the cost function of action penalty in the current step, and the second term is the state penalty in current state(by the current action, the current state goes to the next step, the penalty is denoted by the state of the next step). Bellman's function satisfies Markovian process, which means the cost function of current step is only affected by the action and state of the current step.

$$J^*(x_k) = \min_{u_k}\big(r(x_k, u_k) + J(x_{k+1})\big) \qquad (36)$$

where $r(x_k, u_k) = x^T(k)Qx(k) + u^T(k)Ru(k)$ is the action penalty of the system, and $J(x_{k+1})$ is the state penalty of the system.

For linear discrete-time system, Watkins [39], in his PhD thesis, proposed a cost function with respect to both state and action for a stable control policy *K*. The *Q*-function is

$$Q(x_k, u_k) = r(x_k, u_k) + Q(x_{k+1}, Kx_{k+1}) \tag{37}$$

The value of $r(x_k, u_k)$ implies the cost of taking any admissible action $u_k$ at state $x_k$, and $J(x_k)$ is the cost at state $x_k$. Q-function consists of both action cost and state cost. Similar to the principle of LQR, the optimal control rule is achieved by using greedy policy under Q-function, which can be written as:

$$u_k = K'x_k = arg\min_{u_k}\big(r(x_k, u_k) + Q(x_{k+1}, Kx_{k+1})\big) \tag{38}$$

where $K'$represents optimal *K* after optimization iteration(s).

It is assumed that the Q function can be represented by some complex combination of gains and traffic information. And somehow, it is estimated that the Q function can be denoted by a linear combination of some traffic features. Therefore, the relationship between features and cost function is represented by (33). Once approximator is trained by least square error method, optimal input can be determined so that it can lead to minimum cost function. Here the linear neural network is employed to approximate Q-function:

$$Q(x_k, u_k) = \omega^T\varphi(x_k, u_k) \tag{39}$$

Where the $\varphi(x_k, u_k)$ is the basis function and the ***w*** is the parameter vector. As the cost function of the linear quadratic regulator problem is the quadratic polynomial function with respect to the states and actions, the so called one node neural network( which can also be called nonlinear feature linear regression) can approximate the Q-function well if the basis function is chosen as

$$\varphi(x, u) = [x_0, x_1^2, x_1x_2, x_1x_3, x_1u, x_2^2, x_2x_3, x_2u, x_3^2, x_3u, u^2]^T \tag{40}$$

and

$$\omega = [\omega_0, \omega_1, \omega_2, \omega_3, \omega_4, \omega_5, \omega_6, \omega_7, \omega_8, \omega_9, \omega_{10}]^T \tag{41}$$

In order to find optimal $\boldsymbol{u}$, we take partial derivation of *Q*-function with respect to *u*, then *u* is derived as follows:

$$\frac{\partial Q(x_k,u_k)}{\partial u} = 0 \tag{42}$$

$$u = Kx = -\frac{\omega_4 x_1 + \omega_7 x_2 + \omega_9 x_3}{2\omega_{10}} \tag{43}$$

Where the control policy *K* is

$$K = -\frac{1}{2\omega_{10}}[\omega_4 \ \omega_7 \ \omega_9] \tag{44}$$

In order to achieve optimal solution online, we apply linear regression method, in which we take use of updated basis function $\varphi(x,u)$ and cost function $r(x,u)$. Here $\omega = [\omega_0, \omega_1, \omega_2, \omega_3, \omega_4, \omega_5, \omega_6, \omega_7, \omega_8, \omega_9, \omega_{10}]^T$ denotes the parameter vector. $\varphi(x,u) = [x_0 = 1, x_1^2, x_1x_2, x_1x_3, x_1u, x_2^2, x_2x_3, x_2u, x_3^2, x_3u, u^2]^T$ is the augmented input with the constant 1 introduced as an additional dimension $x_0$. The weight $\omega_0$ in ***w*** is the bias term which serves as an offset. The learning problem is to find the best ***w*** according to some performance measure using the training set *S* where *S* is denoted as:

$$S = \left(\varphi^{(l)}, r^{(l)}\right)_{l=1}^{N} \tag{45}$$

In this problem, according to () and (), update rule between Q-function and cost function is

$$\omega_i^T \varphi(x_k,u_k) = r(x_k,u_k) + \omega^T \varphi(x_{k+1},u_{k+1}) \tag{46}$$

It can also be rewritten as:

$$\omega_i^T\left(\varphi(x_k,u_k) - \varphi(x_{k+1},u_{k+1})\right) = r(x_k,u_k) \tag{47}$$

By solving the equation using Least Mean Square method(LMS), the ***w*** is as follows:

$$\omega_i = (\Phi^T\Phi)^{-1}\Phi^T Y \tag{48}$$

where

$$\Phi = [\nabla\varphi_{K-N}, \nabla\varphi_{K-N+1}, \nabla\varphi_{K-N+2}, \cdots, \nabla\varphi_K] \tag{49}$$

$$\nabla\varphi_K = \varphi(x_k,u_k) - \varphi(x_{k+1},u_{k+1}) \tag{50}$$

$$Y = [r(x_{K-N},u_{K-N}), r(x_{K-N+1},u_{K-N+1}), \cdots, r(x_K,u_K)]^T \tag{51}$$

According to current optimal ***w***, optimal *K* is given by

$$K_{i+1} = -\frac{1}{2\omega_{10}}[\omega_i(4)\ \omega_i(7)\ \omega_i(9)] \tag{52}$$

Following is the proof of LSM method:

We first define regularizer

$$L(w;S) = \left|\left|\varphi w - r\right|\right|^2 \tag{53}$$

which can be rewritten as

$$L(w;S) = w^T\varphi^T\varphi w - 2r^T\varphi w + r^T r \tag{54}$$

By differentiating $L$ with respect to $\boldsymbol{w}$, and set the derivative to the zero vector $\boldsymbol{0}$, we obtain:

$$2\varphi^T\varphi w - 2\varphi^T r = 0 \tag{55}$$

$$\Longrightarrow \varphi^T\varphi w = \varphi^T r \tag{56}$$

$$\Longrightarrow w = (\varphi^T\varphi)^{-1}\varphi^T r \tag{57}$$

The solution of ***w*** is also called close form solution, where the calculation is deterministic, instead of iterative.

## Feature Scaling and Regularization

As we can observe in calculation of ***w***, we cannot avoid matrix inverse, and the prerequisite of doing matrix inverse is that the matrix is nonsingular( even though we use pseudo-inverse $\varphi^T\varphi$ to ensure non-singularity before matrix inverse, but it is always better to ensure that matrix $\varphi$ itself is nonsingular.

One way to guarantee that the matrix is invertible is to make sure that the number of samples $N$ is no less than the length of vector ***w***. Another way is to introduce regularization method to modify the matrix singularity. The principle of regularization method is to introduce a penalty for using relatively large ***w***, it is mathematically proven that large ***w*** will increase matrix collinearity and lead to regression overfitting. Therefore, by introducing appropriate regularization method, basic function ***w*** shrinks, and therefore, not only the matrix is guaranteed nonsingular, overfitting is also avoided.

It should be emphasized that preventing overfitting is extremely important in training online parameters. As in each training loop, due to limited sampling time period, there are very limited samples, which means we cannot make too much processing on these data. In this occasion, therefore, preventing solution overfitting by using regularization is necessary. Here we introduce *L2* regularization (Ridge Regularization), in which the regularizer is modified as

$$L_\lambda(w;S) = L(w;S) + \lambda\left|\left|w\right|\right|^2 \tag{58}$$

$$\Rightarrow L_\lambda(w;S) = \left||\varphi w - r|\right|^2 + \lambda\left||w|\right|^2 \tag{59}$$

$$\Rightarrow L_\lambda(w;S) = w^T\varphi^T\varphi w - 2r^T\varphi w + r^T r + \lambda\left||w|\right|^2 \tag{60}$$

By differentiating $L_\lambda$ with respect to $\boldsymbol{w}$, and set the derivative to the zero vector $\boldsymbol{0}$, we obtain:

$$2\varphi^T\varphi w - 2\varphi^T r + 2\lambda w = 0 \tag{61}$$

$$\Rightarrow (\varphi^T\varphi + \lambda I)w = \varphi^T r \tag{62}$$

$$\Rightarrow w = (\varphi^T\varphi + \lambda I)^{-1}\varphi^T r \tag{63}$$

It is noted that, by using LMS method, the close form solution is very similar to the previous one. But with the introduction of $\lambda I$, $\varphi^T\varphi + \lambda I$ is always invertible. It is guaranteed that we do not need to worry about the singularity issue that may bother us in pure linear regression without using regularization. The only necessary condition is that the number of samples is no less than the number of predictors. Once satisfied, the control policy can be updated as follows:

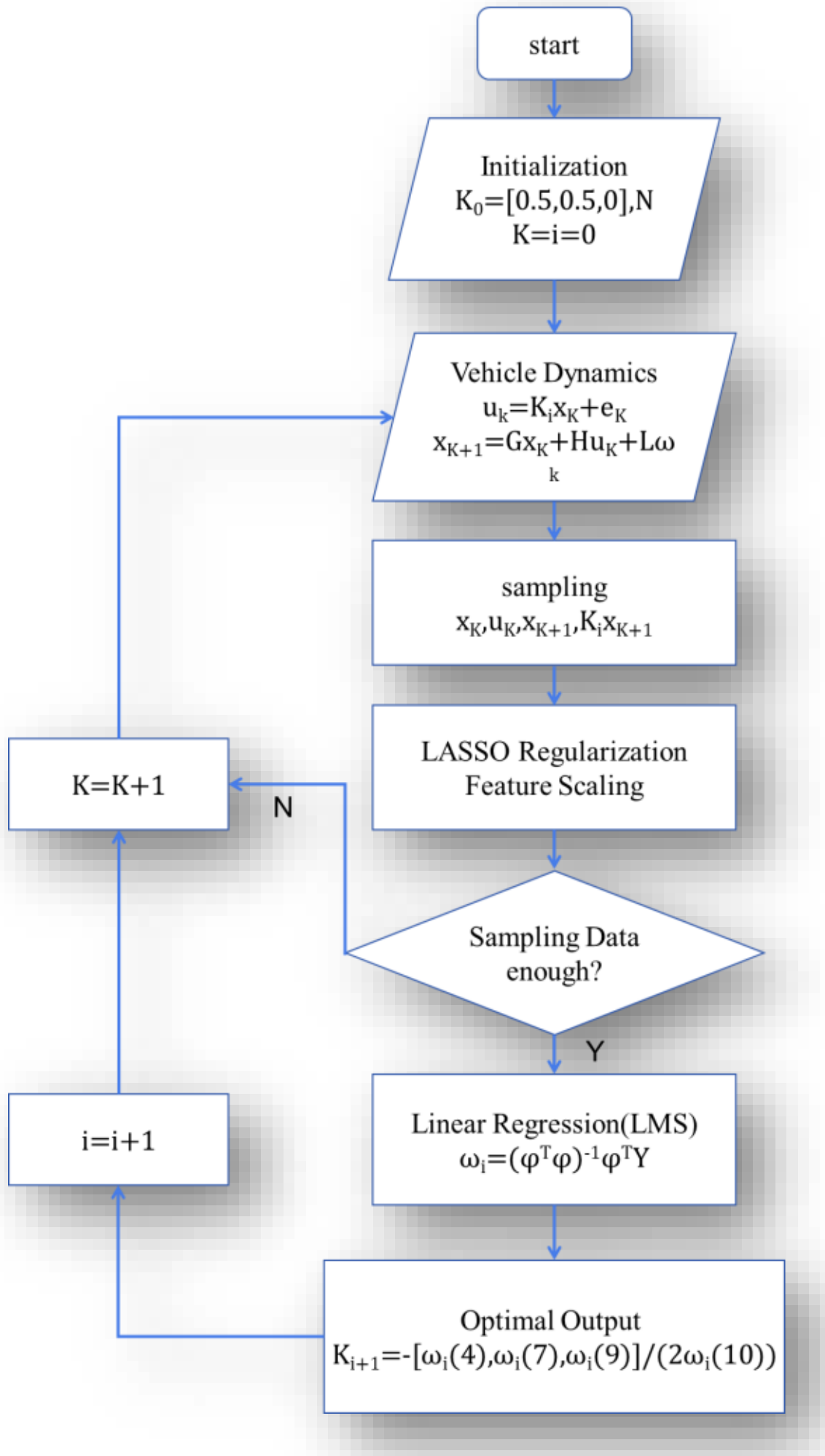

Another very popular regularization method: least absolute shrinkage and selection operator(LASSO) can also be applied as follows:

$$L_\lambda(w;S) = L(w;S) + \lambda|w| \tag{64}$$

$$subject\ to\ \ \textstyle\sum_{j=1}^{N}\left|w_j\right| \leq (t)^+$$

In data processing, one of the most important preprocessing technics is feature scaling, which increases convergence rate when we apply gradient descent method in linear regression. When dealing with vehicle data, i.e. distance error, relative velocity, and host vehicle acceleration, the variation among the three parameter is big, which means convergence rate in gradient descent can be greatly influenced by those parameters with bigger magnitude. Therefore, feature scaling is necessary in data preprocessing, through which the magnitude of every feature is constrained between -1 and 1. Standardization is recommend as following:

$$x' = \frac{x-\bar{x}}{\sigma} \tag{65}$$

where $x'$is updated $x$ after feature scaling, $\bar{x}$ is mean value of dataset of $\boldsymbol{x}$, and $\sigma$ is standard deviation.

## Lower Controller Design and Optimization

Till now, we have illustrated elaborately the details of the upper controller design, but in the design process, we have assumed that the desired acceleration can be achieved ideally, which means we do not consider the time delay and noise from the real vehicle dynamics. Therefore, in the next step, we have to design a lower controller which can achieve the desired acceleration command from the upper controller accurately and quickly. In a vehicle system, the objective of lower controller is to achieve a desired longitudinal acceleration by regulating the throttle position or the brake pressure.

In this paper, a hybrid PID control strategy is applied in the vehicle acceleration controller which includes both throttle controller and brake controller. This approach has been chosen because the complex inverse model of the vehicle need not be considered when designing the controller. Conventionally, PID is simple and efficient in most cases, while in vehicle acceleration control, a simple PID may not work well with nonlinear vehicle dynamics where there are time varying parameters and uncertainties in tire and engine. Therefore, some manipulation is necessary for PID so that it can adapt to the nonlinearity of vehicle system. One popular method is fuzzy-PID, where the three parameters of PID vary with time based on the fuzzy rules developed in fuzzy logic control. Since the PID parameters varies with time, it can better adapt to the nonlinearity of the vehicle system. The approach integrating PID control and fuzzy control together has the merits of both and is used in the throttle control. For a brake controller design, things become much more easier. As i-Booster from Bosch is used as the actuator of brake, it provides linear control on brake, and its response time and linearity are well design. Therefore, the brake controller deals with a quasi-linear system, which reduces greatly the difficult in controller design.

## Throttle Controller

The throttle control is put into use when the desired acceleration is located at the throttle control zone. The objective of the throttle control is to make the actual vehicle acceleration track the desired acceleration by adjusting the throttle position. A block diagram of the throttle control algorithm is shown in Fig. 10. A low pass filter of second order with a cut-off frequency $\omega_n$ of 10 *rad* / *s* and damping ratio $\xi$ of 1 is applied to avoid frequent action on the throttle. Then the desired acceleration is converted to throttle position, which is in the unit of percentage, therefore, we apply a saturation block to constrain that the output varies between 0% to 100%.

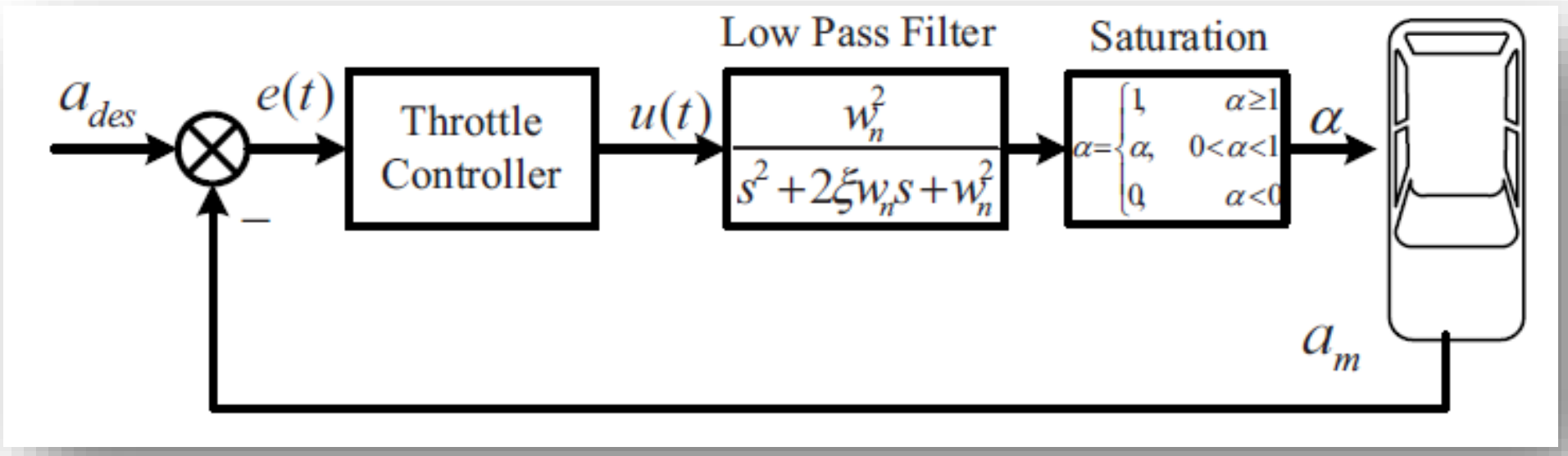

Fig. 10 Throttle control diagram

The PID control law is given:

$$u(t) = k_p e(t) + k_i \int e(t)dt + k_d \frac{de(t)}{dt} \tag{66}$$

Where $k_p$, $k_i$, and $k_d$ are the proportional, integral, and derivative gains, respectively. The tracking error *e(t)*= $a_{desire} - a_{real}$ is defined and $u(t)$ is the output of controller. The PID controller with the fixed parameters is simple but cannot always have an effective control on the nonlinear and time-varying systems. The PID controller with parameters tuned online are suited for these systems. Fig.11 shows the control system with a fuzzy gain scheduling PID controller, where $k_{f,e}$ and $k_{f,ec}$ are the input scaling factors, and $k_{f,p}$ and $k_{f,i}$ are the output scaling gains. The approach is to make use of the fuzzy rules and reasoning to generate the PID controller's parameters online. It takes *e(t)* and $\frac{de(t)}{dt}$ as inputs, and outputs $u_{f,p}$ and $u_{f,i}$ which are utilized to tune the parameters $k_p$ and $k_i$ respectively. Here the derivative gain $k_d$ is fixed.

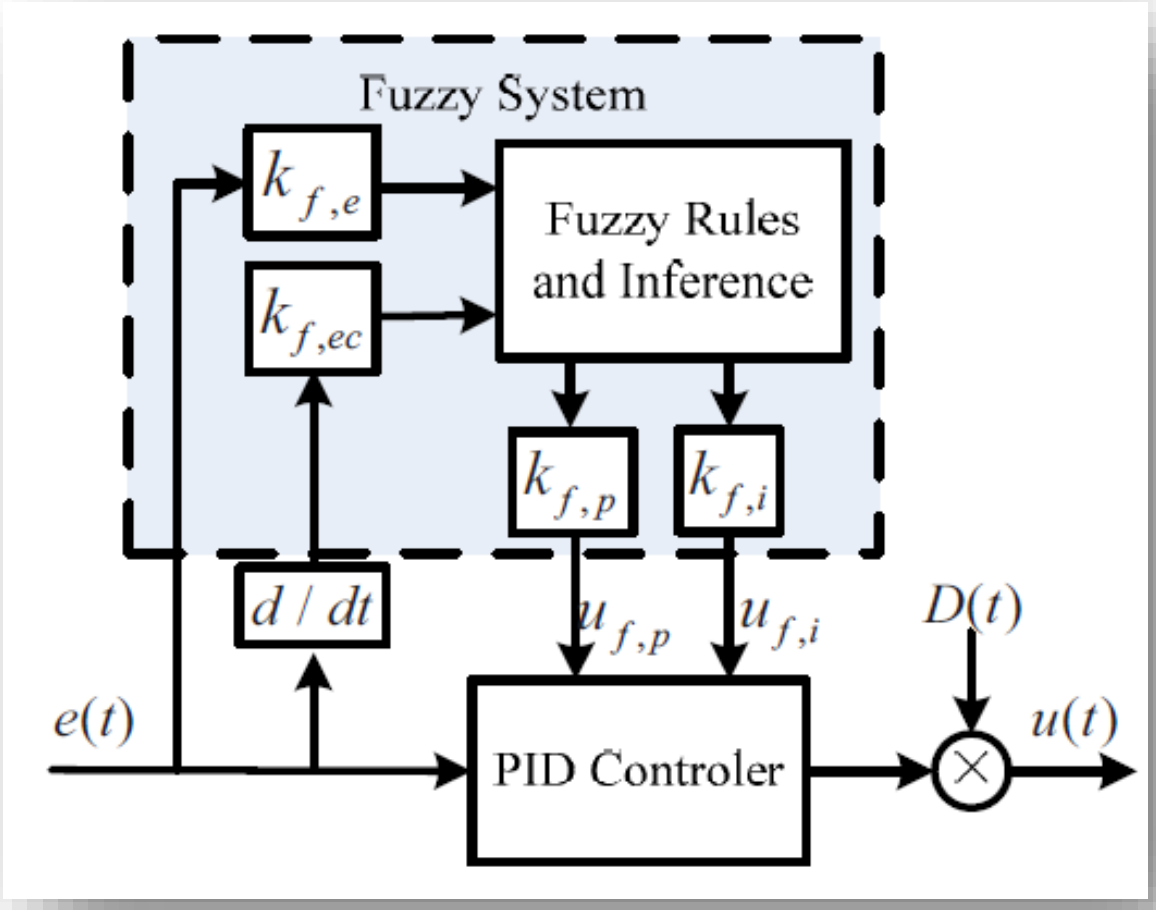

Fig.11 Fuzzy gain scheduling of PID controller

In the fuzzy gain scheduling scheme, the outputs are $u_{f,p}$ and $u_{f,i}$, which are determined based on the current error *e(t)* and its first derivative $\frac{de(t)}{dt}$. The values of the fuzzy variables are taken from asset of fuzzy partitions, represented by membership functions with a variety of shapes: triangular, trapezoidal, Gaussian and so on. Here all the input and output membership functions are chosen as triangular. The membership functions of *e(t)* and $\frac{de(t)}{dt}$ are shown in Fig.12 and Fig.13 respectively. $u_{f,p}$ and $u_{f,i}$ have the same functions shown in Fig. xx. Where *N, ZO, P* represent *Negative*, *Zero*, and *Positive*, and *S, M, B* represents *Small*, *Medium*, and *Big*. The fuzzy rules are determined heuristically based on the step time response of the controller. It follows the principle that:

1. when the tracking error *e(t)* is big, the proportional gain $k_p$ should be big so as to generate a big signal and the integral $k_i$ should be small to avoiding integral saturation;
2. when the tracking error *e(t)* is small, the proportional gain $k_p$ should be small to avoid overshooting and the integral $k_i$ should be big to erase steady state error and improve steady precision.

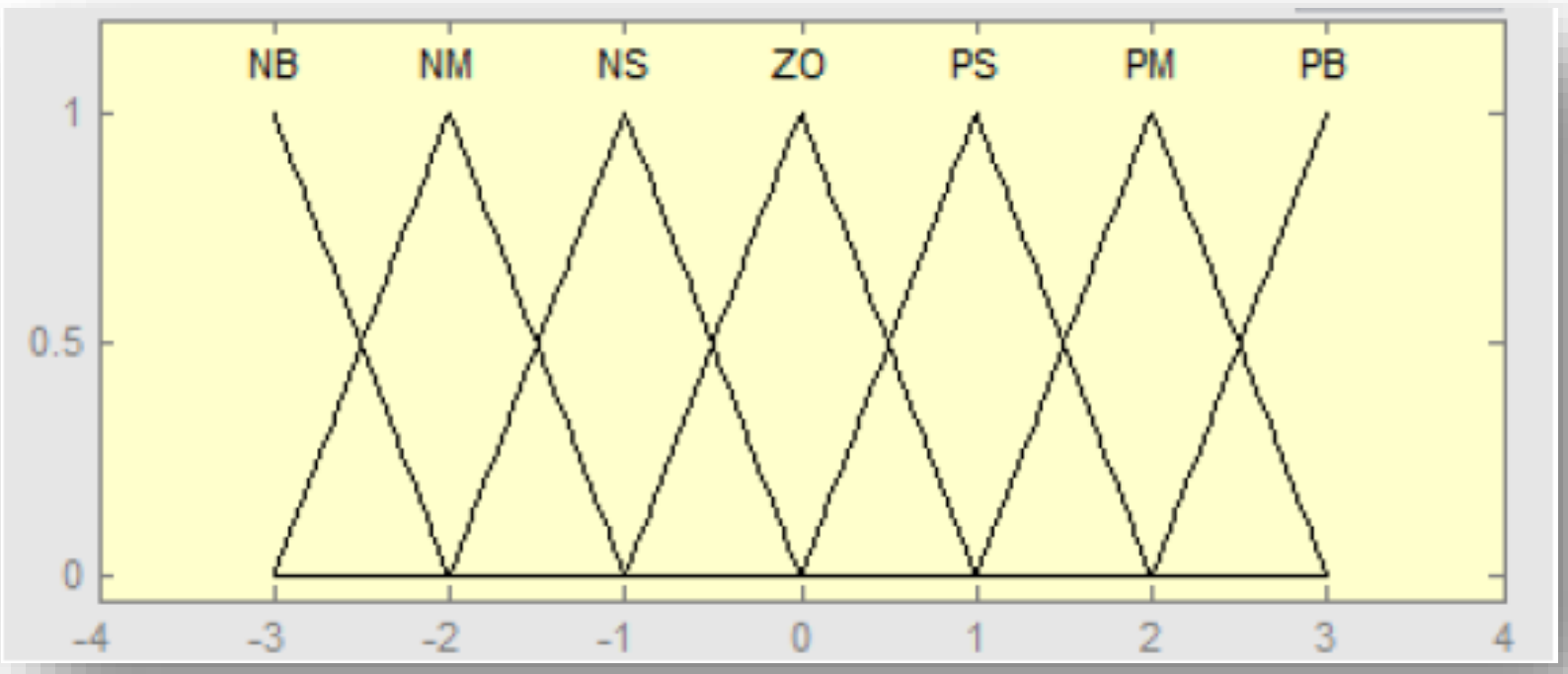

Fig.12 The membership function of *e(t)*

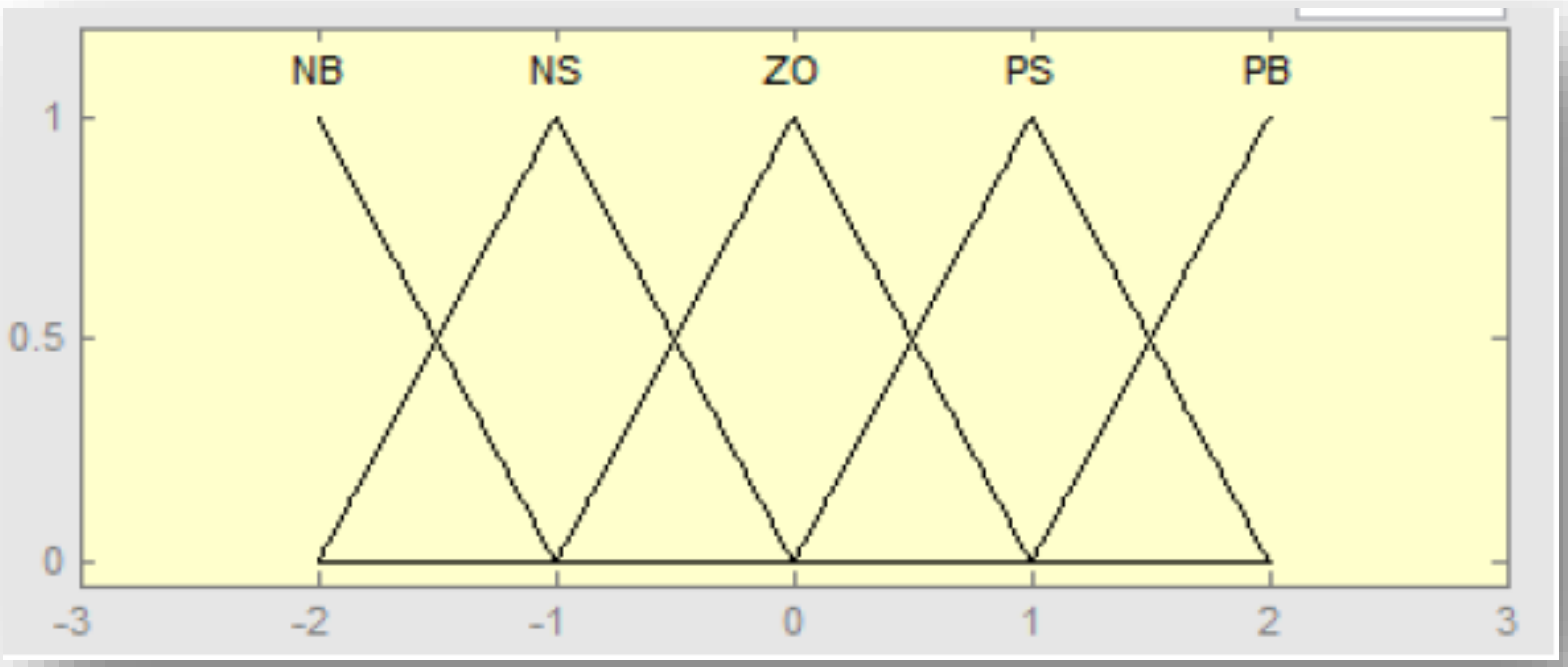

Fig.12 The membership function of $\frac{de(t)}{dt}$

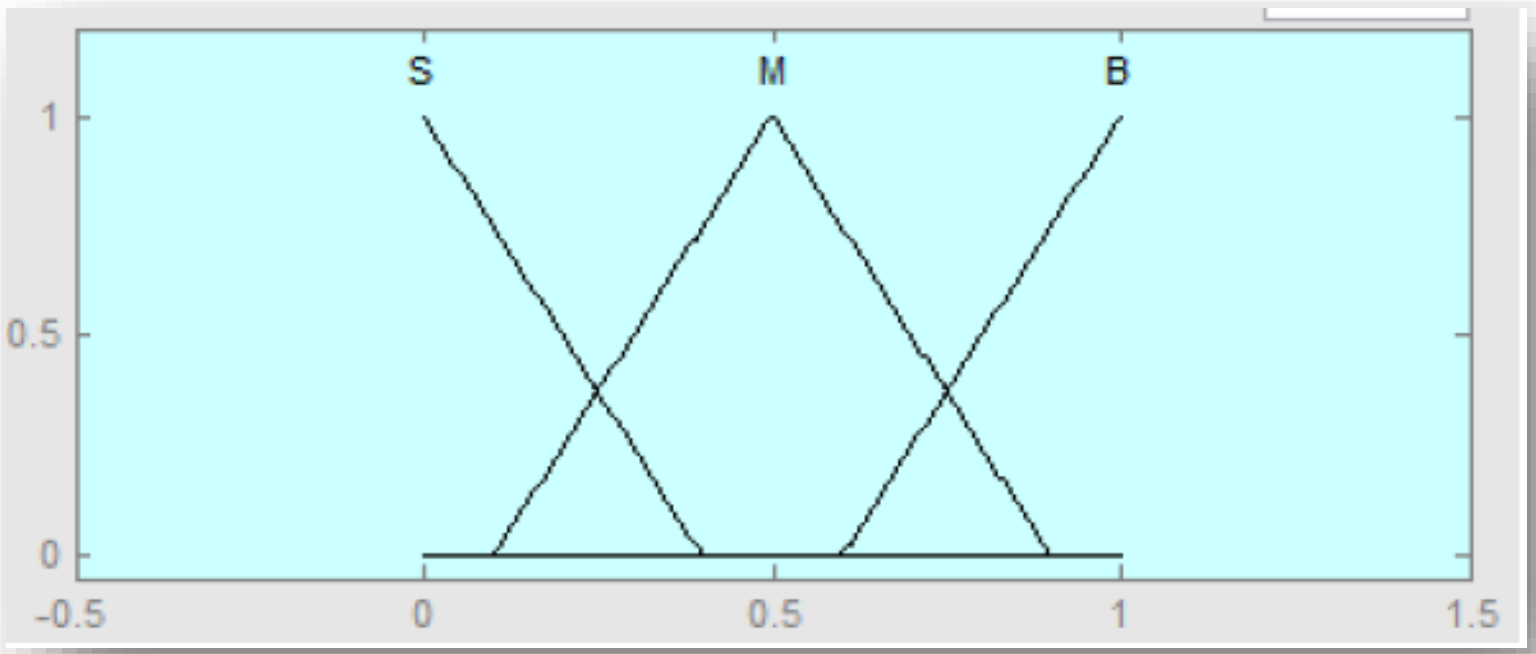

Fig.13 The membership function of outputs

Table 1: Rule table for proportional gain $u_{f,p}$

| | | $de(t)/dt$ | | | | |
|---|---|---|---|---|---|---|
| | | NB | NS | ZO | PS | PB |
| $e(t)$ | NB | B | B | B | B | B |
| | NM | M | B | B | B | M |
| | NS | S | M | B | M | S |
| | ZO | S | S | M | S | S |
| | PS | S | M | B | M | S |
| | PM | M | B | B | B | M |
| | PB | B | B | B | B | B |

Table 2: Rule table for integral gain $u_{f,i}$

| | | $de(t)/dt$ | | | | |
|---|---|---|---|---|---|---|
| | | NB | NS | ZO | PS | PB |
| $e(t)$ | NB | S | S | S | S | S |
| | NM | M | S | S | S | M |
| | NS | B | M | S | M | B |
| | ZO | B | B | M | B | B |
| | PS | B | M | S | M | B |
| | PM | M | S | S | S | M |
| | PB | S | S | S | S | S |

It is assumed that parameters $k_p$ and $k_i$ of PID controller are tuned in prescribed ranges, $[k_{p,\min}, k_{p,\max}]$ and $[k_{\mathrm{i},\min}, k_{\mathrm{i},\max}]$ respectively. The parameters of PID controller can be gained based on the outputs $(u_{f,p}, u_{f,\mathrm{i}})$ of the fuzzy system as:

$$k_p = k_{p,\min} + u_{f,p}(k_{p,\max} - k_{p,\min}) \tag{67}$$

$$k_i = k_{\mathrm{i},\min} + u_{f,\mathrm{i}}(k_{i,max} - k_{i,min}) \tag{68}$$

Based on the Ziegler-Nichols formula, a rough rule is used to determine the range of $k_p$ and $k_i$ as:

$$K_{p,min} = 0.3k_u, \qquad K_{p,max} = 0.7k_u$$

$$K_{i,min} = K_u/T_u, \qquad K_{i,max} = K_u/0.25T_u$$

Where $K_u$ and $T_u$ are the gain and the period of oscillation at the stability limit under P control respectively.

The changing in the scaling gains at the input and output of the fuzzy inference system have a significant impact on the performance of the fuzzy control system, hence they are often convenient parameters for tuning. To determine the feasible values of $k_{f,e}$, $k_{f,ec}$, $k_{f,p}$, $k_{f,i}$, one category of modified particle swarm optimizer is used. A fitness function associated with the performance indices of step response – the rise time (tr), the maximum overshoot(OS), and the integral of absolute error (IAE) is introduced so that we can put different weights on different parameters. For example, if we are more concerned about tr compared to OS and IAE in the system, more weight can be put on the parameter tr than OS and IAE.

## Brake Controller

Braking system aims to improve driving safety emergency situation. In the vehicle's deceleration process, the brake torque is applied only when the outside resistance is not sufficient to follow the desired acceleration. Here when the desired acceleration for a given vehicle velocity is smaller than the switching line, the brake control is activated. A brake control law is derived under a nonslip condition of the braking wheel. The desired acceleration is related to brake torque, and brake torque is proportional to the brake pressure in the disk pressure. Fortunately, the IBC(Intelligent Braking Controller) we choose allows us to directly control the disk pressure. As it quickly responds to the desired disk pressure, accordingly, the desired acceleration can be quickly achieved as well. Therefore, by using feedback incremental PID control, we are able to realize desired acceleration by controlling disk pressure. Fig xx indicates the diagram of increment PID controller.

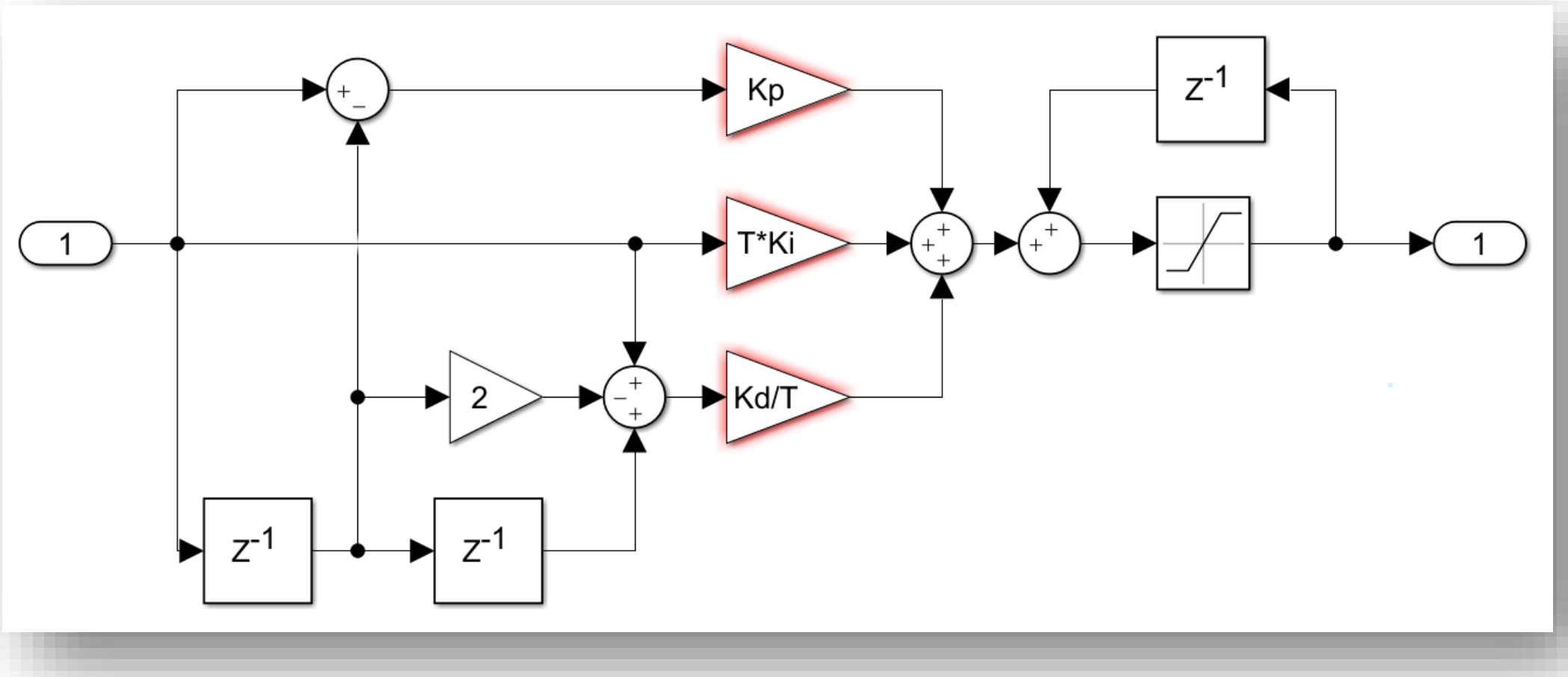

Fig. 14 diagram of increment PID controller

This principle of PID controller is as follows:

$$u_k = K_p(e_k + \frac{T}{T_i}\sum_{j=1}^{k} e_j + \frac{T_d}{T}(e_k - e_{k-1}) \quad (69)$$

where

$$u_k = P_p(k) + P_i(k) + P_d(k) \quad (70)$$

$$P_p(k) = K_p e_k \quad (71)$$

$$P_i(k) = K_p \frac{T}{T_i}\sum_{j=1}^{k} e_j \quad (72)$$

$$P_d(k) = K_p \frac{T_d}{T}(e_k - e_{k-1}) \quad (73)$$

For an incremental PID controller, the input is the difference between the error in last step and in current step, therefore $\Delta u_k$ is defined as:

$$\Delta u_k \triangleq u_k\text{-}u_{k-1} \quad (74)$$

$$\Delta u_k = K_p(e_k - e_{k-1} + \frac{T}{T_i}e_k + \frac{T_d}{T}(e_k - 2e_{k-1} + e_{k-2})) \quad (75)$$

Expect for the main part of PID controller, in real cases(Fig. 15) we also need to apply

1. saturation function which filter the signal noise;
2. rate limiter which constrains the rate change of the commands so that the acceleration change is not abrupt;
3. anti-windup function which stops integration part when the output reaches its boundary.

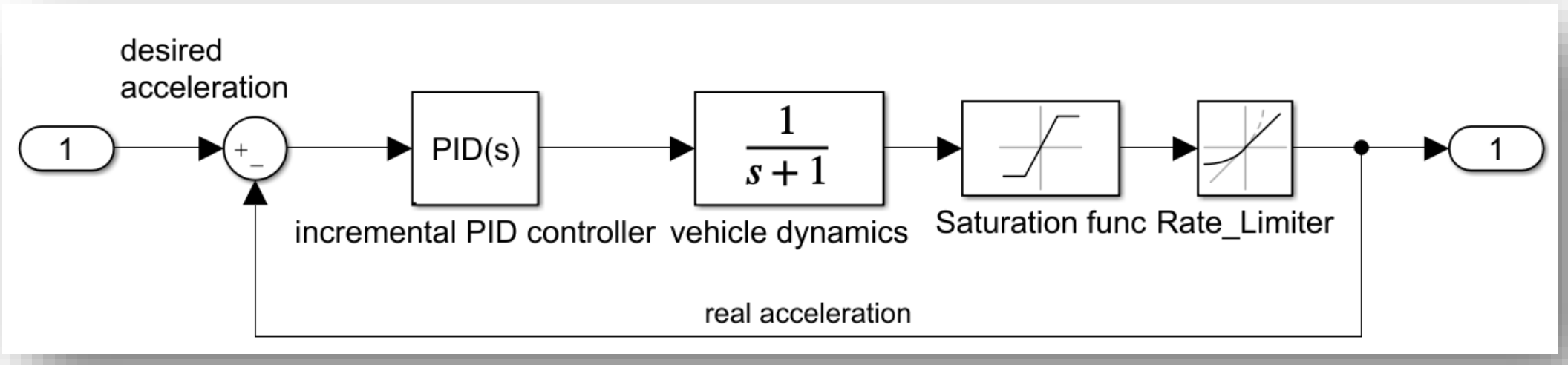

Fig. 15 Real vehicle system structure

## Switching logic

### Switching logic between throttle and brake

The objective of the acceleration control is to achieve a desired longitudinal acceleration $a_{des}$ by regulating the throttle position α, the brake pressure $p_{brake}$ or switching between them. Then throttle position is transformed to motor toque, and brake pressure is translated to brake pedal position, which is linear to brake cylinder pressure. Cylinder pressure can also be represented as a brake torque. The diagram is shown in Fig. 16.

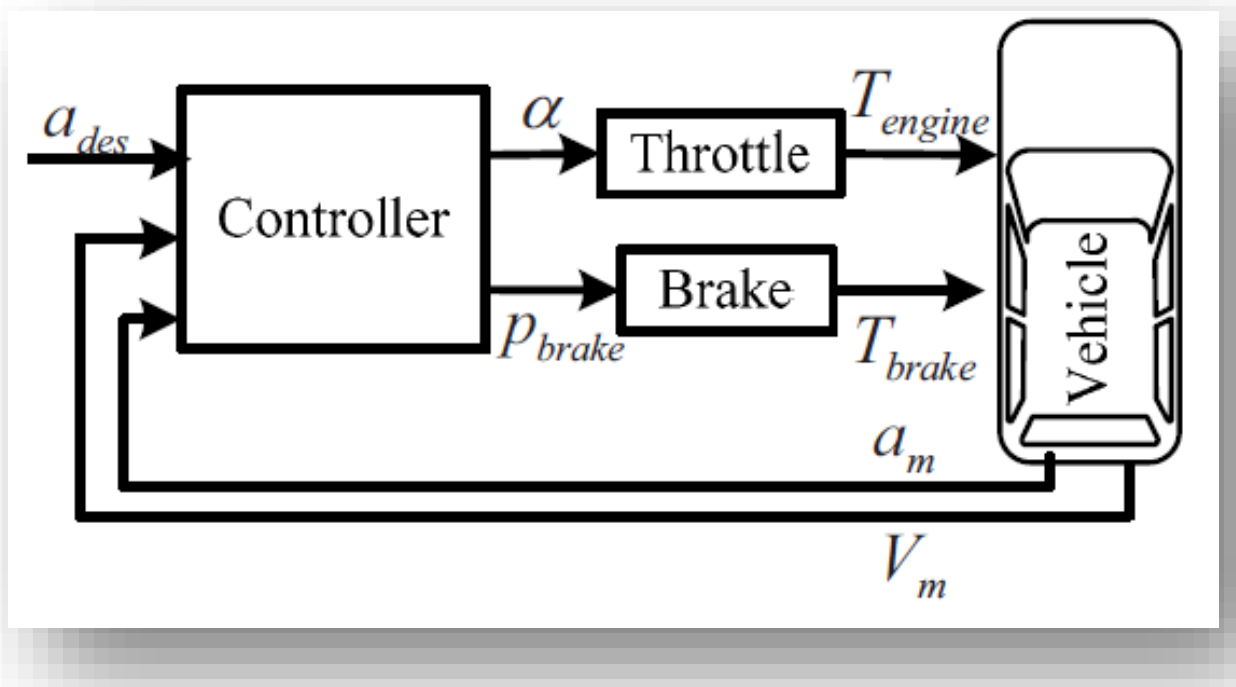

Fig. 16 General control structure

The throttle controller and the brake controller cannot work simultaneously. Fig.17 shows the natural deceleration at a given velocity when the throttle is closed ($\alpha = 0$). In order to avoid frequent switching between the throttle control and the brake control which will bring strong oscillation for vehicle, a buffer zone with width *h* is added. The throttle control is applied when $a_{desire} > a_{real} + h$, and the brake control is applied when *desire real* $a_{desire} < a_{real} - h$. When at the dead zone, the brake control and the throttle control are laid off at the same time. After testing on vehicle, we design the *h* with an absolute value of *0.05g*, which means the dead zone has a buffer range of *0.1g*.

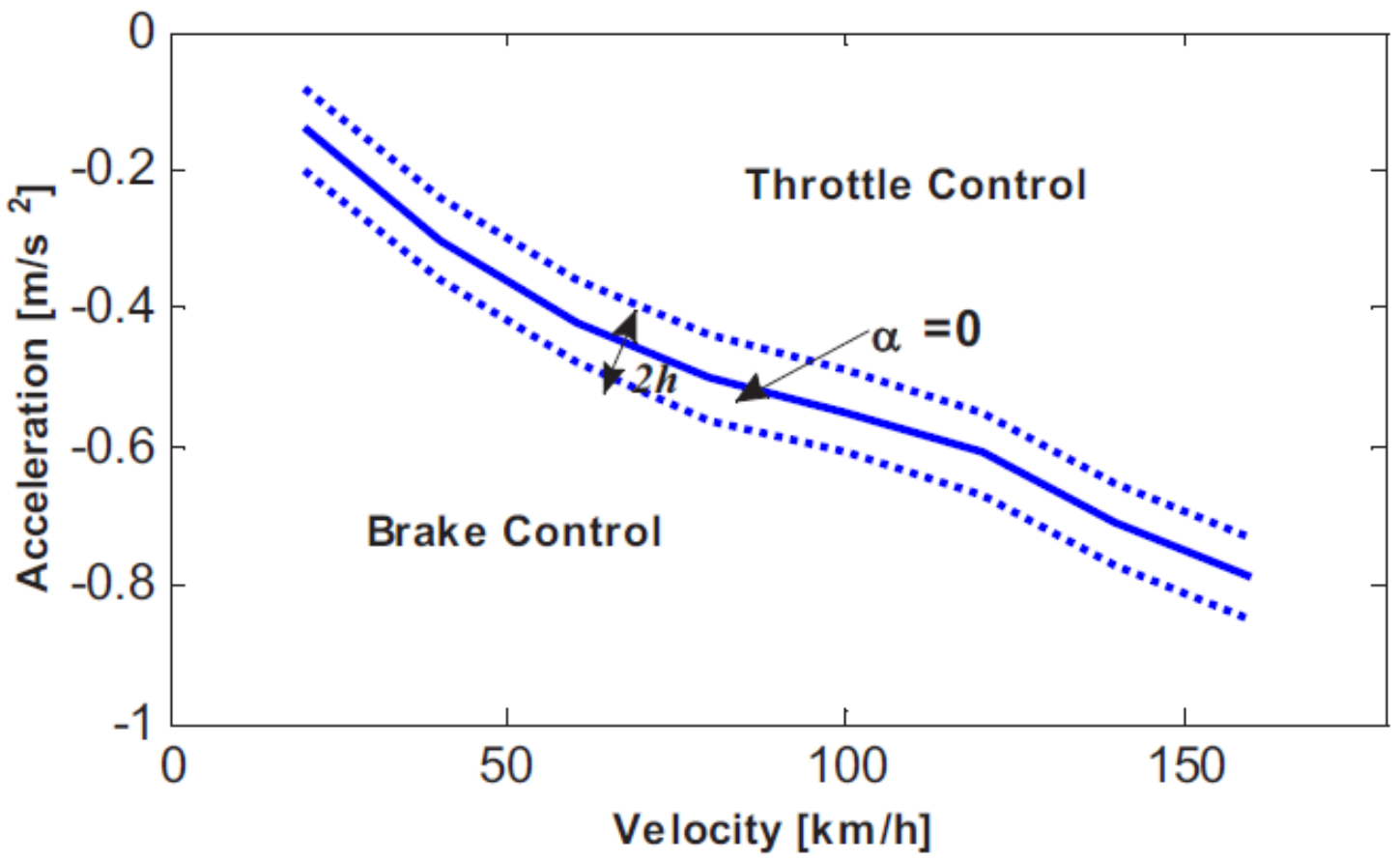

Fig. 17 Natural deceleration under different velocity

**Switching logic between CCS and ACC**

While headway time based distance control(ACC upper controller) and velocity control(CCS controller) are successful, in real scenarios the repetitive switching phenomena occurs. While the vehicle approaches another vehicle under the constant speed controller, at the desired distance the system keeps switching back and forth between velocity and space controllers (In real cases, the vehicle vibrates at a high frequency). Therefore, we still need to develop a switching logic so that the vehicle, based on the current relative state with the preceding vehicle, can automatically switching between the CCS and ACC states smoothly (Fig.18).

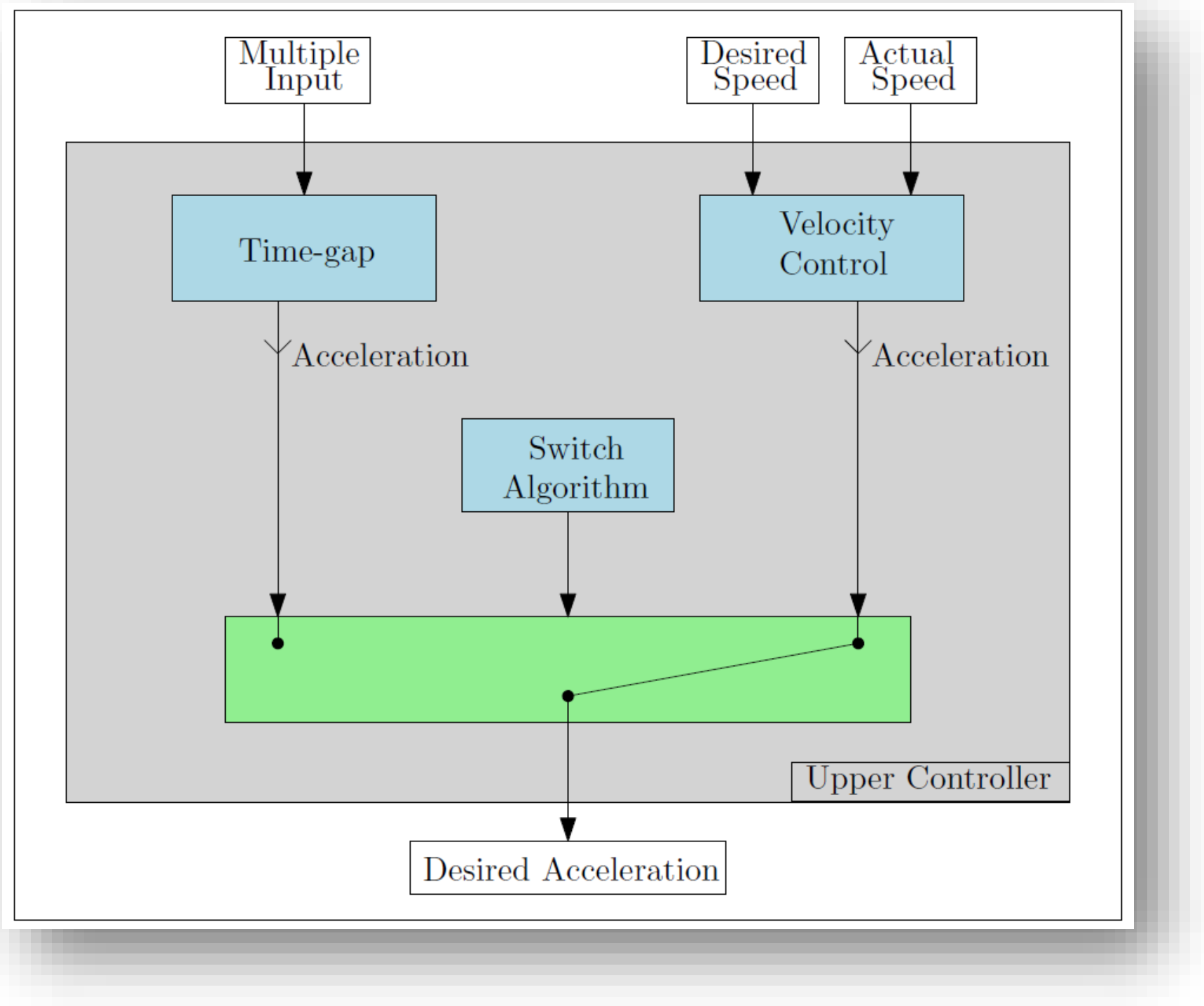

Fig. 18 Upper controller in detail

The switching logic ensures that ACC system shifts between the two modes according to the current traffic situations, avoiding too much shifting oscillation in between. Conventionally, the most common way is to compare the desired distance and real distance and activate distance control when the distance error is positive and shifts back to speed control mode when the distance error is negative. Even though in theory it works well, however, in real cases, you can image that when the distance error undershoots even for a very short time, the system switches to velocity control. Under the influence of the velocity control mode, the vehicle speeds up, and again switches back to distance control mode. Therefore, to solve for the problem, it is required that when the controller switches to distance control mode from velocity control mode, a constant($t_{safety}$) close to 1 is used to increase the desired distance which results in:

$$L_{desire} = t_{safety} \times d_{desire}$$

Note that this desired distance $L_{desire}$ is not used in the controller. This term is used just at the proposed algorithm presented in Fig.19. If the vehicle switches back to velocity control mode, then the additional distance is removed to prevent the vehicle to switch to distance control mode too early.

To conclude, the switching logic consists of two different criteria, which are distance criteria and velocity criteria. When shifting from cruise control mode to distance control mode, we change the desire distance $d_{desire}$ to $L_{desire}$ by multiplying a coefficient $t_{safety}$, which in real cases we decide as $t_{safety}$= 1.12. by elongating the desired distance, we allow a little bit overshooting of real distance, and therefore, the vehicle can stay more time in distance control mode instead of immediately switching back to velocity control mode. On the other hand, we still need to consider velocity criteria. In distance control mode, when the host vehicle velocity decrease below the velocity of the preceding vehicle, it is possible that the host vehicle velocity may even decrease, since in the distance upper controller we consider not only distance error, but also relative velocity, and host vehicle acceleration. Therefore, it is even when host vehicle velocity is smaller than the preceding one, host vehicle may still slowdown, which is not what we do not expect. But once we introduce velocity criteria, we can solve this problem. In distance control mode, when the preceding vehicle velocity is higher than that of the host vehicle, the host vehicle can go to velocity control mode as it is safe to accelerate even when the real distance is smaller than $L_{desire}$.

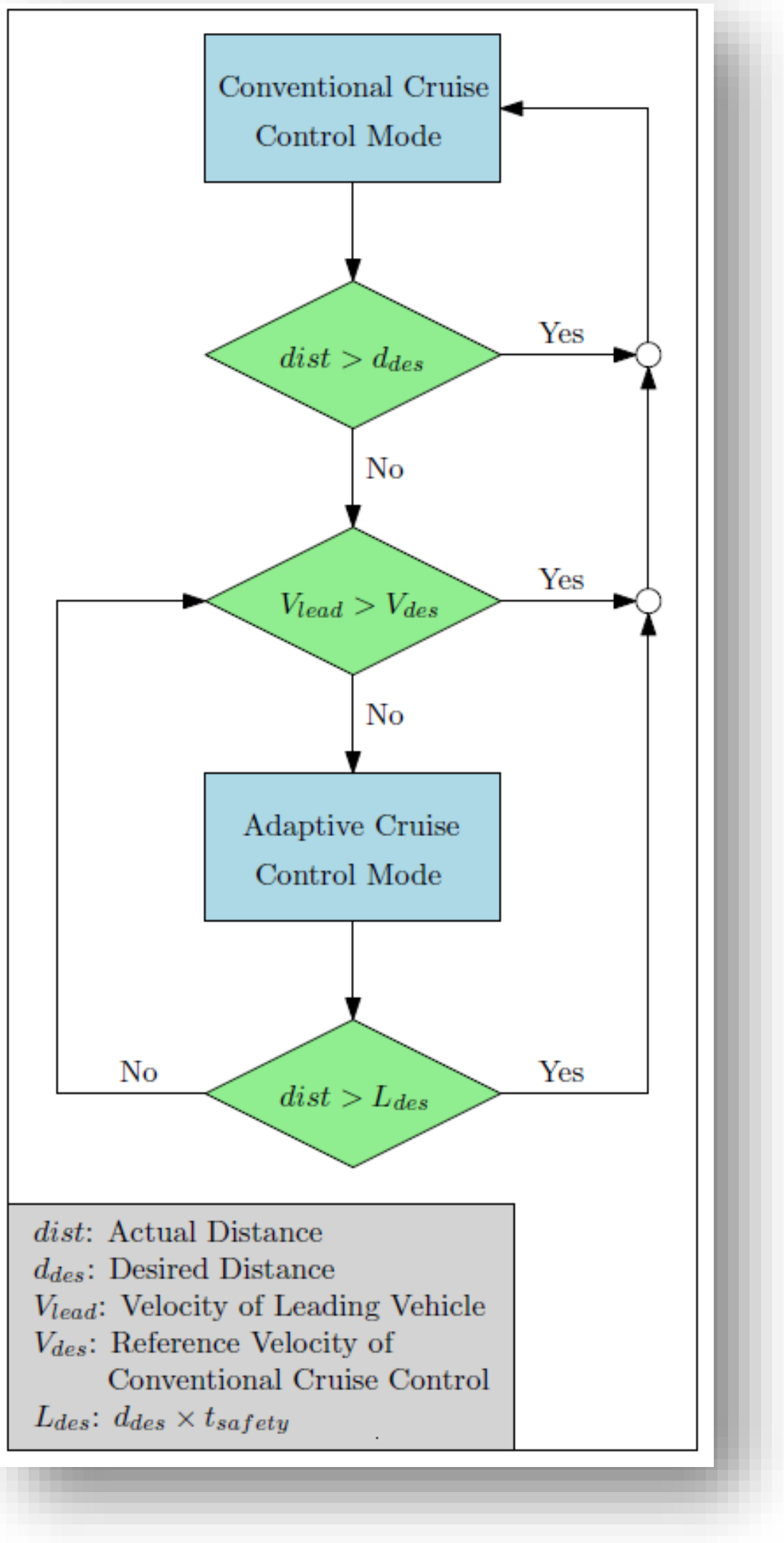

Fig. 19 Adaptive switching logic

## Simulation results

In this section, system performances of different controller, under different working scenarios, are studied and verified with *MATLAB/Simulink* and *Carsim*. Distance error, relative velocity and following vehicle acceleration are selected as the main criteria to judge the performance of different controllers. In the first subsection, LQR, ALQG, and MPC are selected. First, ideal scenarios where there is no noise are studies. Then, in real-situation scenarios, high frequency Gaussian white noise is introduced so that we can see how different controllers react to system noise. In both scenarios, preceding vehicle velocity is designed to vary in a sinusoid wave, which is shown in Fig.20. in the second subsection, comparisons between (1) LQR and ALQG and (2) MPC and ALQG are studied.

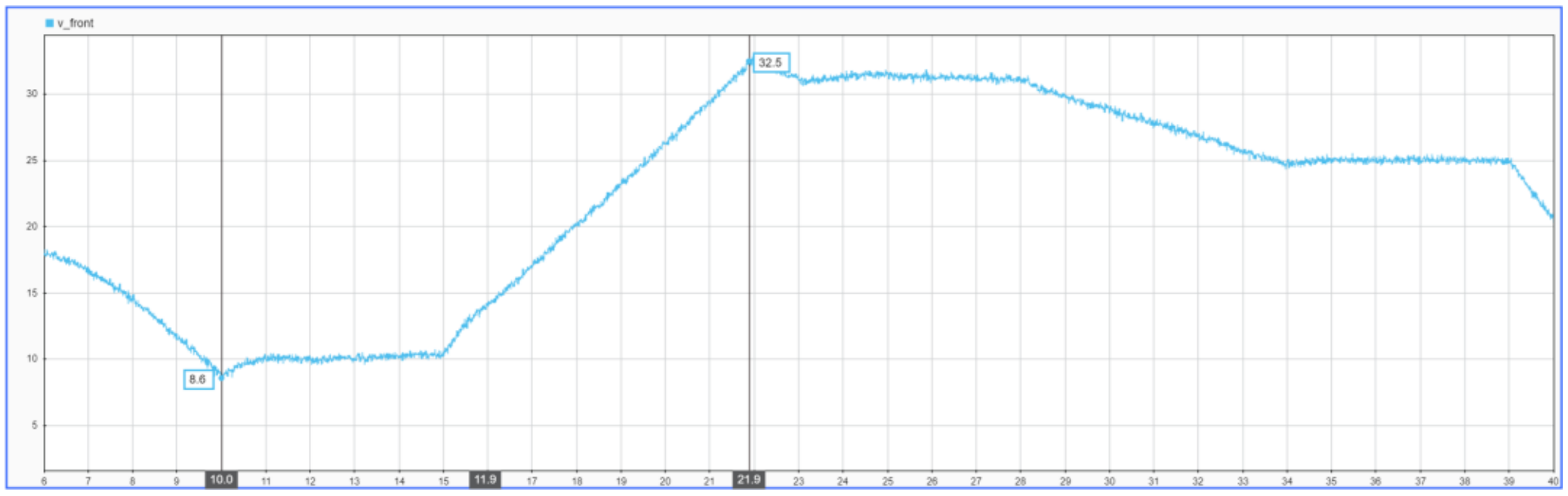

Fig.20: preceding vehicle velocity

## 1.1 Ideal and Real Scenarios

In ideal scenarios, noise is ignored. LQR, ALQG and MPC controllers are individually studied under the same initial condition. Distance error, relative velocity and following vehicle velocity are separately displayed in three figures. Fig.21 is system performance of LQR. Fig.22 is system performance of MPC, and Fig.23 is system performance of ALQG.

(a) distance error

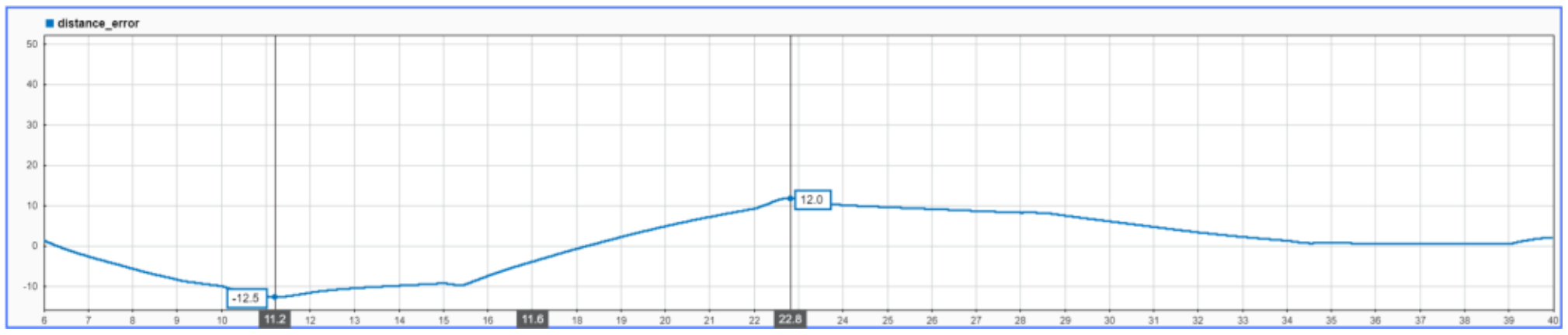

(b) following vehicle acceleration

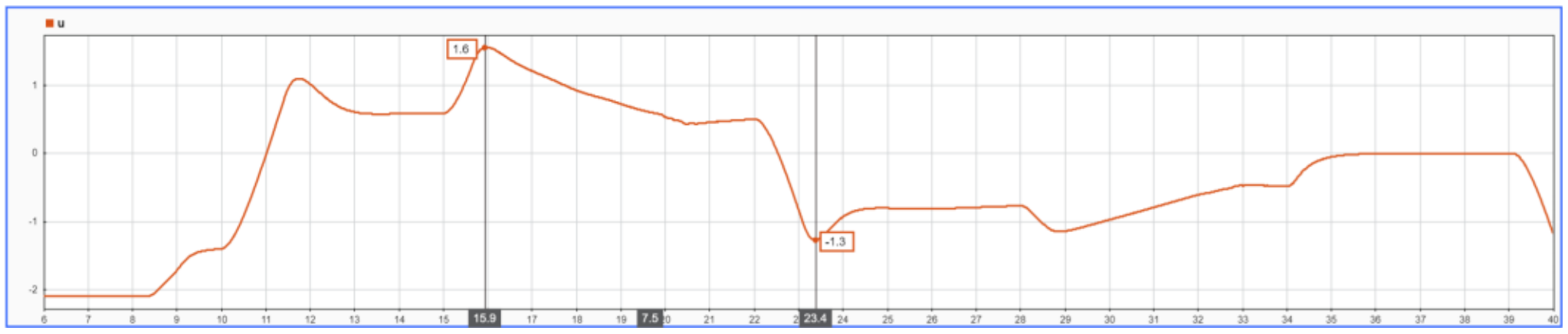

(c) relative velocity

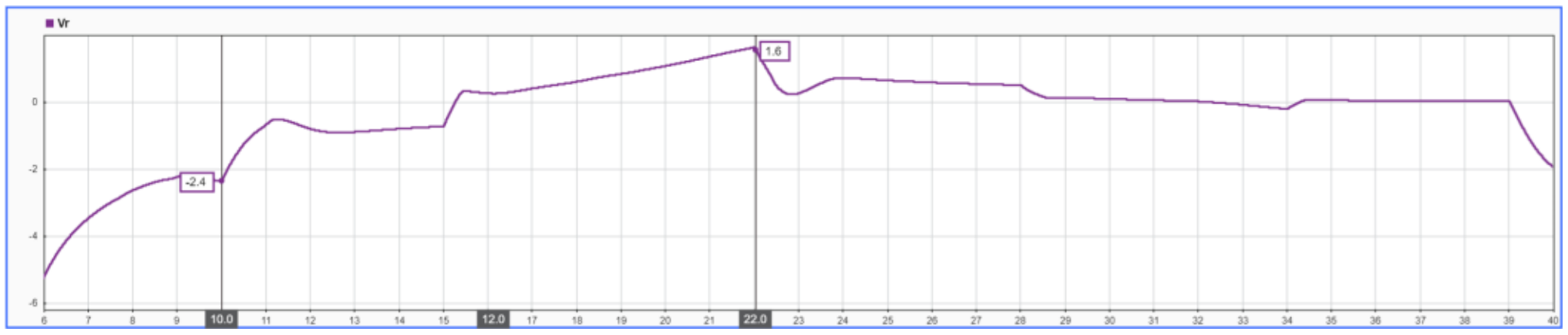

Fig.21: System performance of LQR under Ideal Scenarios: (a) distance error; (b) following vehicle acceleration; (c) relative velocity

(a) distance error

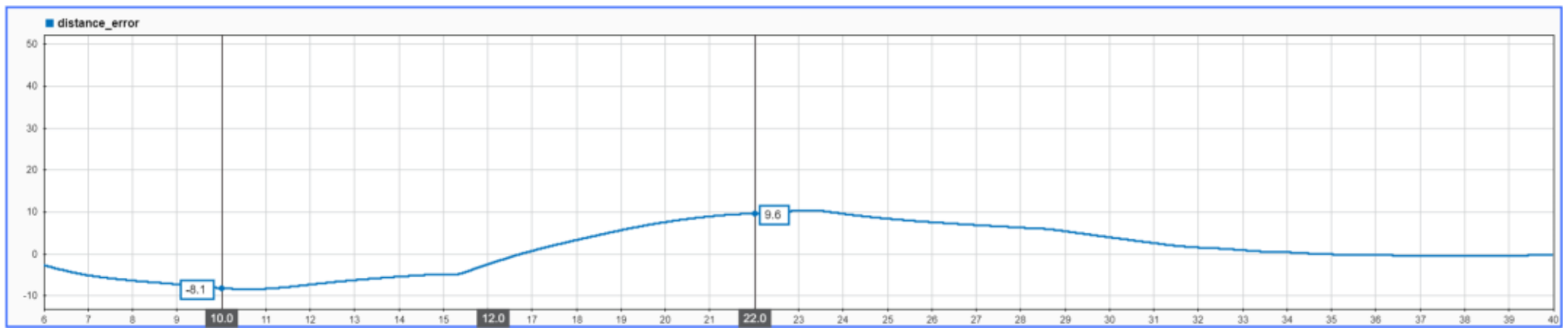

(b) following vehicle acceleration

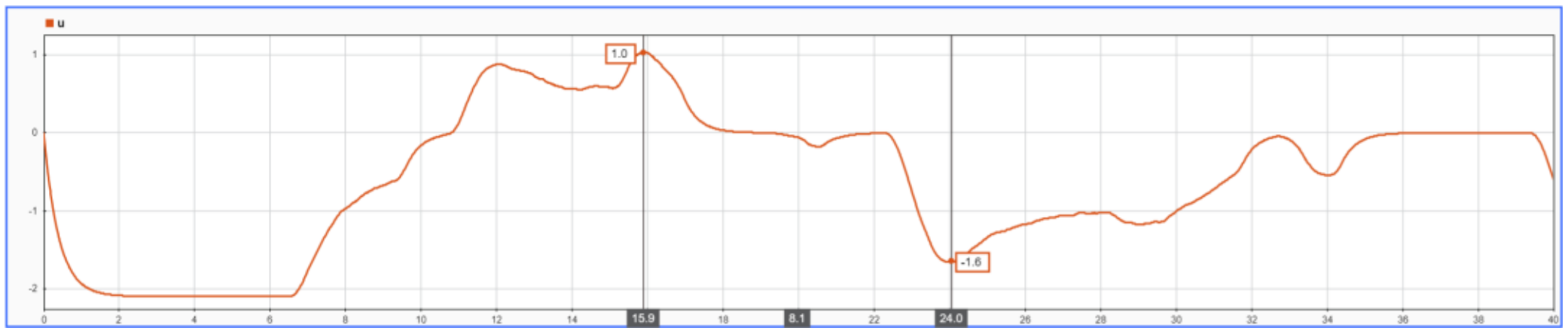

(c) relative velocity

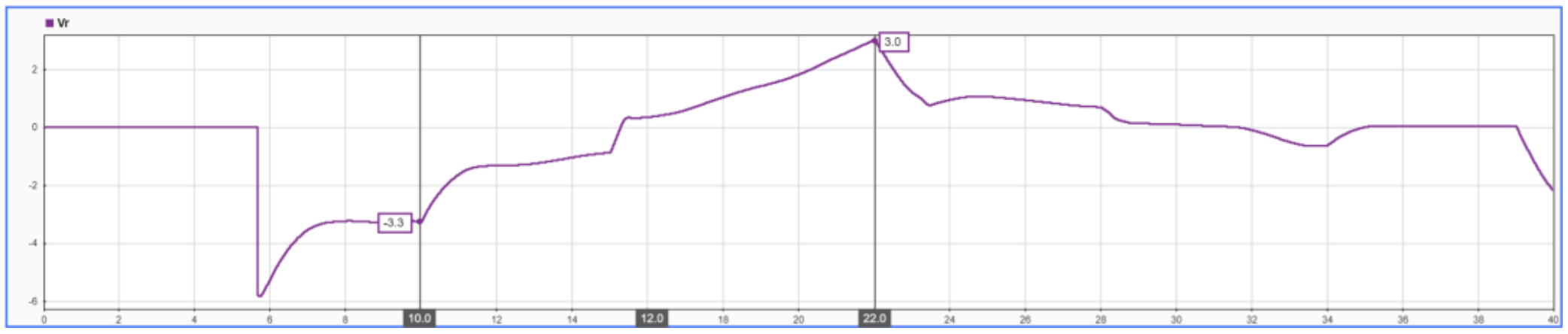

Fig.22: System performance of MPC under Ideal Scenarios: (a) distance error; (b) following vehicle acceleration; (c) relative velocity

(a) distance error

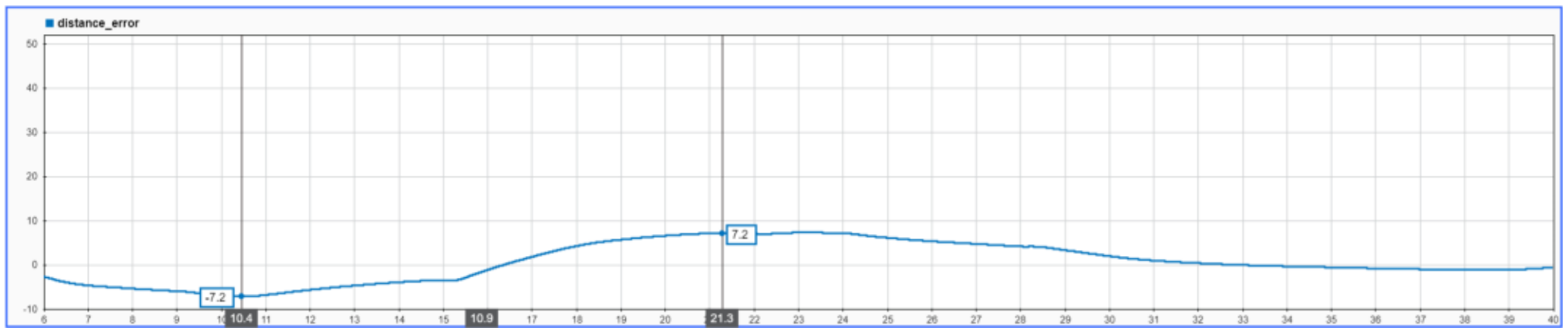

(b) following vehicle acceleration

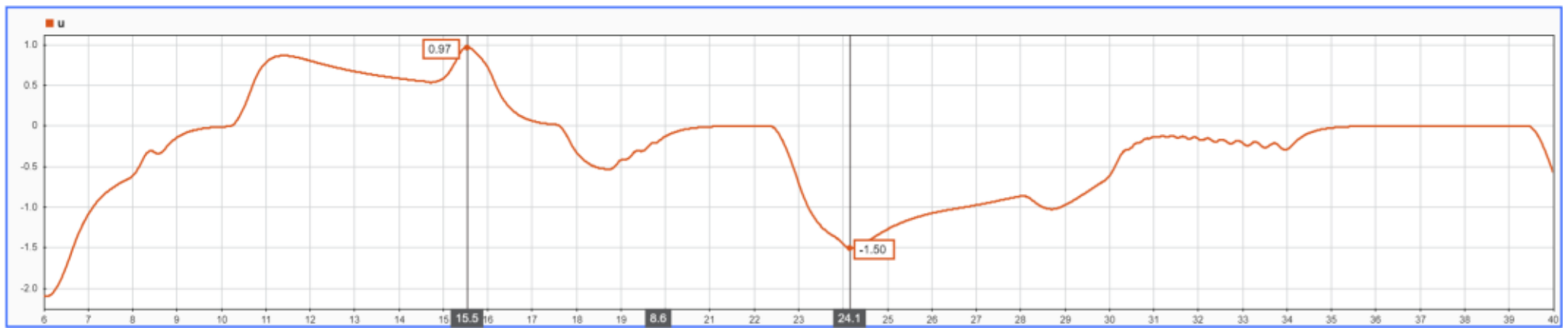

(c) relative velocity

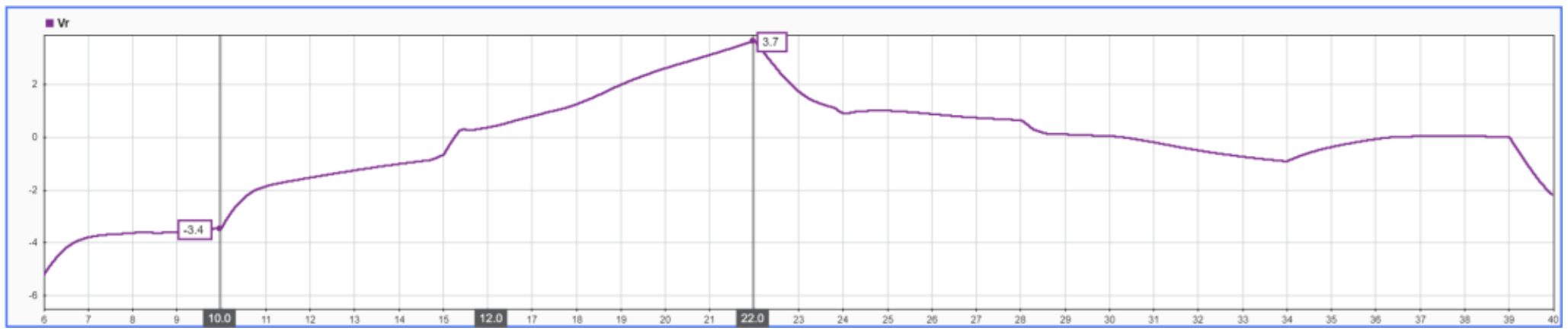

Fig.23 System performance of ALQG under Ideal Scenarios: (a) distance error; (b) following vehicle acceleration; (c) relative velocity

Similarly, under real scenarios where noise is introduced, system performances of the three controllers are also separately studied. Due to the high frequency noise that we deliberately introduced, following vehicle acceleration varies more frequently when comparing with the ideal scenarios. Fig.24 shows system performance of LQR, Fig.25 MPC, and Fig.26 ALQG. According to Fig.24, LQR cannot well tackle with disturbance, following vehicle acceleration is interfered with noise. According to Fig.25 and Fig.26, even though MPC, to some extent, compensate for noise effect by prediction, ALQG still has better performance by using Kalman filter and automatically tuning parameters.

(a) distance error

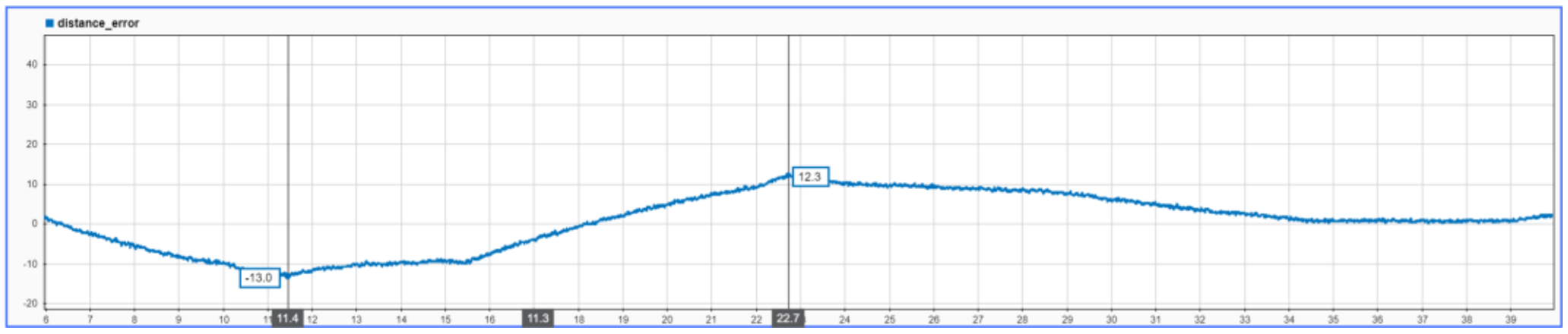

(b) following vehicle acceleration

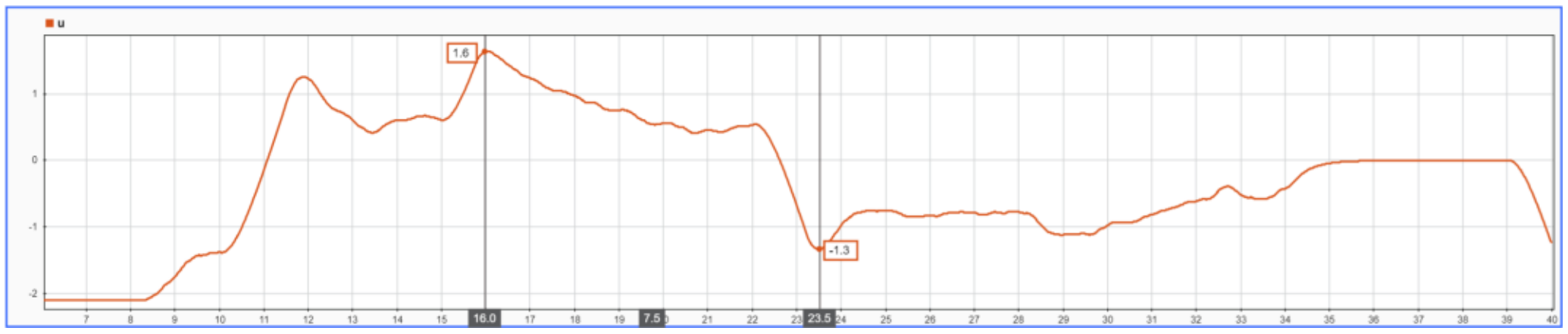

(c) relative velocity

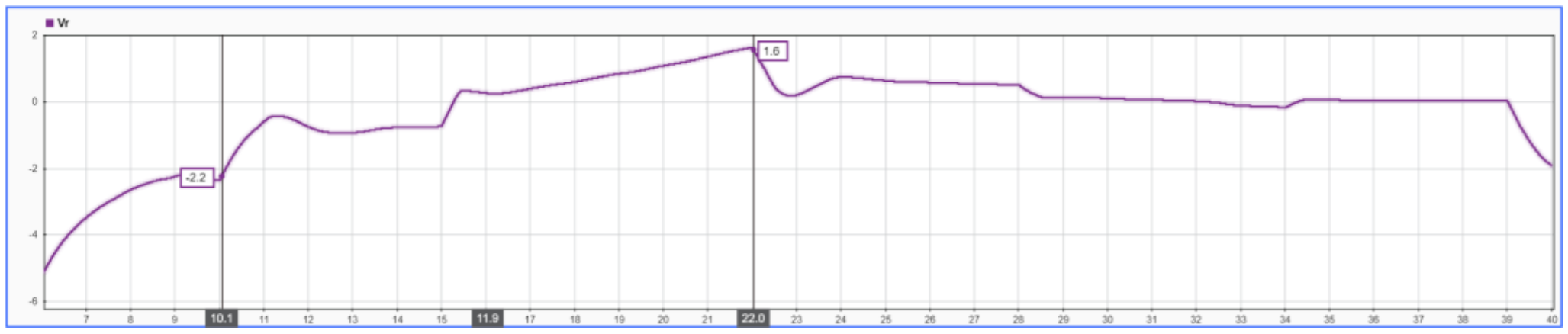

Fig.24: System performance of LQR under Real Scenarios: (a) distance error; (b) following vehicle acceleration; (c) relative velocity

(a) distance error

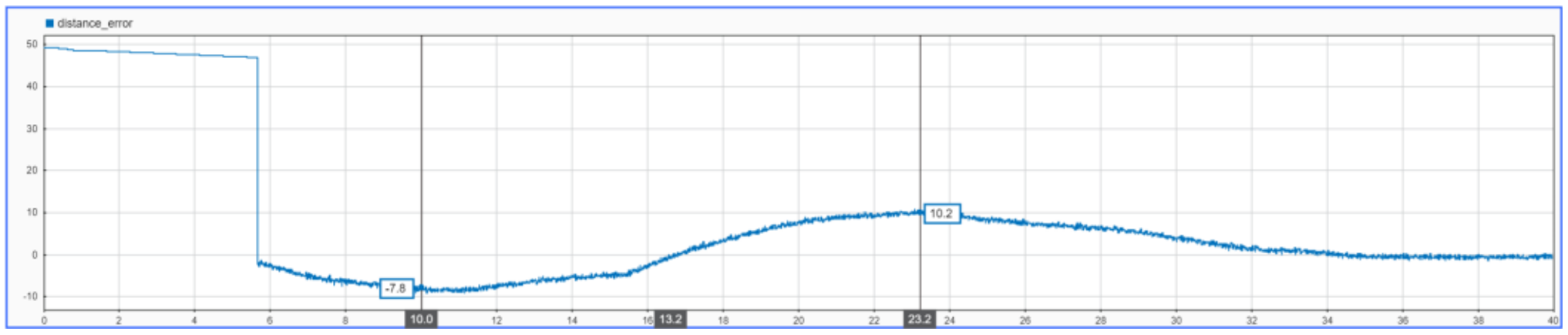

(b) following vehicle acceleration

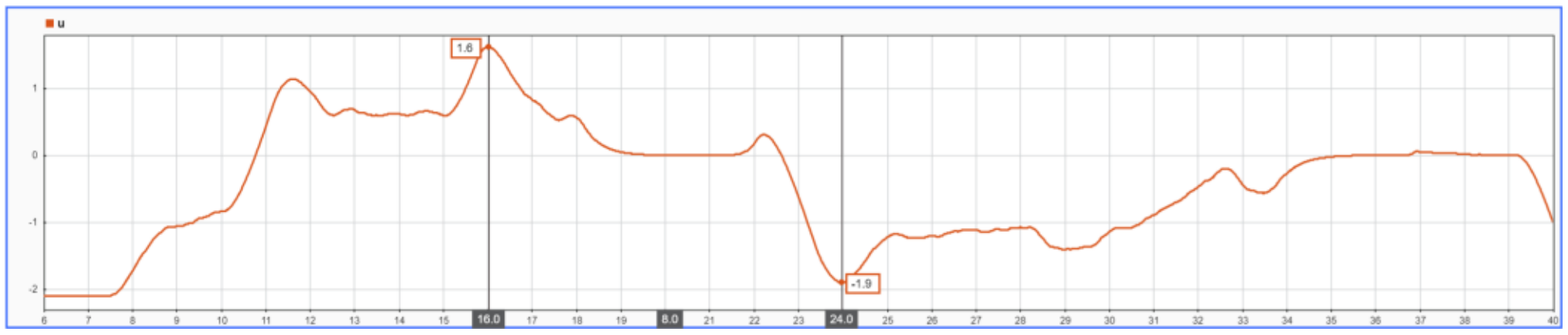

(c) relative velocity

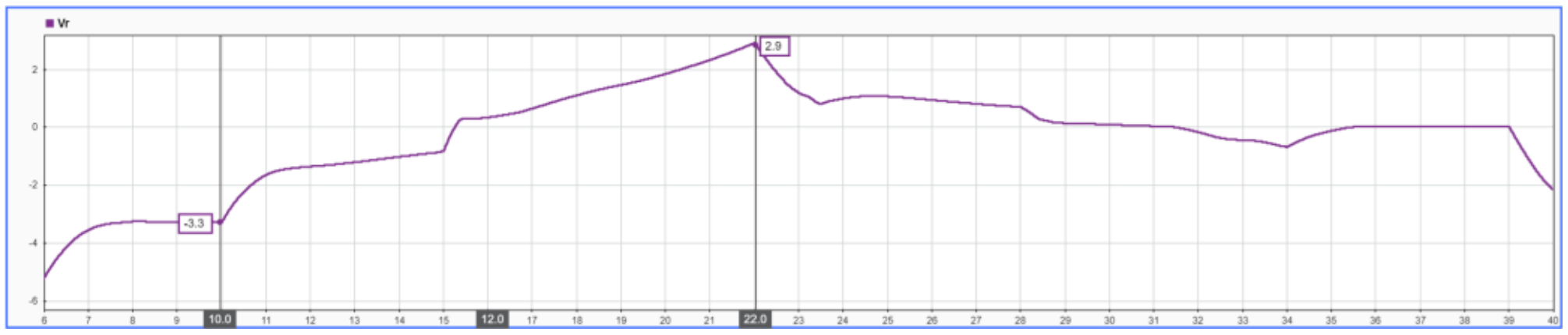

Fig.25: System performance of MPC under Real Scenarios: (a) distance error; (b) following vehicle acceleration; (c) relative velocity

(a) distance error

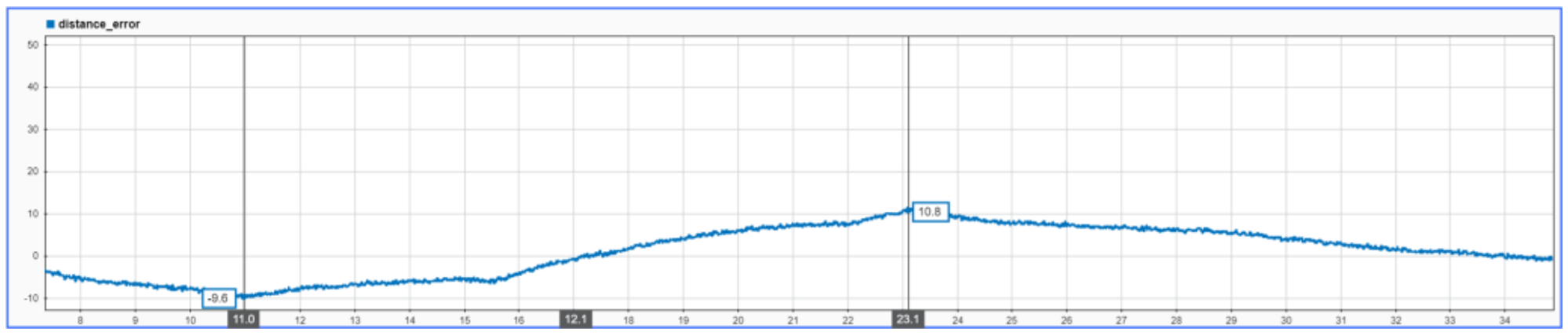

(b) following vehicle acceleration

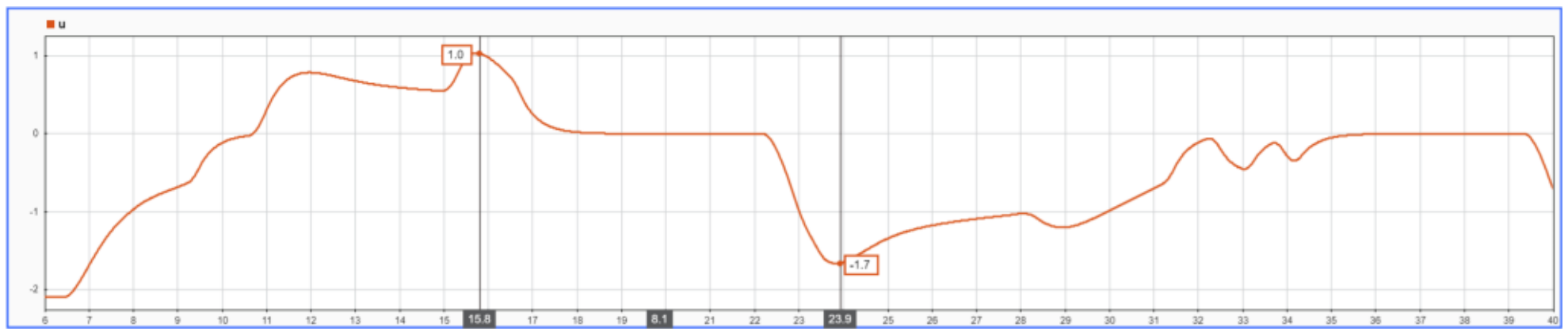

(c) relative velocity

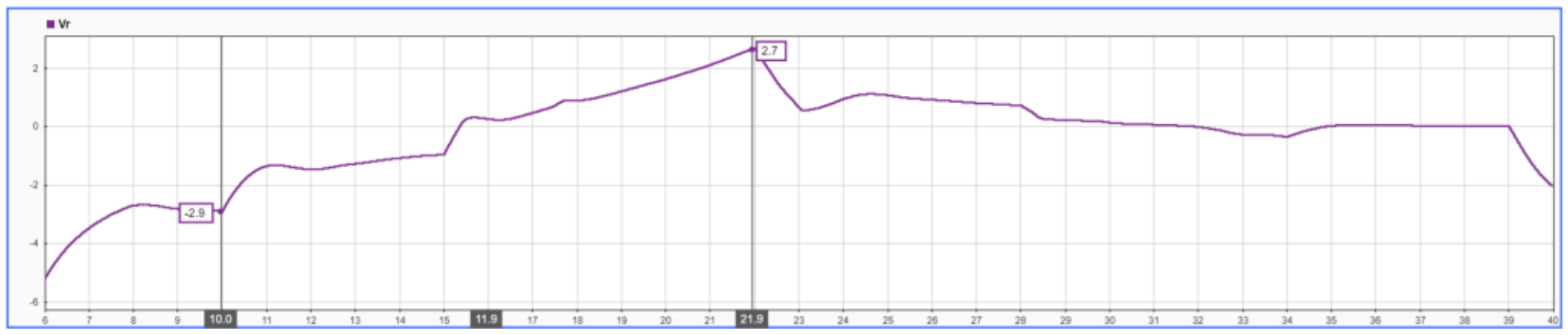

Fig.26 System performance of ALQG under Real Scenarios: (a) distance error; (b) following vehicle acceleration; (c) relative velocity.

## 1.2 Comparison of Different Controllers

In this part, only real scenarios with noise are selected for comparison. We first compare system performance between LQR and ALQG in Fig.11, then compare system performance between MPC and ALQG in Fig.12.

(a) distance error

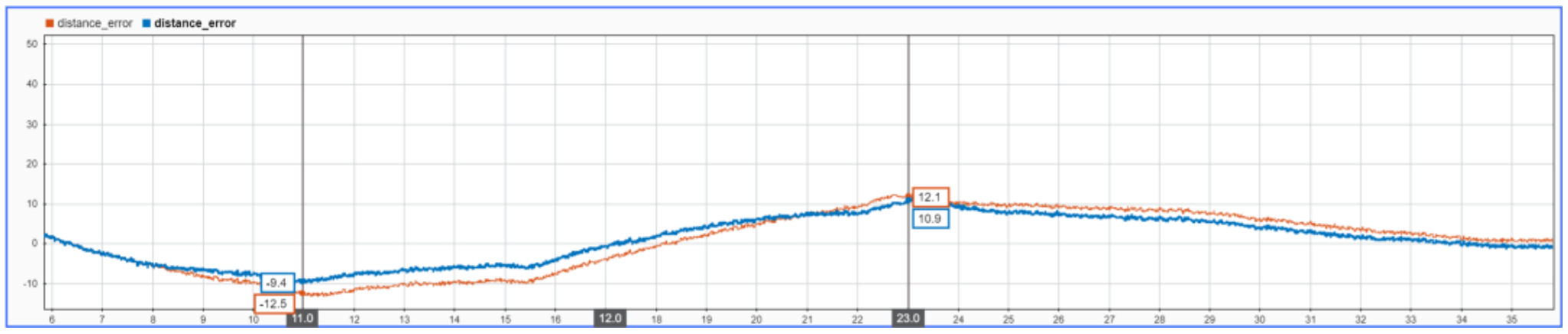

(b) following vehicle acceleration

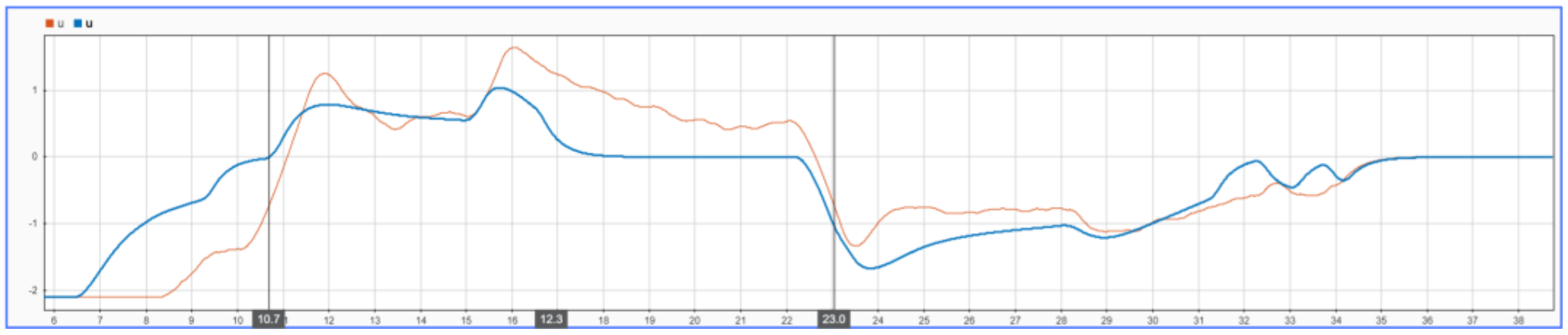

(c) relative velocity

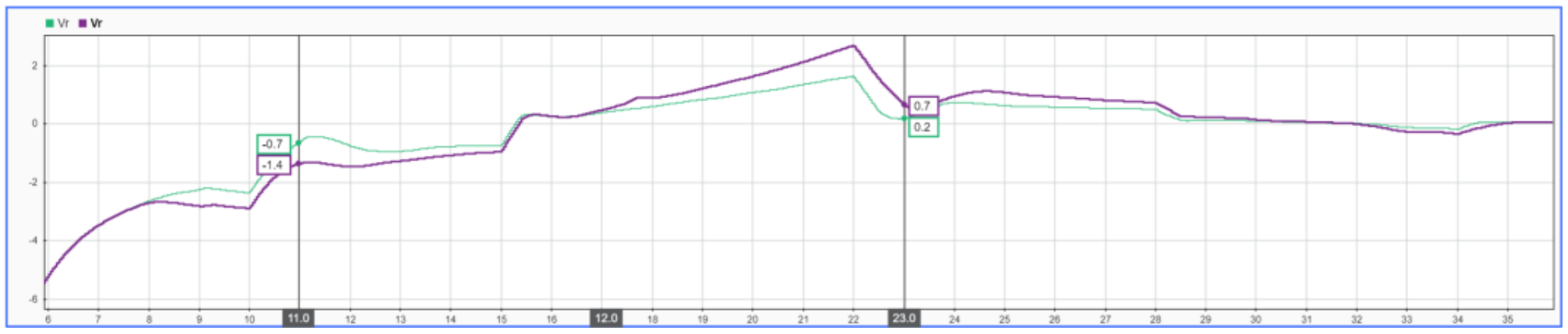

Fig. 27: system performance between LQR and ALQG: (a) distance error; (b) following vehicle acceleration; (c) relative velocity

In Fig.27(a), ALQG has less distance error than LQR. As preceding vehicle velocity varies, ALQG can quickly responds to the distance error, and constrains the error in a relatively small range. Basically, ALQG has 2m less distance error than LQR. In terms of following vehicle acceleration shown in Fig.27(b), ALQG also has better performance. The absolute value of following vehicle acceleration and acceleration change rate of ALQG are smaller than that of LQR. Meanwhile, due to Kalman filter, the effect of real situation noise is greatly reduced. Therefore, the following vehicle acceleration of ALQG is much smoother than LQR. Basically, by using ALQG, the following vehicle acceleration is better constrained around zero. As can be seen in the figure, the maximum acceleration value by using ALQG is around 0.1*g*, while the maximum value by using LQR reaches up to 0.16*g*. As for relative velocity shown in Fig.27(c), LQR, in most of the time, is at least twice as big as ALQG.

(a) distance error

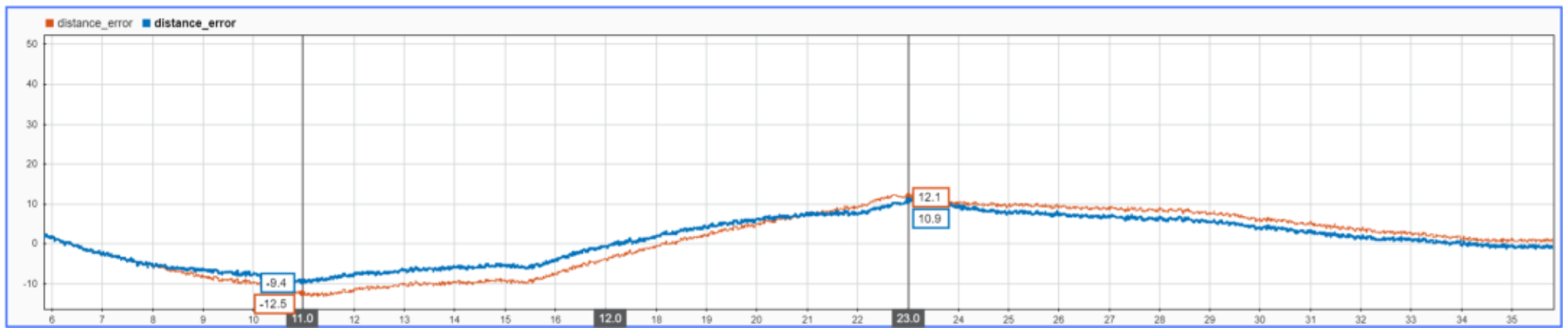

(b) following vehicle acceleration

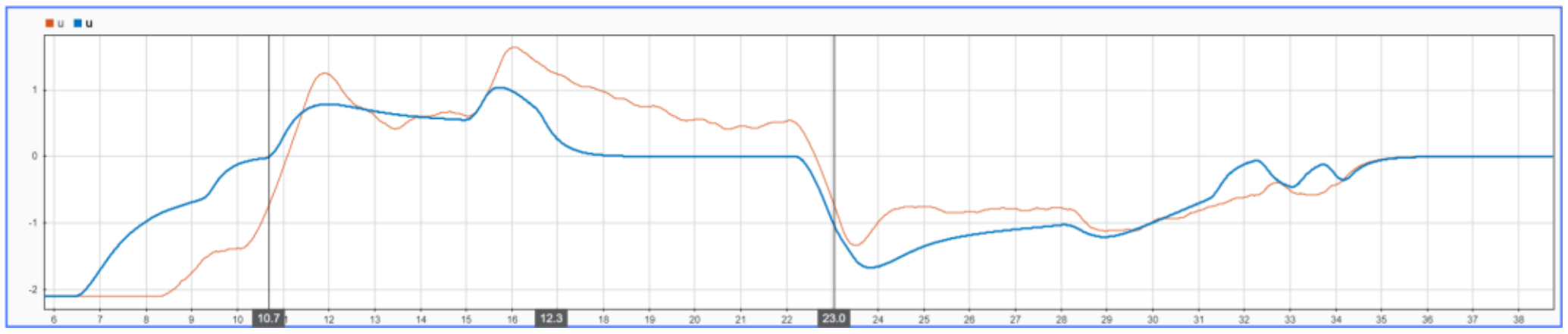

(c) relative velocity

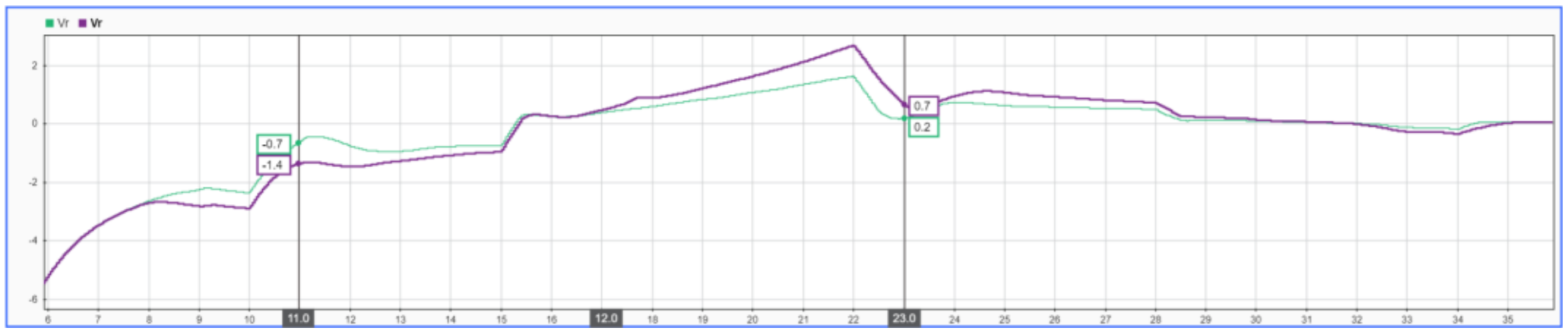

Fig. 28: system performance between MPC and ALQG: (a) distance error; (b) following vehicle acceleration; (c) relative velocity

In Fig.28(a), ALQG still has slightly better performance than MPC, but usually, ALQG has only 1m less distance error than MPC. In terms of following vehicle acceleration shown in Fig.28(b), MPC output is very similar with the output of ALQG. The value of following vehicle acceleration of ALQG and MPC are almost in the same range. But it is obvious that MPC has bigger overshoots. According to Fig.28(b), at time 11.5s and 16.0s, there are two obvious overshoots in MPC. While in ALQG, at the same time period, the following vehicle acceleration varies much smoother. As for relative velocity, ALQG also keeps a relatively smaller value than MPC.

## Vehicle test results

According to the simulation results, we decide that ALQG controller is used as ACC upper controller, and incremental PID controller is used as ACC lower controller. Based on this structure, we also applied above mentioned switching logics to design ACC lower controller. After considering some other issues like signal diagnose, signal backup, anti-windup, saturation, we finally designed an ACC system for vehicle test. The follow is the results of vehicle test. Typical scenarios are designed for testing, including:

Table.3 vehicle test scenarios

| ***classification*** | ***description*** |
|---|---|
| #1 Host vehicle cruise control | Moving at 30kph |
| | Moving at 40kph |
| | Moving at 50kph |
| | Moving at 60kph |
| #2 State shifting from CCS to ACC | Moving at 30kph, preceding vehicle cutting in at 40kph |
| | Moving at 40kph, preceding vehicle cutting in at 30kph |
| | Moving at 50kph, preceding vehicle cutting in at 40kph |
| #3 Preceding vehicle decelerate | Steady tracking at 30kph, then the preceding vehicle brakes till stops |
| | Steady tracking at 40kph, then the preceding vehicle brakes till stops |
| #4 Stop&Go mode | Steady tracking at 30kph, then the preceding vehicle brakes till stops, then starts again |
| | Steady tracking at 40kph, then the preceding vehicle brakes till stops, then starts again |
| #5 State shifting from ACC to CCS | Steady tracking at 30kph, then preceding vehicle disappears, and the host vehicle goes to cruise control mode at 40kph |
| | Steady tracking at 40kph, then preceding vehicle disappears, and the host vehicle goes to cruise control mode at 50kph |

Fig. x shows the test result of scenario where the host vehicle moves at a speed of 30kph, then a preceding vehicle cuts in at the speed of 40kph. Here we still consider the three parameters: distance error, relative velocity, and host vehicle acceleration to judge the performance of ACC upper controller. Meanwhile, we also have a look at the volt and pressure to see whether the switching logic between throttle and brake works. As can been seen in the figure, distance error is constrained around 0.8m, relative velocity is constrained between 0 to 2m/s, and host vehicle acceleration is around 0. The vehicle testing result is very similar to simulation results, and the outcome is satisfactory. As for the switching logic, volt and braking pressure do not appear simultaneously, and shifting state vibration is also avoided successfully. We can see in the figure that there is no frequently shifting between pressure and volt. We apply the same analysis rules to the other figures, and we can conclude that under different traffic scenarios, the ACC system can always deliver good system performance.

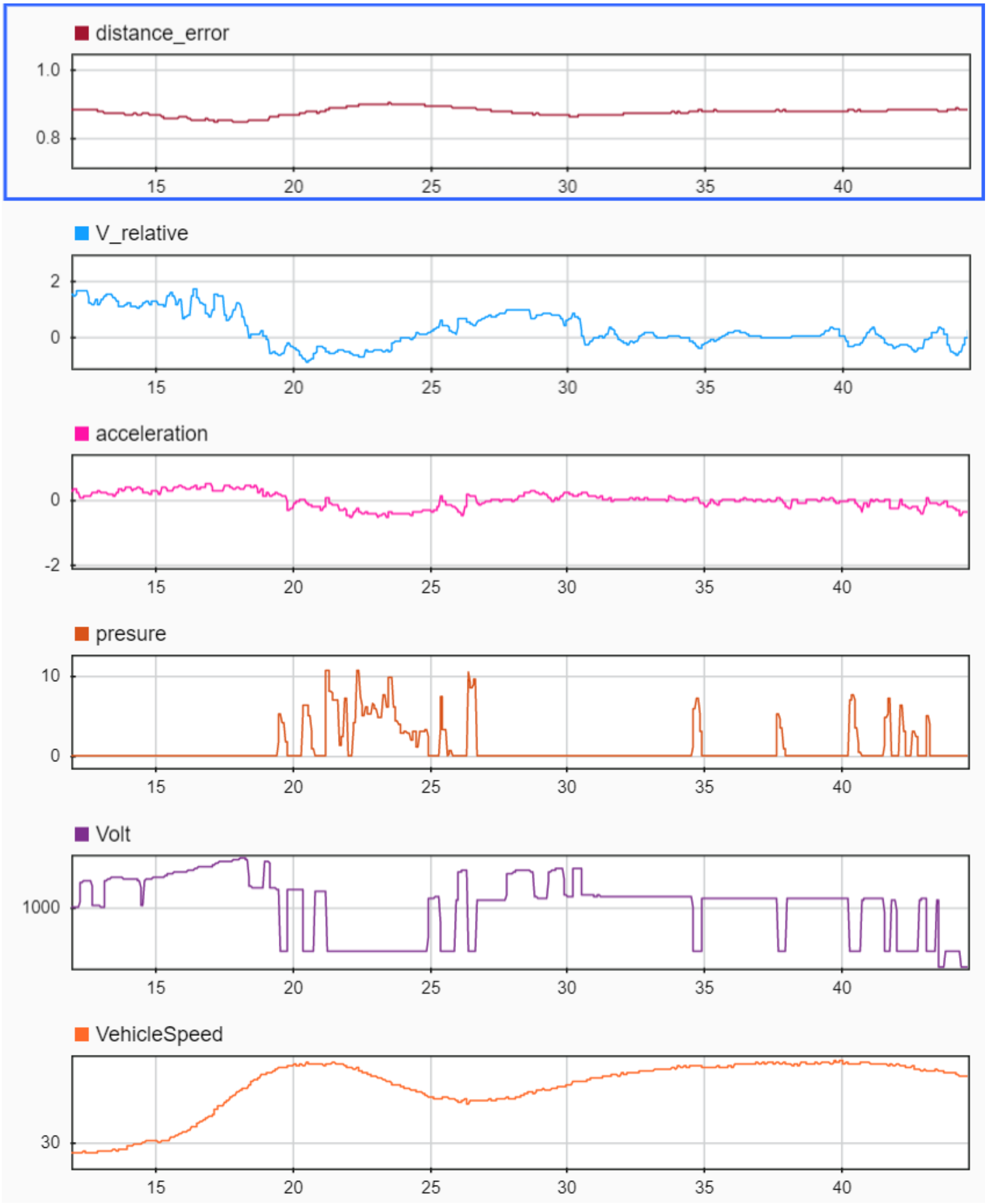

Fig.29 vehicle test #2-1

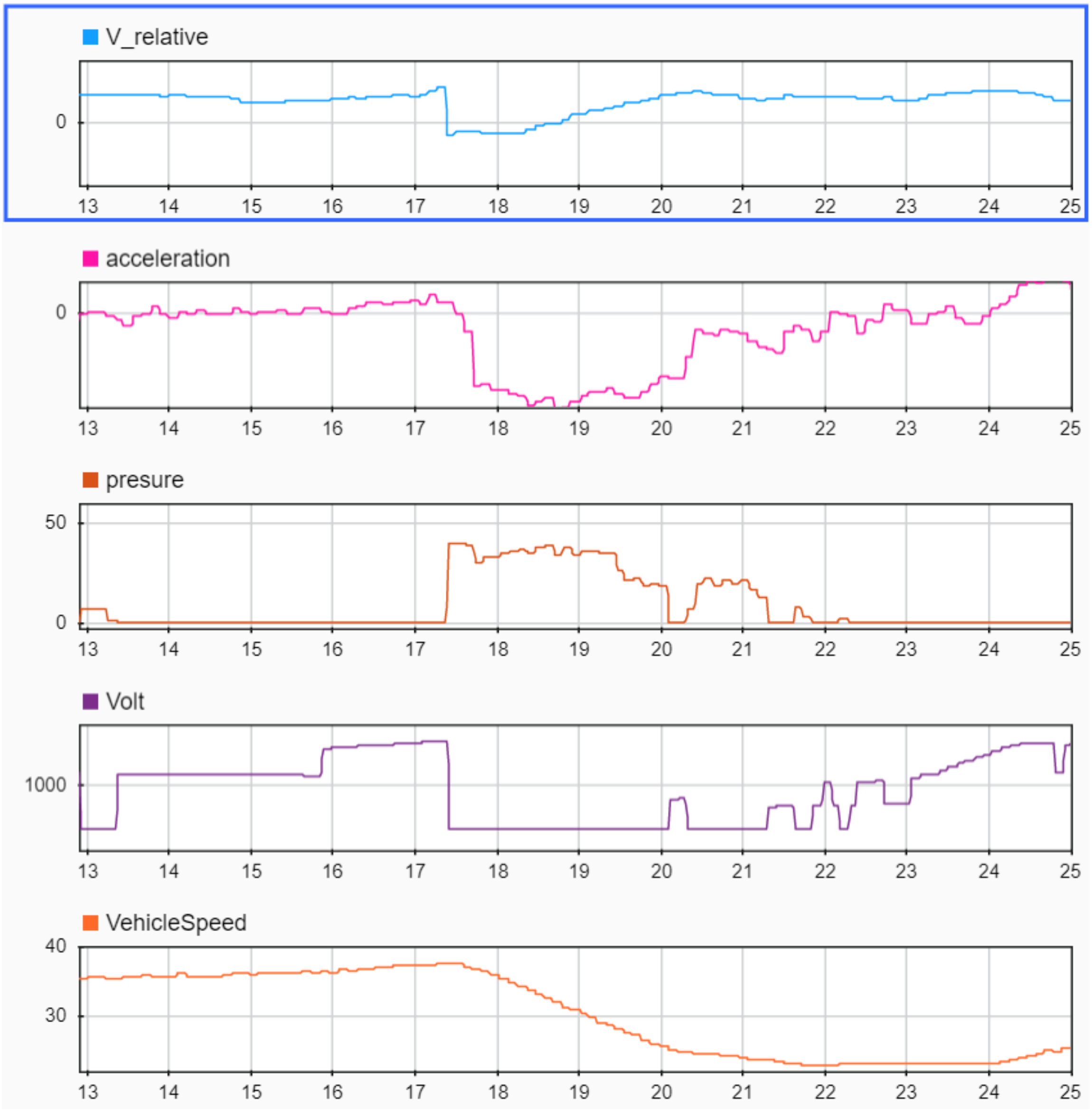

Fig. 30 vehicle test #2-2

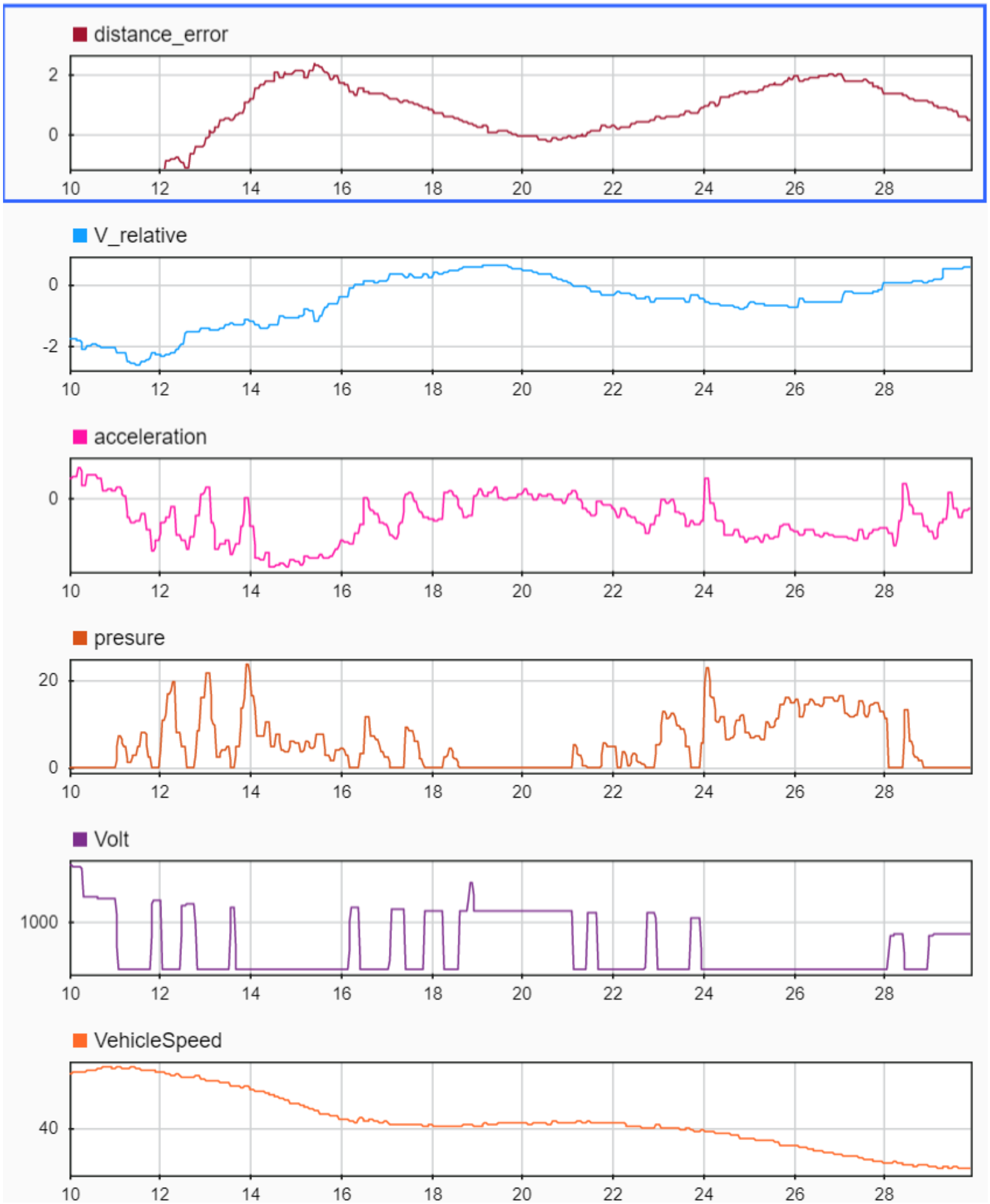

Fig.31 vehicle test #2-3

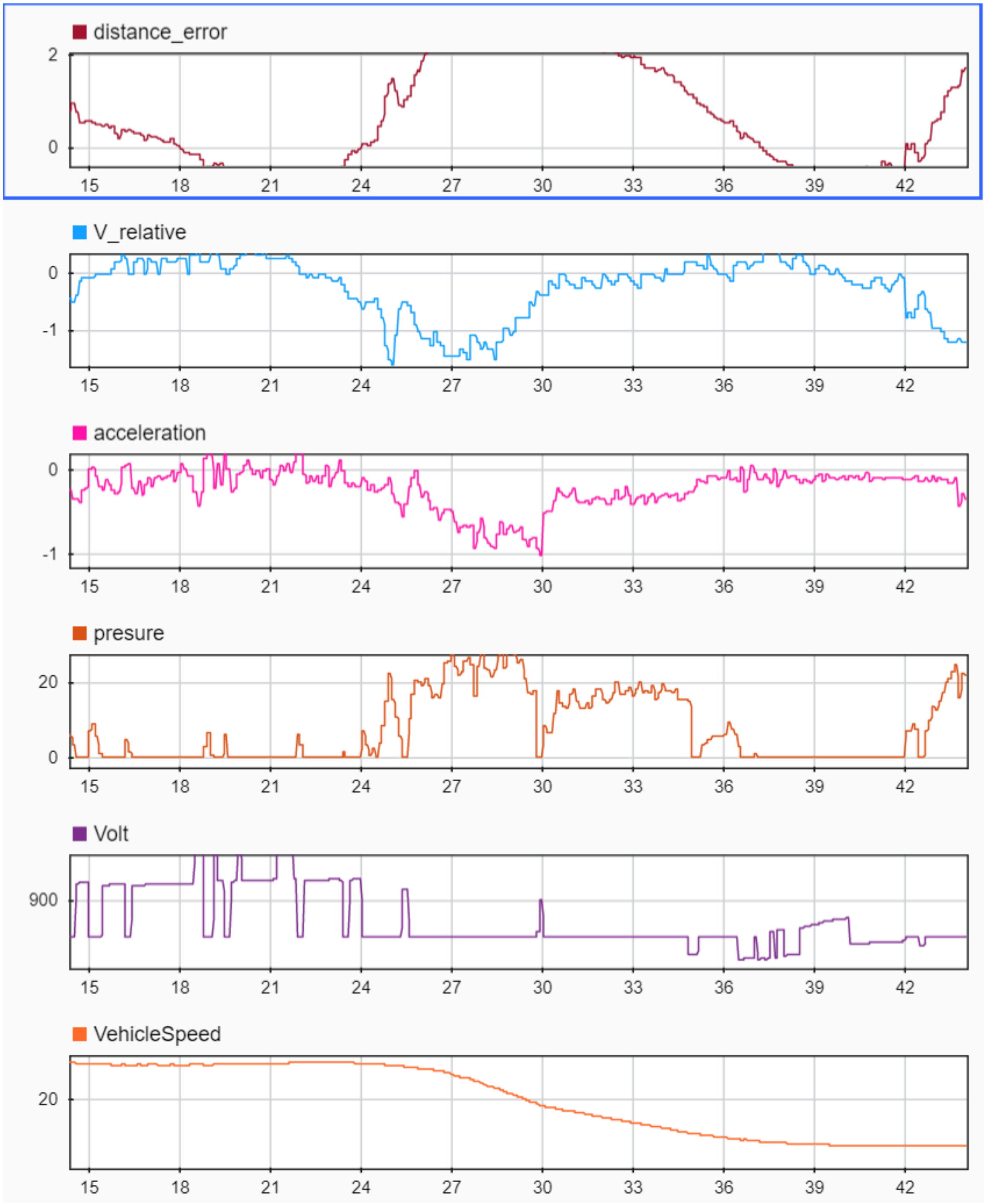

Fig. 32 vehicle test #3-1

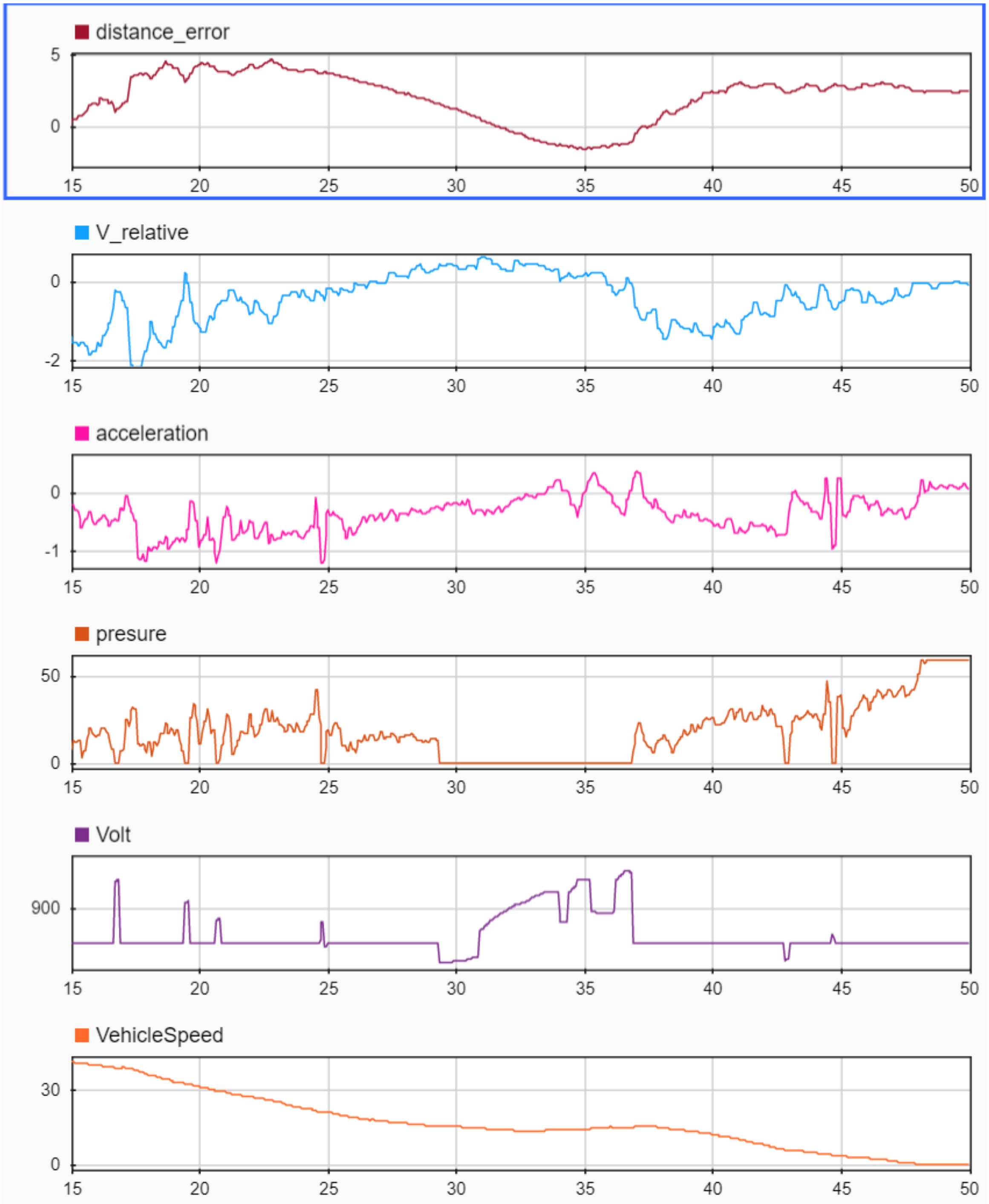

Fig. 33 vehicle test #3-2

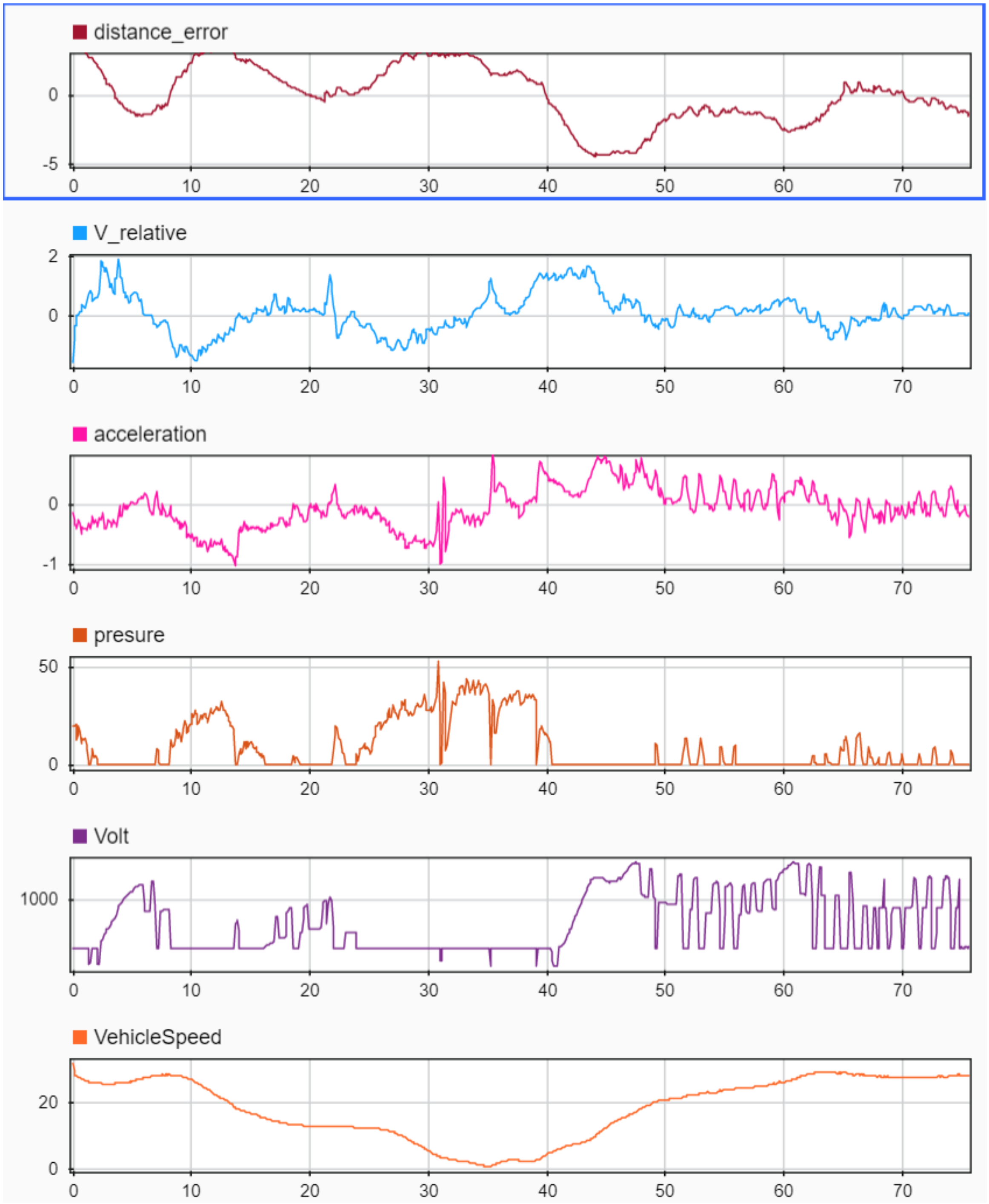

Fig. 34 vehicle test #4-1

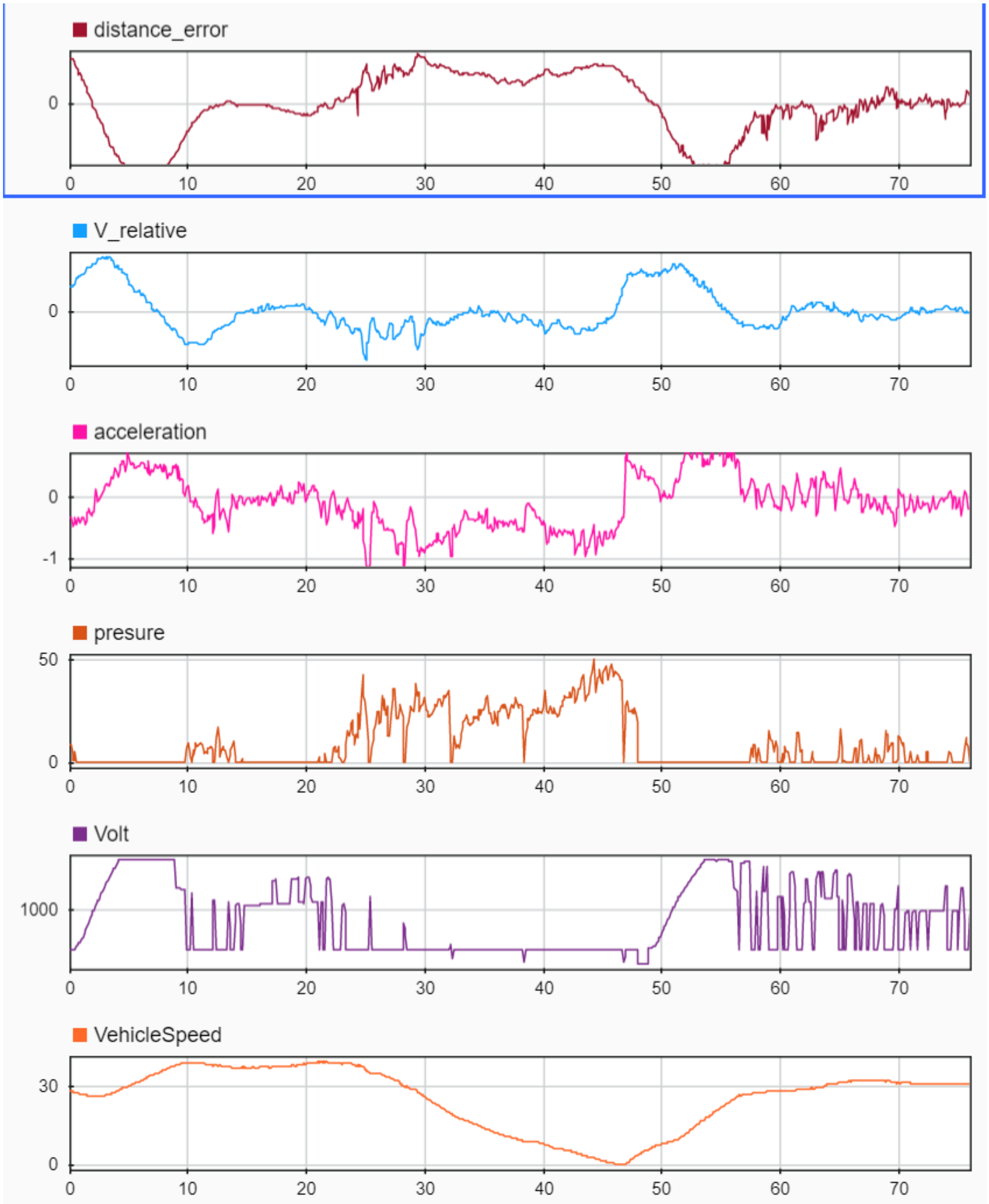

Fig. 35 vehicle test #4-2

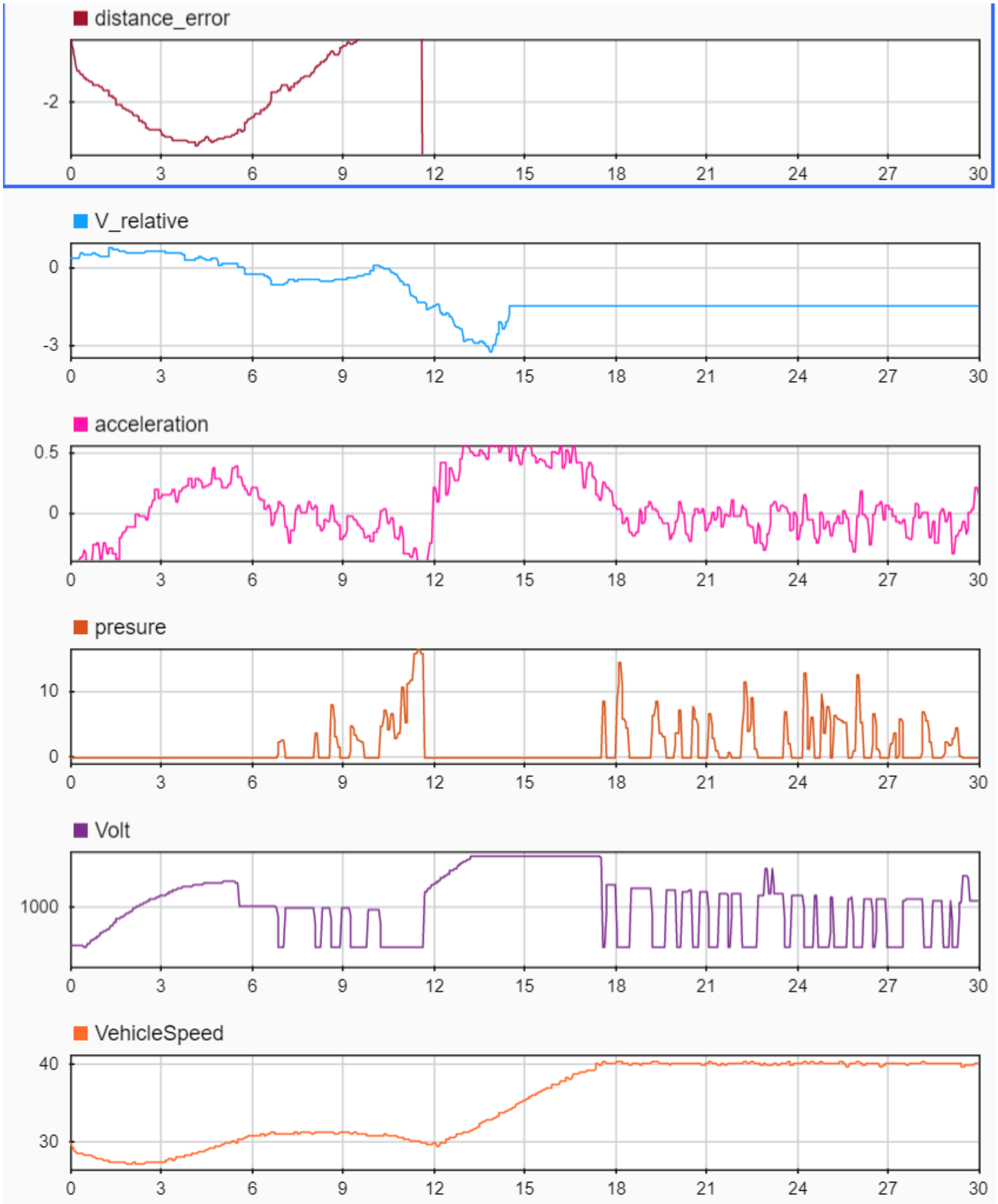

Fig. 36 vehicle test #5-1

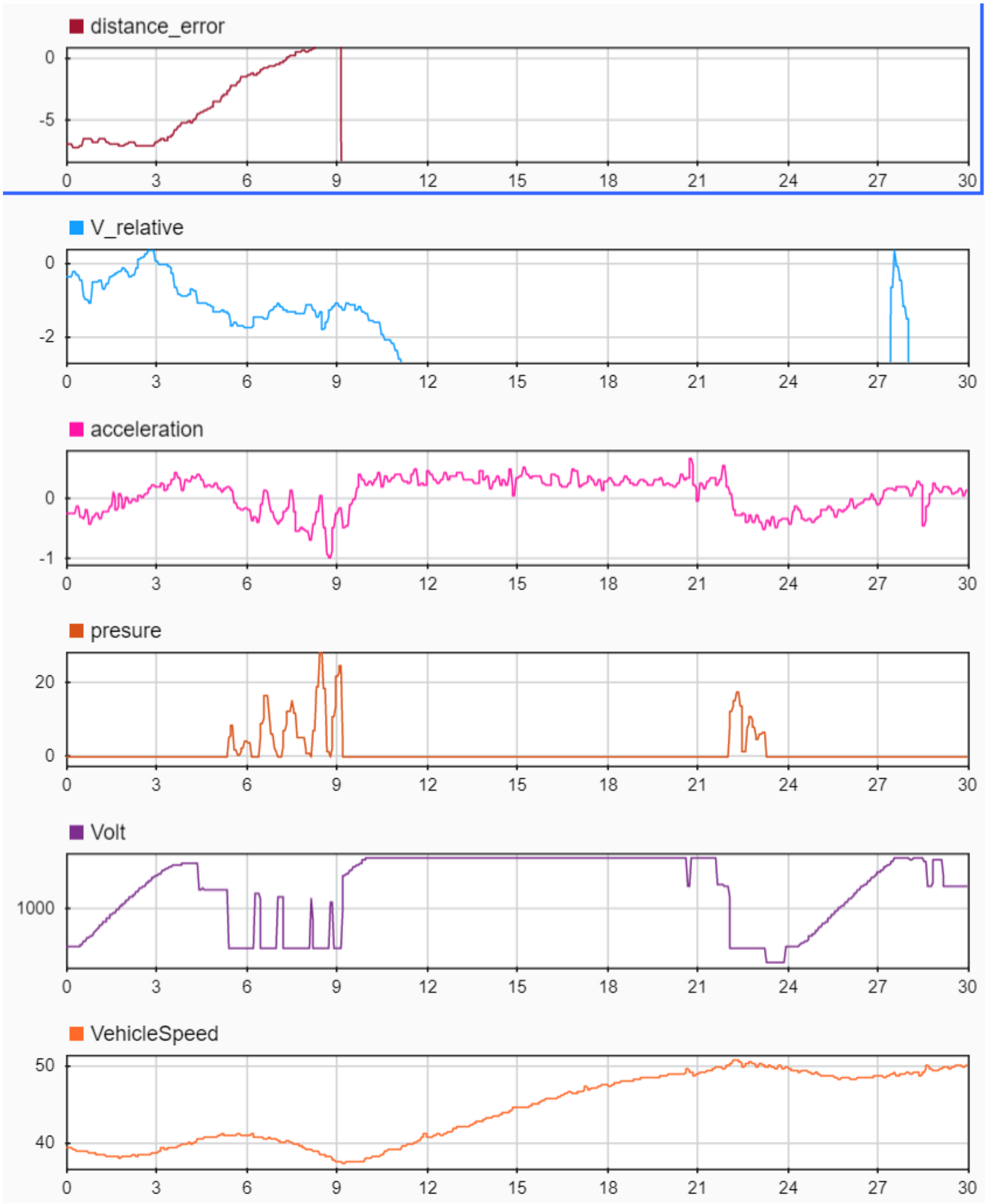

Fig. 37 vehicle test #5-2